\documentclass[preprint,tightenlines,a4paper,aps,prd,superscriptaddress,nofootinbib,amssymb,amsmath,alignedleftspaceno]{revtex4-2} 
\pdfoutput=1
\usepackage[utf8]{inputenc}
\usepackage[T1]{fontenc}
\usepackage[USenglish]{babel}
\usepackage[group-digits=integer,quantity-product=\ ,uncertainty-descriptors={stat, syst}]{siunitx}
\usepackage{csquotes}
\usepackage[all]{foreign}
\defasforeign[ansaetze]{ansätze}
\renewcommand{\cf}{\xperiodafter{{\foreignabbrfont{cf}}}}

\usepackage{tikz}
\usepackage{braket}
\DeclareMathOperator{\tr}{tr}

\newcommand{\pd}[2]
{
    \frac{\partial #1}{\partial #2}
}
\usepackage[pdfusetitle,pdfauthor={Miguel Salg},bookmarksopen=true,bookmarksopenlevel=1]{hyperref}
\usepackage[english]{cleveref}
\usepackage{graphicx}
\usepackage{tikz}
\usetikzlibrary{decorations}
\pgfdeclaredecoration{complete sines}{initial}
{
    \state{initial}[
        width=+0pt,
        next state=sine,
        persistent precomputation={\pgfmathsetmacro\matchinglength{
            \pgfdecoratedinputsegmentlength / int(\pgfdecoratedinputsegmentlength/\pgfdecorationsegmentlength)}
            \setlength{\pgfdecorationsegmentlength}{\matchinglength pt}
        }] {}
    \state{sine}[width=\pgfdecorationsegmentlength]{
        \pgfpathsine{\pgfpoint{0.25\pgfdecorationsegmentlength}{0.5\pgfdecorationsegmentamplitude}}
        \pgfpathcosine{\pgfpoint{0.25\pgfdecorationsegmentlength}{-0.5\pgfdecorationsegmentamplitude}}
        \pgfpathsine{\pgfpoint{0.25\pgfdecorationsegmentlength}{-0.5\pgfdecorationsegmentamplitude}}
        \pgfpathcosine{\pgfpoint{0.25\pgfdecorationsegmentlength}{0.5\pgfdecorationsegmentamplitude}}
    }
    \state{final}{}
}
\usepackage{longtable}
\usepackage[protrusion=true,expansion,tracking,kerning]{microtype}
\microtypecontext{spacing=nonfrench}
\newdimen\defaultaddspace
\begin{document}
\title{Electromagnetic form factors of the nucleon from \texorpdfstring{$N_f = 2 + 1$}{Nf=2+1} lattice QCD}
\author{Dalibor Djukanovic}
 \affiliation{Helmholtz Institute Mainz, Staudingerweg 18, 55128 Mainz, Germany}
 \affiliation{GSI Helmholtzzentrum für Schwerionenforschung, 64291 Darmstadt, Germany}
\author{Georg von Hippel}
 \affiliation{\texorpdfstring{PRISMA${}^+$}{PRISMA+} Cluster of Excellence and Institute for Nuclear Physics, Johannes Gutenberg University Mainz, Johann-Joachim-Becher-Weg 45, 55128 Mainz, Germany}
\author{Harvey B. Meyer}
 \affiliation{Helmholtz Institute Mainz, Staudingerweg 18, 55128 Mainz, Germany}
 \affiliation{\texorpdfstring{PRISMA${}^+$}{PRISMA+} Cluster of Excellence and Institute for Nuclear Physics, Johannes Gutenberg University Mainz, Johann-Joachim-Becher-Weg 45, 55128 Mainz, Germany}
\author{Konstantin Ottnad}
 \affiliation{\texorpdfstring{PRISMA${}^+$}{PRISMA+} Cluster of Excellence and Institute for Nuclear Physics, Johannes Gutenberg University Mainz, Johann-Joachim-Becher-Weg 45, 55128 Mainz, Germany}
\author{Miguel Salg}
 \email{msalg@uni-mainz.de}
 \affiliation{\texorpdfstring{PRISMA${}^+$}{PRISMA+} Cluster of Excellence and Institute for Nuclear Physics, Johannes Gutenberg University Mainz, Johann-Joachim-Becher-Weg 45, 55128 Mainz, Germany}
\author{Hartmut Wittig}
 \affiliation{Helmholtz Institute Mainz, Staudingerweg 18, 55128 Mainz, Germany}
 \affiliation{\texorpdfstring{PRISMA${}^+$}{PRISMA+} Cluster of Excellence and Institute for Nuclear Physics, Johannes Gutenberg University Mainz, Johann-Joachim-Becher-Weg 45, 55128 Mainz, Germany}
\begin{abstract}
    There is a long-standing discrepancy between different measurements of the electric and magnetic radii of the proton.
Lattice QCD calculations are a well-suited tool for theoretical investigations of the structure of the nucleon from first principles.
However, all previous lattice studies of the proton's electromagnetic radii have either neglected quark-disconnected contributions or were not extrapolated to the continuum and infinite-volume limit.
Here, we present results for the electromagnetic form factors of the proton and neutron computed on the $(2 + 1)$-flavor Coordinated Lattice Simulations (CLS) ensembles including both quark-connected and -disconnected contributions.
From simultaneous fits to the $Q^2$-, pion-mass, lattice-spacing, and finite-volume dependence of the form factors, we determine the electric and magnetic radii and the magnetic moments of the proton and neutron.
For the proton, we obtain as our final values $\langle r_E^2 \rangle^p = \qty{0.672(14)(18)}{fm^2}$, $\langle r_M^2 \rangle^p = \qty{0.658(12)(8)}{fm^2}$, and $\mu_M^p = \num{2.739(63)(18)}$.
The magnetic moment is in good agreement with the experimental value, as is the one of the neutron.
On the one hand, our result for the electric (charge) radius of the proton clearly points towards a small value, as favored by muonic hydrogen spectroscopy and the recent $ep$-scattering experiment by PRad.
Our estimate for the magnetic radius, on the other hand, is well compatible with that inferred from the A1 $ep$-scattering experiment.

\end{abstract}
\date{\today}
\preprint{MITP-23-044}

\maketitle
\newpage
\section{Introduction}
Despite the fact that protons and neutrons, collectively referred to
as nucleons, make up the largest fraction of the mass of the visible
matter in the universe \cite{Wilczek2012}, there are many open
problems relating to their internal structure, which are the subject
of a broad research effort in subatomic physics.
In particular, the question whether discrepant measurements of the
electric and magnetic radii can be reconciled has been
vigorously debated \cite{Karr2020}.

Regarding the electric radius, the value reported by the A1
collaboration based on $ep$-scattering data [$\sqrt{\langle r_E^2 \rangle^p} = \qty{0.879(5)(6)}{fm}$ \cite{Bernauer2014}], while in
good agreement with hydrogen spectroscopy [$\sqrt{\langle r_E^2 \rangle^p} = \qty{0.8758(77)}{fm}$ \cite{Mohr2012}] at the time of publication, is
incompatible with the most precise determination, which comes from the
spectroscopy of muonic hydrogen [$\sqrt{\langle r_E^2 \rangle^p} = \qty{0.84087(39)}{fm}$ \cite{Pohl2010,Antognini2013}].
This significant inconsistency between measurements using either
electrons or muons has been dubbed the \enquote{proton radius puzzle}
\cite{Carlson2015}. It has triggered many additional efforts to
explain or resolve the discrepancy.

The most recent experiments using electronic hydrogen spectroscopy
favor the lower value \cite{Beyer2017,Bezginov2019,Grinin2020}, with
the exception of Ref.\@ \cite{Fleurbaey2018} which reports a larger value in agreement with older measurements \cite{Mohr2012}.
The latest determinations from $ep$ scattering yield differing results as well:
while the A1 collaboration has essentially confirmed their previous result using the initial-state radiation technique [$\sqrt{\langle r_E^2 \rangle^p} = \qty{0.878(11)(31)}{fm}$ \cite{Mihovilovic2021}], the PRad experiment at Jefferson Lab has reported a smaller value [$\sqrt{\langle r_E^2 \rangle^p} = \qty{0.831(7)(12)}{fm}$ \cite{Xiong2019}].
It is worth pointing out that dispersive analyses had already favored a smaller proton radius for a long time \cite{Mergell1996,Belushkin2007}, and continue to do so \cite{Hoferichter2016,Lin2021a,Lin2021,Lin2022}.
This applies in particular to the dispersive analysis of the data taken by the A1 experiment \cite{Lorenz2015,Alarcon2020}.

In an effort to resolve the still existing tensions, several new experimental efforts are underway:
A new $ep$-scattering experiment, MAGIX \cite{Grieser2018}, is being prepared at the Mainz-based accelerator MESA, which is currently under construction.
An upgrade of PRad, dubbed PRad-II, has been approved \cite{Gasparian2022}, while the ULQ${}^2$ experiment at ELPH in Tohoku, Japan, is already taking data \cite{Suda2022}.
To complement the results from electronic and muonic hydrogen spectroscopy and $ep$ scattering with a result from $\mu p$ scattering, the MUSE collaboration aims to measure the $\mu p$ cross section to subpercent precision at PSI \cite{Cline2021}.
Furthermore, the AMBER experiment at CERN plans to determine the electric proton radius to a precision on the order of \qty{0.01}{fm} using a similar method \cite{Quintans2022}.

For the magnetic radius, an analysis based on the $z$-expansion obtains two different numbers, depending on whether just the A1 data is analyzed [$\sqrt{\langle r_M^2 \rangle^p} = \qty{0.776(34)(17)}{fm}$ \cite{Bernauer2014,Lee2015}] or the rest of the $ep$-scattering world data excluding A1 [$\sqrt{\langle r_M^2 \rangle^p} = \qty{0.914(35)}{fm}$ \cite{Lee2015}].
The tension is not as large as for the electric radius, but still, these two numbers are not compatible with each other.
For this reason, the magnetic proton radius has received somewhat more attention recently, see, \eg, Ref.\@ \cite{Alarcon2020}.
Of the newly planned experiments devoted to the electric proton radius, only MAGIX will address the magnetic radius as well \cite{Ledroit2019}, though.

In order to understand whether these discrepancies can be traced to
the experimental data and their analyses, or whether they are perhaps an indication for physics beyond the standard model \cite{Carlson2015}, a precise standard-model prediction of the proton's radii is required.
Since the inner dynamics of the nucleon is governed by the strong
interaction, it is mandatory to apply a nonperturbative methodology,
such as lattice QCD, since the QCD coupling is large at typical hadronic scales \cite{Fritzsch1973,Wilson1974}.

In lattice QCD as in the context of scattering experiments,
radii are extracted from the derivative of the electromagnetic form
factors $G_{E, M}(Q^2)$ at $Q^2 = 0$. To distinguish between proton
and neutron requires the calculation of quark-disconnected diagrams,
which are notorious for their unfavorable signal-to-noise ratio.
Previous lattice calculations of electromagnetic form factors and
radii published in Refs.\@ \cite{Goeckeler2005,Yamazaki2009,Syritsyn2010,Bratt2010,Alexandrou2013,Bhattacharya2014,Shanahan2014a,Shanahan2014,Green2014,Capitani2015,Alexandrou2017,Hasan2018,Ishikawa2018,Shintani2019,Alexandrou2019,Alexandrou2020,Djukanovic2021,Ishikawa2021},
with the exception of Refs.\@ \cite{Alexandrou2017,Alexandrou2019},
have neglected quark-disconnected contributions due to this technical
complication.

In this paper we present our results for the electromagnetic form
factors of the proton and neutron computed from a set of CLS ensembles
with $N_f = 2 + 1$ flavors of $\mathcal{O}(a)$-improved Wilson quarks
\cite{Bruno2015} at four different lattice spacings and pion masses
between 130 and \qty{290}{MeV}.  Our study improves on all previous
calculations by explicitly evaluating both quark-connected and
-disconnected contributions and, at the same time, taking into account
all relevant systematic effects due to excited-state contamination,
finite-volume effects and the extrapolation to the physical point. To
assess the influence of excited states, we employ a wide range of
source-sink separations and apply the summation method. In addition to
determining the shape of the form factors at moderate momentum
transfers ($Q^2 \lesssim \qty{0.6}{GeV^2}$), we extract the electric
and magnetic radii and magnetic moments in the isovector and
isoscalar channels, as well as those of the proton and neutron. For
this purpose, we perform simultaneous fits to the $Q^2$-, pion-mass,
lattice-spacing, and finite-volume dependence of the form factors to
the expressions resulting from covariant baryon chiral perturbation
theory (B$\chi$PT), including vector mesons and amending the expressions by models for
lattice artefacts. Systematic errors are quantified using a model
average with weights derived from the Akaike Information Criterion
(AIC). Our model-averaged results at the physical point reproduce the
experimental values of the magnetic moments within our quoted
uncertainties and favor a small value both for the electric and the
magnetic radius of the proton. Our main findings and conclusions are presented in
the companion letter \cite{Djukanovic2024}.

This paper provides the full details of our calculation and is organized as follows:
\Cref{sec:setup} describes our lattice setup and computational details, as well as the extraction of the effective form factors from our lattice observables, while \cref{sec:excited_states} is dedicated to the treatment of excited states.
In \cref{sec:q2_dependence}, a comprehensive account of the
parametrization of the $Q^2$-dependence of the form factors and their
extrapolation to the physical point is presented.
\Cref{sec:model_average} discusses the model average and our final results, and \cref{sec:conclusions} draws some conclusions.
Details on our treatment of the pion and nucleon masses relevant to
the analysis are contained in \cref{sec:appendix_masses}.
Additional crosschecks on our excited-state analysis can be found in \cref{sec:appendix_twostate}, and a closer
examination of the form factors on our near-physical pion mass
ensemble in \cref{sec:appendix_E250}.
For completeness and ease of reference, we provide tables of all computed form factors in \cref{sec:appendix_formfactors}, of the results of all direct B$\chi$PT fits in \cref{sec:appendix_bchpt_fits}, of the radii and magnetic moments extracted using $z$-expansion fits in \cref{sec:appendix_zexp}, and of their extrapolation to the physical point in \cref{sec:appendix_CCF}.

\section{Lattice setup}
\label{sec:setup}
To compute the electromagnetic form factors of the nucleon, we consider the vector current insertion for quark flavor $f$,
\begin{equation}
    V^\mu_f(x) = \bar{\Psi}_f(x) \gamma^\mu \Psi_f(x) ,
    \label{eq:local_vector_current}
\end{equation}
and its nucleon matrix elements,
\begin{equation}
    \braket{N(\mathbf{p}', s') | V^\mu_f(x) | N(\mathbf{p}, s)} = e^{iq \cdot x} \bar{u}^{s'}(\mathbf{p'}) \mathcal{V}^\mu_f(q) u^s(\mathbf{p}) .
    \label{eq:NME}
\end{equation}
Here, $N(\mathbf{p}, s)$ denotes a nucleon state with three-momentum \textbf{p} and spin $s$, $u^s(\mathbf{p})$ the corresponding Dirac spinor, and $q_\mu = p'_\mu - p_\mu$ the four-momentum transfer.
The quantities $\mathcal{V}^\mu_f(q)$ are defined by
\begin{equation}
    \mathcal{V}^\mu_f(q) = \gamma^\mu F_1^f(Q^2) + i \frac{\sigma^{\mu\nu} q_\nu}{2m_N} F_2^f(Q^2) ,
    \label{eq:NME_Vmu}
\end{equation}
where $m_N$ is the nucleon mass, and $Q^2 = -q^2 > 0$ in the spacelike region.
Furthermore, we have introduced the Dirac and Pauli form factors $F_1(Q^2)$ and $F_2(Q^2)$, respectively, which are connected to the electric and magnetic Sachs form factors $G_E(Q^2)$ and $G_M(Q^2)$ via
\begin{align}
    \label{eq:GE_from_F12}
    G_E(Q^2) &= F_1(Q^2) - \frac{Q^2}{4m_N^2}F_2(Q^2) , \\
    \label{eq:GM_from_F12}
    G_M(Q^2) &= F_1(Q^2) + F_2(Q^2) .
\end{align}
The electric form factor at zero momentum transfer yields the nucleon's electric charge, \ie $G_E^p(0) = 1$ and $G_E^n(0) = 0$, whereas the magnetic form factor at zero momentum transfer is identified with the magnetic moment, $G_M(0) = \mu_M$.
The corresponding radii are given by the derivative of the form factors at zero momentum transfer,
\begin{equation}
    \langle r^2 \rangle = -\frac{6}{G(0)} \left.\pd{G(Q^2)}{Q^2}\right|_{Q^2 = 0} .
    \label{eq:radii}
\end{equation}
The only exception to this definition is the electric radius of the neutron, where the normalization factor is omitted,
\begin{equation}
    \langle r_E^2 \rangle^n = -6 \left.\pd{G_E^n(Q^2)}{Q^2}\right|_{Q^2 = 0} .
    \label{eq:electric_radius_neutron}
\end{equation}

For our lattice determination of these quantities, we use the CLS ensembles \cite{Bruno2015} which have been generated with $2 + 1$ flavors of non-perturbatively $\mathcal{O}(a)$-improved Wilson fermions \cite{Sheikholeslami1985,Bulava2013} and a tree-level improved Lüscher-Weisz gauge action \cite{Luescher1985}.
Only ensembles following the chiral trajectory characterized by $\tr M_q = 2m_l + m_s = \text{const.}$ are employed.
In order to prevent topological freezing \cite{Schaefer2011}, the fields obey open boundary conditions (oBC) in time \cite{Luescher2011,Luescher2013}, with the exception of the ensembles E250, D450, and N451, which use periodic boundary conditions (pBC) in time.
\Cref{tab:ensembles} displays the set of ensembles entering the analysis:
they cover four lattice spacings in the range from \qty{0.050}{fm} to \qty{0.086}{fm}, and several different pion masses, including one slightly below the physical value (E250).
We note that data is available on additional (heavier) ensembles, but only the ones shown in \cref{tab:ensembles} are included in this analysis.

\begingroup 
\squeezetable 
\begin{table*}[htb]
    \caption{Overview of the ensembles used in this study.
      $N_\mathrm{meas, HP}$ and $N_\mathrm{meas, LP}$ denote the aggregated number of high-precision (HP) and low-precision (LP) solves used for the computation of the connected and the disconnected contributions, respectively.
      For the connected contribution, $N_\mathrm{meas}^\mathrm{conn, max}$ refers to the number of measurements used for the largest value of $t_\mathrm{sep}$, while for the smaller values, the number of measurements is scaled down in stages.}
    \label{tab:ensembles}
    \begin{ruledtabular}
        \begin{tabular}{lccccccccccccc}
            ID                   & $\beta$ & $t_0^\mathrm{sym}/a^2$ & $T/a$ & $L/a$ & $M_\pi$ [MeV] & $N_\mathrm{cfg}^\mathrm{conn}$ & $N_\mathrm{cfg}^\mathrm{disc}$ & $N_\mathrm{meas, HP}^\mathrm{conn, max}$ & $N_\mathrm{meas, LP}^\mathrm{conn, max}$ & $N_\mathrm{meas, HP}^\mathrm{disc}$ & $N_\mathrm{meas, LP}^\mathrm{disc}$ & $t_\mathrm{sep}/a$ \\ \hline
            C101                 & 3.40    & 2.860(11)              & 96    & 48    & 227           & 1988                           & 994                            & 1988                                     & \num{63616}                              & \num{7951}                          & \num{237965}                        & 4 -- 17            \\
            N101\footnotemark[1] & 3.40    & 2.860(11)              & 128   & 48    & 283           & 1588                           & 1588                           & 1588                                     & \num{50816}                              & \num{3176}                          & \num{406518}                        & 4 -- 17            \\
            H105\footnotemark[1] & 3.40    & 2.860(11)              & 96    & 32    & 283           & 1024                           & 1024                           & 4096                                     & \num{49152}                              & \num{25585}                         & \num{248331}                        & 4 -- 17            \\[\defaultaddspace]
            D450                 & 3.46    & 3.659(16)              & 128   & 64    & 218           & 498                            & 498                            & 3984                                     & \num{63744}                              & \num{3984}                          & \num{63744}                         & 4 -- 20            \\
            N451\footnotemark[1] & 3.46    & 3.659(16)              & 128   & 48    & 289           & 1010                           & 1010                           & 8080                                     & \num{129280}                             & \num{8080}                          & \num{129280}                        & 4 -- 20 (stride 2) \\[\defaultaddspace]
            E250                 & 3.55    & 5.164(18)              & 192   & 96    & 130           & 398                            & 796                            & 3184                                     & \num{101888}                             & \num{6368}                          & \num{203776}                        & 4 -- 22 (stride 2) \\
            D200                 & 3.55    & 5.164(18)              & 128   & 64    & 207           & 1996                           & 998                            & 1996                                     & \num{63872}                              & \num{8982}                          & \num{271258}                        & 4 -- 22 (stride 2) \\
            N200\footnotemark[1] & 3.55    & 5.164(18)              & 128   & 48    & 281           & 1708                           & 1708                           & 1708                                     & \num{22828}                              & \num{13664}                         & \num{406016}                        & 4 -- 22 (stride 2) \\
            S201\footnotemark[1] & 3.55    & 5.164(18)              & 128   & 32    & 295           & 2092                           & 2092                           & 2092                                     & \num{66944}                              & \num{4181}                          & \num{96279}                         & 4 -- 22 (stride 2) \\[\defaultaddspace]
            E300                 & 3.70    & 8.595(29)              & 192   & 96    & 176           & 569                            & 569                            & 569                                      & \num{18208}                              & \num{1138}                          & \num{163872}                        & 4 -- 28 (stride 2) \\
            J303                 & 3.70    & 8.595(29)              & 192   & 64    & 266           & 1073                           & 1073                           & 1073                                     & \num{17168}                              & \num{3219}                          & \num{145872}                        & 4 -- 28 (stride 2)
        \end{tabular}
    \end{ruledtabular}
    \footnotetext[1]{These ensembles are not used in the final fits but only to constrain discretization and finite-volume effects.}
\end{table*}
\endgroup 

In order to compensate for the twisted mass introduced for the light quarks \cite{Palombi2009,Luescher2013} and the rational approximation used for the dynamical strange quark \cite{Horvath1999,Clark2007} during the gauge field generation, all observables need to be reweighted.
We employ the reweighting factors computed in Ref.\@ \cite{Kuberski2023} with exact low-mode deflation on all ensembles except E300, where the standard stochastic CLS run \cite{Bruno2015} is used.
In all cases, we correct for the treatment of the strange-quark determinant following the procedure outlined in Ref.\@ \cite{Mohler2020}.

We measure the two- and three-point functions of the nucleon,
\begin{align}
    \label{eq:C2}
    \Braket{C_2(\mathbf{p}'; t_\mathrm{sep})} &= \sum_{\mathbf{y}} e^{-i \mathbf{p}' \cdot \mathbf{y}} \Gamma^p_{\beta\alpha} \Braket{N_\alpha(\mathbf{y}, t_\mathrm{sep}) \bar{N}_\beta(0)} , \\
    \label{eq:C3}
    \Braket{C_{3, O}(\mathbf{p}', \mathbf{q}; t_\mathrm{sep}, t)} &= \sum_{\mathbf{y}, \mathbf{z}} e^{i \mathbf{q} \cdot \mathbf{z}} e^{-i \mathbf{p}' \cdot \mathbf{y}} \Gamma^p_{\beta\alpha} \Braket{N_\alpha(\mathbf{y}, t_\mathrm{sep}) O(\mathbf{z}, t) \bar{N}_\beta(0)} .
\end{align}
Here, $\Gamma^p$ denotes the polarization matrix of the nucleon and we have set the source position $x$ to zero for simplicity.
In our setup, the nucleon at the sink is at rest, \ie for a momentum transfer \textbf{q} the initial and final states have momenta $\mathbf{p} = -\mathbf{q}$ and $\mathbf{p}' = \mathbf{0}$, respectively.
Our interpolating operator for the proton is the same as in Refs.\@ \cite{Djukanovic2021,Djukanovic2022,Agadjanov2023}, which is built using Gaussian-smeared \cite{Guesken1989} quark fields with spatially APE-smeared \cite{Albanese1987} gauge links and tuning the parameters so that a smearing radius $r_G \approx \SI{0.5}{fm}$ \cite{Hippel2013} is realized.

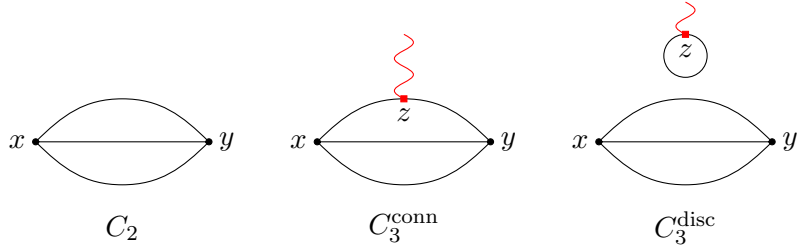
\begin{figure*}[ht]
    \begin{tikzpicture}[scale=0.57]
	\node[left] at (-2,0) {$x$};
	\node[right] at (2,0) {$y$};

	\filldraw (-2,0) circle (0.07) -- (2,0) circle (0.07);
	\draw (-2,0) to [out=45, in=180] (0,1) to [out=0, in=135] (2,0);
	\draw (-2,0) to [out=315, in=180] (0,-1) to [out=0, in=225] (2,0);

	\node[left] at (4.5,0) {$x$};
	\node[right] at (8.5,0) {$y$};
	\node[below] at (6.5,1) {$z$};

	\filldraw (4.5,0) circle (0.07) -- (8.5,0) circle (0.07);
	\draw (4.5,0) to [out=45, in=180] (6.5,1) to [out=0, in=135] (8.5,0);
	\filldraw[color=red] (6.43,0.93) rectangle ++(0.14,0.14);
	\draw (4.5,0) to [out=315, in=180] (6.5,-1) to [out=0, in=225] (8.5,0);

	\node[left] at (11,0) {$x$};
	\node[right] at (15,0) {$y$};
	\node[below] at (13,2.5) {$z$};

	\filldraw (11,0) circle (0.07) -- (15,0) circle (0.07);
	\draw (11,0) to [out=45, in=180] (13,1) to [out=0, in=135] (15,0);
	\draw (13,2) circle (0.5);
	\filldraw[color=red] (12.93,2.43) rectangle ++(0.14,0.14);
	\draw (11,0) to [out=315, in=180] (13,-1) to [out=0, in=225] (15,0);

	\begin{scope}[color=red, decoration={complete sines, segment length=0.375cm, amplitude=0.25cm}]
	    \draw[decorate] (6.5,1) -- (6.5,2.5);
	    \draw[decorate] (13,2.5) -- (13,3.25);
	\end{scope}

	\node at (0,-2) {$C_2$};
	\node at (6.5,-2) {$C_3^\mathrm{conn}$};
	\node at (13,-2) {$C_3^\mathrm{disc}$};
\end{tikzpicture}
    \caption{Diagrammatic representation of the two- and three-point functions of the nucleon.
      Only quark lines are shown, while all gluon lines are suppressed.
      The red squares in the three-point functions represent the operator insertion, and the wavy red lines the external photons.}
    \label{fig:C2_3}
\end{figure*}

For the three-point functions \cref{eq:C3}, the pertinent Wick contractions yield a connected and a disconnected contribution, $\braket{C_{3,O}} = \braket{C_{3,O}^\mathrm{conn}} + \braket{C_{3,O}^\mathrm{disc}}$.
They are depicted diagrammatically in \cref{fig:C2_3}.
To calculate the connected part, we employ the \enquote{fixed-sink} variant of the extended propagator method.
This requires additional inversions for each source-sink separation while allowing the momentum transfer to be varied via a phase factor at the point of the current insertion \cite{Martinelli1989}.
The disconnected part of the three-point functions is constructed from the quark loops and the two-point functions according to
\begin{align}
    \label{eq:C3_disconnected}
    \Braket{C_{3, O}^\mathrm{disc}(\mathbf{p}', \mathbf{q}; t_\mathrm{sep}, t)} &= \Braket{L^{O, \mathrm{disc}}(\mathbf{q}; t) C_2(\mathbf{p}'; t_\mathrm{sep})} , \\
    \label{eq:loops}
    L^{O, \mathrm{disc}}(\mathbf{q}; z_0) &= -\sum_{\mathbf{z}} e^{i \mathbf{q} \cdot \mathbf{z}} \tr[D_q^{-1}(z, z) \Gamma^O] .
\end{align}
The all-to-all propagator $D_q^{-1}(z, z)$ appearing in the quark loops \cref{eq:loops} is computed via stochastic estimation using a variation of the frequency-splitting technique \cite{Giusti2019}.
To that end, we employ a generalized hopping-parameter expansion \cite{Guelpers2014} combined with hierarchical probing \cite{Stathopoulos2013} for one heavy quark flavor and subsequently apply a split-even estimator, \ie the one-end trick \cite{McNeile2006,Boucaud2008,Giusti2019}, for the remaining flavors.
For further details, we refer the interested reader to Ref.\@ \cite{Ce2022}.

To reduce the cost of the inversions, we apply the truncated-solver method \cite{Bali2010,Blum2013,Shintani2015} with bias correction.
Details on our setup of sources are contained in \cref{tab:ensembles}, alongside the available source-sink separations $t_\mathrm{sep}$.
On ensembles with open boundary conditions in time, the sources are generally placed on a single timeslice in the middle of the lattice.
Additional measurements of the two-point function are used on all oBC ensembles except S201 to extend the statistics for the disconnected contribution.
For these additional measurements, we put the nucleon sources on different timeslices in the bulk of the lattice.
We have checked explicitly that the observables studied in this work are not significantly influenced by boundary effects for the chosen source positions.
No such issues arise for ensembles with periodic boundary conditions in time.
Here, one can distribute the sources randomly on edges of sub-blocks of the entire lattice volume, as dictated by even-odd \cite{DeGrand1988} and Schwarz \cite{Luescher2003} preconditioning.
On all ensembles, we employ iterative statistics for the different source-sink separations.
This means that with rising $t_\mathrm{sep}$, the number of sources used for the computation of the connected part is increased.
The scaling of measurements is tuned in such a way that the behavior of the effective statistics as a function of $t_\mathrm{sep}$ more closely resembles a constant instead of showing an exponential decay of the signal-to-noise ratio.
For the disconnected part, the highest statistics at our disposal is always utilized, in order to obtain the best possible signal.

Instead of the local current \cref{eq:local_vector_current}, we use the conserved vector current in the same way as in Ref.\@ \cite{Djukanovic2021}, so that no renormalization is required.
The $\mathcal{O}(a)$-improvement is also performed analogously to Ref.\@ \cite{Djukanovic2021}, with the improvement coefficient computed in Ref.\@ \cite{Gerardin2019a}.

As a first step towards extracting the effective form factors, the nucleon two-point functions from \cref{eq:C2} are averaged over equivalent momentum classes.
We call all three-momenta \textbf{p} which share the same modulus $|\mathbf{p}|$ equivalent and assign them the equivalence class $\mathfrak{p} = \{\tilde{\mathbf{p}} \in \mathbb{P}^3 : |\tilde{\mathbf{p}}| = |\mathbf{p}|\}$.
Here, $\mathbb{P}^3$ is the set of possible lattice momenta, $\mathbb{P}^3 = \{ \mathbf{p} = \frac{2\pi}{L} \mathbf{n}_p : \mathbf{n}_p \in \mathbb{Z}^3 \} \subset \mathbb{R}^3$.
The momentum-averaged two-point functions are then defined as
\begin{equation}
    \langle \bar{C}_2(\mathfrak{p}; t_\mathrm{sep}) \rangle = \left. \sum_{\tilde{\mathbf{p}} \in \mathfrak{p}} \langle C_2(\tilde{\mathbf{p}}; t_\mathrm{sep}) \rangle \middle/ \sum_{\tilde{\mathbf{p}} \in \mathfrak{p}} 1 \right. .
    \label{eq:C2_equivalence_averaging}
\end{equation}
Afterwards, we calculate the ratios \cite{Korzec2009}
\begin{equation}
    R_{V_\mu}(\mathbf{0}, \mathbf{q}; t_\mathrm{sep}, t) = \frac{\langle C_{3, V_\mu}(\mathbf{0}, \mathbf{q}; t_\mathrm{sep}, t) \rangle}{\langle C_2(\mathbf{0}; t_\mathrm{sep}) \rangle} \sqrt{\frac{\langle \bar{C}_2(\mathfrak{q}; t_\mathrm{sep} - t) \rangle \langle C_2(\mathbf{0}; t) \rangle \langle C_2(\mathbf{0}; t_\mathrm{sep}) \rangle}{\langle C_2(\mathbf{0}; t_\mathrm{sep} - t) \rangle \langle \bar{C}_2(\mathfrak{q}; t) \rangle \langle \bar{C}_2(\mathfrak{q}; t_\mathrm{sep}) \rangle}} .
    \label{eq:ratio}
\end{equation}
For the connected part, we employ for each value of $t_\mathrm{sep}$ matching statistics in terms of sources for the two- and three-point functions entering \cref{eq:ratio}.
This preserves the full correlation between them, which slightly reduces the statistical fluctuations in the ratio.
For the disconnected part, on the other hand, the highest statistics at our disposal is utilized for all values of $t_\mathrm{sep}$, both for the two-point functions used to construct \cref{eq:C3_disconnected} and for the ones entering \cref{eq:ratio}.
In all cases, the same projection matrix is employed for both the two- and three-point functions entering \cref{eq:ratio}, again to ensure full correlation.
For the connected part, this is only $\Gamma^p_3 = \frac{1}{2} (1 + \gamma_0) (1 + i \gamma_5 \gamma_3)$.
For the disconnected part, we employ all three polarization directions, $\Gamma^p_j = \frac{1}{2} (1 + \gamma_0) (1 + i \gamma_5 \gamma_j)$, $j = 1, 2, 3$, and average the thus obtained effective form factors.
Moreover, we average over the forward- and backward-propagating nucleon for the disconnected part.
For the determination of the nucleon mass (\cf \cref{sec:appendix_masses}), we make use of the unpolarized nucleon, \ie $\Gamma^p = \frac{1}{2}(1 + \gamma_0)$, as this is equivalent to averaging over the three polarization directions in the case of the two-point function.

We use the same estimators for the effective electric and magnetic Sachs form factors as in Ref.\@ \cite{Djukanovic2021}, \ie we do not employ the spatial components of the vector current to compute the electric form factor, as they are more noisy.
Note that for the disconnected contribution to the effective magnetic form factor, one needs to adapt the indices of the momenta and the current insertions in eq.~(11) of Ref.\@ \cite{Djukanovic2021} according to the polarization direction.
We average these estimators over all three-momenta $\tilde{\mathbf{q}}$ belonging to the equivalence class $\mathfrak{q}$ and thus yielding the same $Q^2$ [except for those with $\tilde{q}_1 = \tilde{q}_2 = 0$ (or the components appropriate for $\Gamma^p_j$) in the case of the magnetic form factor].
Furthermore, we assume the relativistic dispersion relation $E_\mathfrak{q} = \sqrt{m_N^2 + |\mathbf{q}|^2}$.
We have checked explicitly that employing the extracted ground-state energies also for non-vanishing momenta instead of the above dispersion relation does not change our results for the ground-state form factors significantly.

We build the effective form factors in the isospin basis, \ie for the isovector ($u - d$) and the connected isoscalar ($u + d$) combinations, as well as for the disconnected contributions of the light and strange quarks.
Since we impose strong SU(2) isospin symmetry ($m_u = m_d$) in our calculation, the disconnected contribution cancels in the isovector case.
The full isoscalar (octet) combination $u + d - 2s$, on the other hand, can be obtained from the connected and disconnected pieces as
\begin{equation}
    G_{E, M}^{\mathrm{eff}, u+d-2s} = G_{E, M}^{\mathrm{eff, conn}, u+d} + 2G_{E, M}^{\mathrm{eff, disc}, l-s} .
    \label{eq:eff_ff_u+d-2s}
\end{equation}
Note that the disconnected part only requires the difference $l - s$ between the light and strange contributions, in which correlated noise cancels and which can be computed very effectively by the one-end trick.
We drop the disconnected contribution $G_E^{\mathrm{eff, disc}, l-s}$ at $Q^2 = 0$, as it has to be zero.
Our data, on the other hand, shows fluctuations around the exact zero due to the stochastic estimation of the quark loops and the application of the truncated-solver method for the calculation of the two-point functions.
Thus, explicitly adding this superfluous term, which is always compatible with zero, would artificially enhance noise in all data points for $G_E$, because we normalize them by $G_E(0)$ (\cf \cref{sec:q2_dependence}).

\begin{figure*}[htb]
    \includegraphics[width=\textwidth]{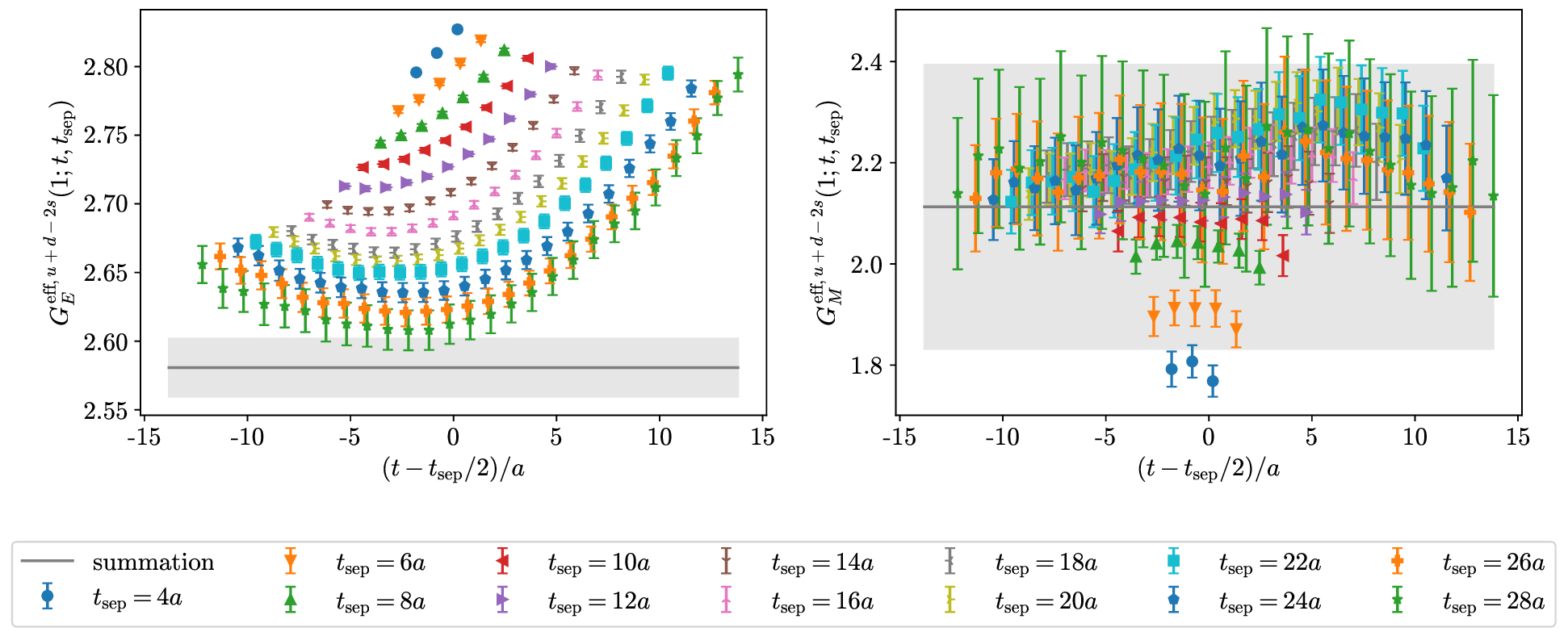}
    \caption{Isoscalar effective form factors at the first non-vanishing momentum on the ensemble E300 ($Q^2 \approx \qty{0.067}{GeV^2}$).
      The left plot shows the electric and the right one the magnetic form factor.
      For each source-sink separation $t_\mathrm{sep}$, the effective form factors are displayed as a function of the operator insertion time $t$, offset to the midpoint between nucleon source and sink.
      The data points are horizontally displaced for better visibility.
      The gray bands and curves depict the results of the summation method using the window average, as detailed in \cref{sec:excited_states}.}
    \label{fig:effective_formfactors}
\end{figure*}

The resulting isoscalar effective form factors are shown in \cref{fig:effective_formfactors} for the first non-vanishing momentum on the ensemble E300.
This illustrates that we obtain a clear signal including the disconnected contributions:
the families of points for the different source-sink separations can be clearly distinguished in $G_E^{\mathrm{eff}, u+d-2s}$, and for the smaller values of $t_\mathrm{sep}$ also in $G_M^{\mathrm{eff}, u+d-2s}$.

Unless otherwise stated, errors are estimated using single-elimination Jackknife.
Autocorrelations are largely absent in the ratios of \cref{eq:ratio}.
Nevertheless, in order to remove any residual autocorrelation, we block our data with a bin size of two, if the spacing between two analyzed configurations in terms of molecular dynamics time does not already account for this factor.
The latter is the case for the disconnected contribution on C101 and D200, for the connected contribution on E250, and for both the connected and the disconnected contributions on E300 and J303.

On some ensembles, we observe that individual measurements on a small number of configurations are located very far outside the distribution of the vast majority of configurations.
Keeping these exceptional configurations in the sample leads to a drastically increased error and, more importantly, to an unexpected scaling of the error with the source-sink separation, \ie the error is inflated strongly only for single values of $t_\mathrm{sep}$.
The most prominent example is D200, where we identify one configuration to be the root cause of the gross overestimation of errors.
Similar observations of outliers have already been reported for previous analyses on CLS ensembles \cite{Agadjanov2023,Bali2023}.
In order to identify the problematic configurations, we first extract the effective form factors using single-elimination Jackknife on unbinned data.
We then scan the isovector, connected isoscalar, and disconnected contributions for all relevant values of $Q^2$, $t_\mathrm{sep}$, and $t$, employing the procedure described in the supplementary material of Ref.\@ \cite{Agadjanov2023}.
The corresponding configurations are omitted during the whole main analysis; the numbers in \cref{tab:ensembles} already reflect this.

We express all dimensionful quantities in units of $t_0$ using the determination of $t_0^\mathrm{sym}/a^2$ from Ref.\@ \cite{Bruno2017}.
Only our final results for the radii are converted to physical units by means of the world-average value of
\begin{equation}
    \sqrt{t_{0, \mathrm{phys}}} = \qty{0.14464(87)}{fm}
    \label{eq:sqrt_t0_phys}
\end{equation}
for $N_f = 2 + 1$ from Ref.\@ \cite{Aoki2021}.
This procedure ensures that the error of the calibration is treated independently of that of the (more precise) pure lattice measurement of $t_0^\mathrm{sym}/a^2$.

\section{Excited-state analysis}
\label{sec:excited_states}
In general, baryonic correlation functions suffer from a strong exponential growth of the relative statistical noise when the distance in Euclidean time between operators is increased \cite{Hamber1983,Lepage1989}.
Therefore, for the typically accessible source-sink separations in current lattice calculations of baryon structure observables, it cannot be guaranteed that contributions from excited states are sufficiently suppressed.
This underlines the necessity to explicitly address the excited-state systematics in order to extract the ground-state form factors from the effective ones \cite{Ottnad2020,Aoki2021}.

In this work, we make use of the summation method \cite{Maiani1987,Doi2009,Capitani2012}.
It takes advantage of the fact that in the ratios of \cref{eq:ratio}, when summed over timeslices in between source and sink, the contributions from excited states are parametrically suppressed.
Accordingly, we sum the effective form factors over the operator insertion time, omitting $t_\mathrm{skip} = 2a$ timeslices at both ends\footnote{We justify this choice in \cref{sec:appendix_twostate}.},
\begin{equation}
    S_{E, M}(Q^2; t_\mathrm{sep}) = \sum_{t = t_\mathrm{skip}}^{t_\mathrm{sep} - t_\mathrm{skip}} G_{E, M}^\mathrm{eff}(Q^2; t_\mathrm{sep}, t) .
    \label{eq:summation}
\end{equation}
In the asymptotic limit with only ground-state contributions, the slope of this quantity as a function of $t_\mathrm{sep}$ is given by the ground-state form factor \cite{Capitani2015,Djukanovic2021},
\begin{equation}
    S_{E, M}(Q^2; t_\mathrm{sep}) \xrightarrow{t_\mathrm{sep} \gg 0} C_{E, M}(Q^2) + \frac{1}{a} (t_\mathrm{sep} + a - 2t_\mathrm{skip}) G_{E, M}(Q^2) + \ldots ,
    \label{eq:summation_asymptotic}
\end{equation}
where the ellipsis denotes exponentially suppressed corrections from excited states.

We perform several fits to \cref{eq:summation_asymptotic} for different starting values $t_\mathrm{sep}^\mathrm{min}$ of the source-sink separation.
Instead of choosing a single value of $t_\mathrm{sep}^\mathrm{min}$ on each ensemble, we perform a weighted average over $t_\mathrm{sep}^\mathrm{min}$, where the weights are given by a smooth window function \cite{Djukanovic2022,Agadjanov2023},
\begin{equation}
    \hat{G} = \frac{\sum_i w_i G_i}{\sum_i w_i} , \qquad w_i = \tanh\frac{t_i - t_w^\mathrm{low}}{\Delta t_w} - \tanh\frac{t_i - t_w^\mathrm{up}}{\Delta t_w} .
    \label{eq:window_average}
\end{equation}
Here, $t_i$ is the value of $t_\mathrm{sep}^\mathrm{min}$ in the $i$-th fit, and we choose $t_w^\mathrm{low} = 6.22\sqrt{t_0} \approx \qty{0.9}{fm}$, $t_w^\mathrm{up} = 7.61\sqrt{t_0} \approx \qty{1.1}{fm}$, and $\Delta t_w = 0.553\sqrt{t_0} \approx \qty{0.08}{fm}$.
Note that the window has been shifted to larger values of $t_\mathrm{sep}^\mathrm{min}$ by \qty{0.1}{fm} compared to the one originally used in Refs.\@ \cite{Djukanovic2022,Agadjanov2023}.
The reason for this is that our data for the electromagnetic form factors are statistically more precise than those for the axial form factor in Ref.\@ \cite{Djukanovic2022} or the sigma term in Ref.\@ \cite{Agadjanov2023}.
Hence, we can resolve excited-state effects for larger values of $t_\mathrm{sep}^\mathrm{min}$, so that the plateau region is expected to start later.
Accordingly, we have observed that the window using larger $t_w^\mathrm{up, low}$ better captures the plateau on the majority of our ensembles.
Thus, we have opted for the more conservative choice, which also yields a slightly larger error.
The two choices are compared in \cref{fig:summation_stability_window_loc} in \cref{sec:appendix_twostate}.

We average over all available values of $t_\mathrm{sep}^\mathrm{min}$, subject to the constraint that at least three values of $t_\mathrm{sep}$ are contained in the underlying fit to \cref{eq:summation}.
It should be stressed that the only quantity that is effectively restricted by this method is the minimal source-sink separation; all fits go up to the largest $t_\mathrm{sep}$ we have computed.
Essentially, the window average merely serves as a smoothing of the lower end of the fit interval.

\begin{figure*}[htb]
    \includegraphics[width=\textwidth]{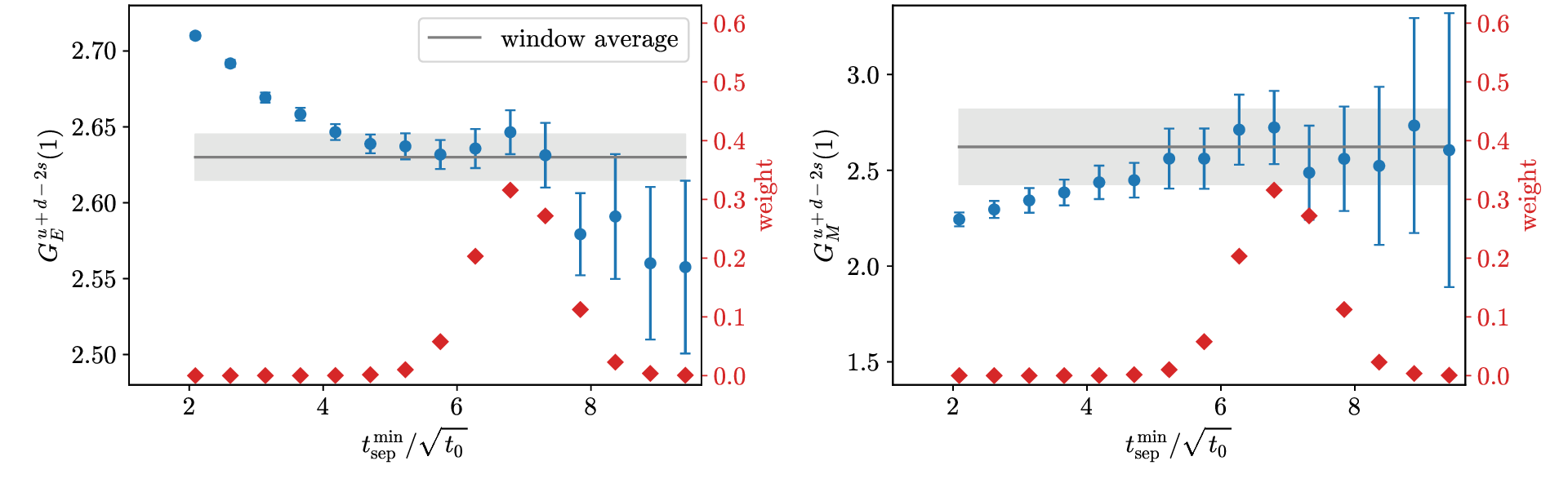}
    \includegraphics[width=\textwidth]{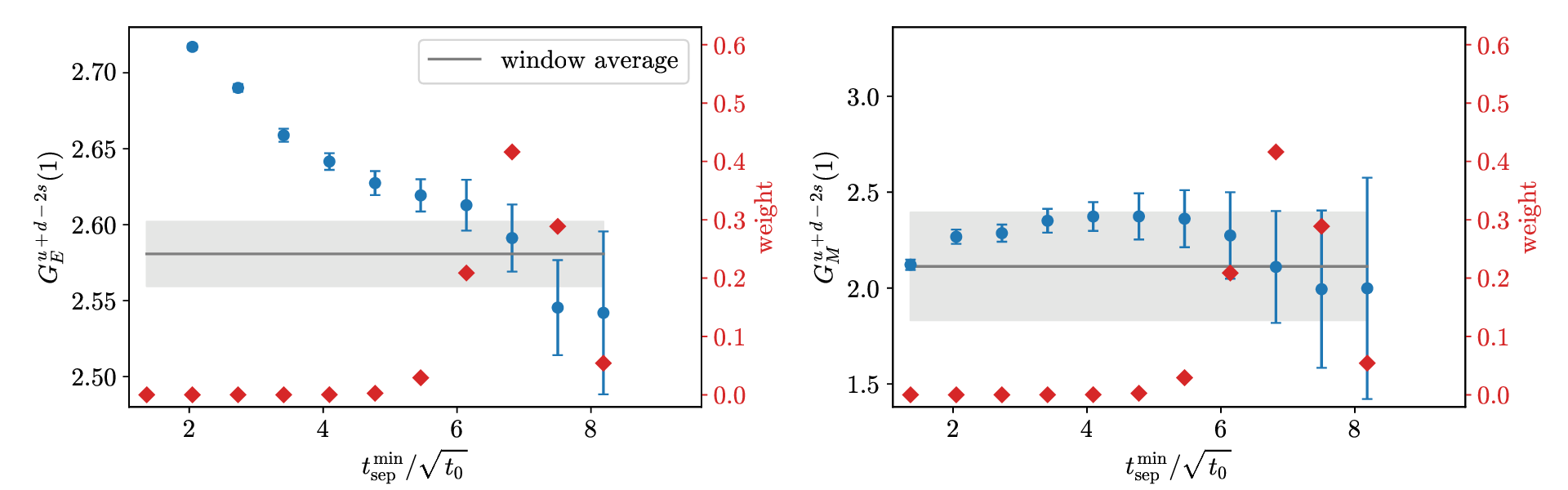}
    \caption{Isoscalar electromagnetic form factors at the first non-vanishing momentum on ensembles D450 (upper panel) and E300 (lower panel) as a function of the minimal source-sink separation entering the fits to \cref{eq:summation}.
      Each blue point represents a single fit starting at the source-sink separation given on the horizontal axis.
      The associated weights derived from \cref{eq:window_average} are shown by the red diamonds, with the gray lines and bands depicting the averaged results.}
    \label{fig:window_average}
\end{figure*}

This strategy is illustrated in \cref{fig:window_average} for the isoscalar combination at the first non-vanishing momentum on the ensembles D450 and E300.
One can see that the window averages agree within their errors with what one might identify as plateaux in the blue points.
This being valid to a similar degree for all other ensembles, flavor combinations, and momenta employed in the analysis, we conclude that the window method reliably identifies the asymptotic value of the effective form factors.
Moreover, it reduces the human bias compared to manually picking one particular value for $t_\mathrm{sep}^\mathrm{min}$ on each ensemble, since we use the same window parameters in units of $t_0$ on all ensembles.
It is important to note that even if a plateau appears to be reached, this does not guarantee ground-state dominance.
The situation is aggravated by the fact that in general, relatively few values of $t_\mathrm{sep}^\mathrm{min}$ are available, and correlated fluctuations in any direction can easily be mistaken for a plateau.
This underlines once more the necessity of an automated strategy such as the window average which can readily be applied to all ensembles and momenta.
The size of the gray error bands in \cref{fig:window_average} furthermore shows that our window average yields, in contrast to error-weighted procedures, an error estimate which is comparable to the errors of the individual points entering the average.
Thus, we are convinced that our error estimates are conservative enough to exclude any systematic bias in the identification of ground-state form factors.

In order to ensure that ground-state dominance is reached by the method described above, we have explored additional variations of it as well as a complementary approach based on two-state fits to the effective form factors.
Details on these crosschecks can be found in \cref{sec:appendix_twostate}.
The form factor values obtained by our preferred method are collected in \cref{sec:appendix_formfactors} for all ensembles and all momenta which we have considered.

\section{Parametrization of the \texorpdfstring{$Q^2$}{Q2}-dependence}
\label{sec:q2_dependence}
As the radii are defined in terms of the slope of the form factors at zero momentum transfer [\cf \cref{eq:radii}], a description of their $Q^2$-dependence is necessary.
Proceeding analogously to Refs.\@ \cite{Capitani2015,Djukanovic2021}, we apply two different methods:
Our preferred strategy is to combine the parametrization of the $Q^2$-dependence with the chiral, continuum, and infinite-volume extrapolation by performing a simultaneous fit to the $Q^2$-, pion-mass, lattice-spacing, and finite-volume dependence of our form factor data directly to the expressions resulting from covariant baryon chiral perturbation theory (B$\chi$PT) \cite{Bauer2012}.
This is explained in detail in \cref{sec:bchpt_fits}.
Alternatively, one can follow the more traditional approach of first extracting the radii on each ensemble from a generic parametrization of the $Q^2$-dependence and subsequently extrapolating them to the physical point.
A crosscheck of our main analysis with this two-step procedure is presented in \cref{sec:zexp}.

\subsection{Direct \texorpdfstring{B$\chi$PT}{BChPT} fits}
\label{sec:bchpt_fits}
For our main analysis using the direct (simultaneous) fits, we fit our data for the form factors to the full expressions of Ref.\@ \cite{Bauer2012} without explicit $\Delta$ degrees of freedom.
The fits are performed for the isovector and isoscalar channels separately, but for $G_E$ and $G_M$ together.
This allows us to take the correlations not only between different $Q^2$, but also between $G_E$ and $G_M$ into account.
The ensembles, on the other hand, are treated as statistically independent.
$G_E(0)$ is fixed by fitting the normalized ratio $G_E(Q^2) / G_E(0)$.
We incorporate the contributions from the relevant vector mesons in the expressions for the form factors.
In the isovector case, this is the $\rho$ meson, while in the isoscalar channel, we include the leading-order terms from the $\omega$ and $\phi$ resonances.
Because the loop diagrams involving $\omega$ or $\phi$ resonances only yield small numerical contributions to the form factors, our fits depend only marginally on them.
This means that the corresponding low-energy constants (LECs) $g_\omega$ and $g_\phi$ are very poorly constrained by our data, so that we neglect these loop diagrams.
The corresponding tree-level diagrams, on the other hand, only depend on the combinations $c_\omega = f_\omega g_\omega$ and $c_\phi = f_\phi g_\phi$, respectively \cite{Bauer2012}.
Thus, we only use the products $c_\omega$ and $c_\phi$ as independent fit parameters.
To summarize, the fit for the isoscalar form factors depends linearly on the LECs $d_7$, $c_7$, $c_\omega$, and $c_\phi$, whereas for the isovector form factors, the relevant LECs are $d_6$, $\tilde{c}_6$, $d_x$, and $G_\rho$ \cite{Bauer2012}.\footnote{The nucleon (average of the proton and neutron) and $\phi$ masses are fixed to their physical values \cite{Workman2022}. Moreover, we replace the pion decay constant and the axial-vector coupling constant in the chiral limit, which appear in the B$\chi$PT formulae, by their physical values, $F_\pi = \qty{92.2}{MeV}$ \cite{Bruno2017} and $g_A = 1.2754$ \cite{Workman2022}, respectively. We also use the KSRF relation \cite{Kawarabayashi1966,Riazuddin1966,Djukanovic2004} $g^2 = M_{\rho, \mathrm{phys}}^2 / (2 F_\pi^2)$. For the numerical evaluation of one-loop integrals, we make use of LoopTools \cite{Hahn1999,Oldenborgh1990}, setting the renormalization scale to $\mu = \qty{1}{GeV}$.}
For further details on the LECs, we refer to table III in Ref.\@ \cite{Capitani2015}.
The full expressions for the form factors which we employ can be found in appendix D.2.1 of Ref.\@ \cite{Bauer2009}.
Starting from the formulae for the Dirac and Pauli form factors given there, we form the appropriate linear combinations for the electric and magnetic Sachs form factors according to \cref{eq:GE_from_F12,eq:GM_from_F12}.

The mass of the $\rho$ meson is set on each ensemble to the value at the corresponding pion mass and lattice spacing.
This is determined from a parametrization of the pion-mass and lattice-spacing dependence of a subset of the values for $M_\rho / M_\pi$ obtained in Ref.\@ \cite{Ce2022},
\begin{equation}
    \frac{M_\rho}{M_\pi} = \frac{M_{\rho, \mathrm{phys}}}{M_{\pi, \mathrm{phys}}} + A \left( \frac{1}{\sqrt{t_0} M_\pi} - \frac{1}{\sqrt{t_{0, \mathrm{phys}}} M_{\pi, \mathrm{phys}}} \right) + C (\sqrt{t_0} M_\pi - \sqrt{t_{0, \mathrm{phys}}} M_{\pi, \mathrm{phys}}) + D \frac{a^2}{t_0} ,
    \label{eq:rho_mass_parametrization}
\end{equation}
with the independent fit parameters $A$, $C$, and $D$.
For the fit to this formula, we disregard ensembles which are not included in our main analysis (\cf \cref{tab:ensembles}), or which are solely used to study finite-volume effects (H105 and S201), since the finite-volume dependence of the $\rho$ masses is not sufficiently constrained by the data.
On each ensemble, we set the $\omega$ mass equal to the $\rho$ mass obtained from \cref{eq:rho_mass_parametrization}, because no lattice data for the $\omega$ masses on our ensembles are available, and the mass splitting between the $\rho$ and $\omega$ mesons is small.
This means that the true $\omega$ mass on our ensembles is probably much closer to the ensemble-dependent $\rho$ mass than to the physical value of the $\omega$ mass.\footnote{The $\phi$ resonance, on the other hand, is much heavier than the $\rho$ and $\omega$ mesons. In the absence of lattice data for the $\phi$ mass on our ensembles, we thus employ the physical value $M_{\phi, \mathrm{phys}}$ \cite{Workman2022} in our fits.}
The physical pion mass $M_{\pi, \mathrm{phys}}$ is fixed in units of $\sqrt{t_0}$ using its value in the isospin limit \cite{Aoki2014},
\begin{equation}
    M_{\pi, \mathrm{phys}} = M_{\pi, \mathrm{iso}} = \qty{134.8(3)}{MeV} ,
    \label{eq:m_pi_phys}
\end{equation}
\ie we employ $\sqrt{t_{0, \mathrm{phys}}} M_{\pi, \mathrm{phys}} = 0.09881(59)$.
Here, we neglect the uncertainty of $M_{\pi, \mathrm{iso}}$ in \unit{MeV} since it is completely subdominant compared to that of $\sqrt{t_{0, \mathrm{phys}}}$, which enters in the unit conversion and is propagated into the fits (see below).
For the pion masses we use on our ensembles, see \cref{sec:appendix_masses}.

Two of the major benefits of the approach presented in this subsection as compared to the two-step procedure described in the next one are the following:
On the one hand, performing a fit across several ensembles significantly decreases the errors on the resulting radii.
On the other hand, it leads to a much larger number of degrees of freedom for the fit.
This increases the stability against lowering the applied momentum cut considerably.
These advantages have already been noticed in our publication on the isovector electromagnetic form factors, Ref.\@ \cite{Djukanovic2021}, and apply in a similar manner to the data presented in this paper.

We perform several such fits, using different models to describe the lattice-spacing and/or finite-volume dependence and, at the same time, applying various cuts in the pion mass ($M_\pi \leq \qty{0.23}{GeV}$ and $M_\pi \leq \qty{0.27}{GeV}$) and momentum transfer ($Q^2 \leq \qtyrange[range-phrase = {, \ldots,  }, range-units = single]{0.3}{0.6}{GeV^2}$), in order to estimate the corresponding systematic uncertainties.
The variations of the results due to the cuts are in most cases much smaller than their statistical errors.
In any case, these variations will be included in our systematic errors by means of a model average (\cf \cref{sec:model_average} below).
Moreover, the p-values of all our direct fits remain on an acceptable level (\cf \cref{sec:appendix_bchpt_fits}).
We conclude that we do not observe any sign of a breakdown of the B$\chi$PT-expansions in the aforementioned range of pion masses and momentum transfers.

We adopt two different models for lattice artefacts, either based on an additive or a multiplicative \ansatz \cite{Djukanovic2021},
\begin{align}
    \label{eq:GE_add}
    G_E^\mathrm{add}(Q^2) &= G_E^\chi(Q^2) + G_E^a a^2 Q^2 + G_E^L t_0 Q^2 e^{-M_\pi L} , \\
    \label{eq:GM_add}
    G_M^\mathrm{add}(Q^2) &= G_M^\chi(Q^2) + G_M^a \frac{a^2}{t_0} + \kappa_L M_\pi \left( 1 - \frac{2}{M_\pi L} \right) e^{-M_\pi L} + G_M^L t_0 Q^2 e^{-M_\pi L} , \\
    \label{eq:GE_mult}
    G_E^\mathrm{mult}(Q^2) &= G_E^\chi(Q^2) + \frac{G_E^a a^2 Q^2 + G_E^L t_0 Q^2 e^{-M_\pi L}}{t_0 (M_\rho^2 + Q^2)} , \\
    \label{eq:GM_mult}
    G_M^\mathrm{mult}(Q^2) &= G_M^\chi(Q^2) + \frac{G_M^a a^2 / t_0 + G_M^L t_0 Q^2 e^{-M_\pi L}}{t_0 (M_\rho^2 + Q^2)} + \kappa_L M_\pi \left( 1 - \frac{2}{M_\pi L} \right) e^{-M_\pi L} .
\end{align}
The precise form of the multiplicative model has been altered compared to the one used in Ref.\@ \cite{Djukanovic2021}, where the correction terms directly multiplied $G_{E, M}^\chi(Q^2)$.
With our updated, more precise data we have found that such terms containing both $G_{E, M}^{a, L}$ and the LECs [via $G_{E, M}^\chi(Q^2)$] lead to instabilities in the determination of $G_{E, M}^{a, L}$.
This is most probably due to the fit becoming non-linear in the fit parameters.
By contrast, our new model in \cref{eq:GE_mult,eq:GM_mult} is, from a technical point of view, also purely additive and thus linear in the fit parameters, while still capturing the essential contribution of $G_{E, M}^\chi(Q^2)$ to the fall-off of the form factors with rising momentum transfer.\footnote{Note that because we employ $M_\rho$ in place of $M_\omega$ on our ensembles, the expressions in \cref{eq:GE_mult,eq:GM_mult} are valid for the isovector and isoscalar channels alike.}
Fits leaving $\kappa_L$ as a free parameter are unstable, and we therefore fix $\kappa_L$ to the value from heavy-baryon chiral perturbation theory \cite{Beane2004},
\begin{equation}
    \kappa_L = -\frac{m_{N, \mathrm{phys}} g_A^2}{4\pi F_\pi^2} \tau_3 .
    \label{eq:kappaL_HBChPT}
\end{equation}
Here, $\tau_3 = +1$ for the proton and $\tau_3 = -1$ for the neutron.
Following \cref{eq:eff_ff_p_n}, this implies that $\tau_3^{u-d} = +2$ and $\tau_3^{u+d-2s} = 0$.
In total, we have seven different fit models: one without any parametrization of lattice artefacts, three including discretization and/or finite-volume effects with the additive model of \cref{eq:GE_add,eq:GM_add}, and three corresponding ones using the multiplicative prescription of \cref{eq:GE_mult,eq:GM_mult}.
The results for all of them are collected in \cref{sec:appendix_bchpt_fits}.

The inclusion of a term describing lattice artefacts requires the use of Gaussian priors for the relevant coefficients to stabilize the fits.
For this purpose, we first perform fits to ensembles at $M_\pi \approx \qty{0.28}{GeV}$ only (N101, H105, N451, N200, S201, and J303; \cf \cref{tab:ensembles}).
Here, we have relatively precise data in a wide range of lattice spacings and volumes.
For these fits, we use a cut in $Q^2$ at \qty{0.6}{GeV^2} and a simultaneous description of the lattice-spacing and finite-volume dependence.
The coefficients for the correction terms as determined from the fits, together with their associated errors, are then employed as priors for the final fits to the ensembles satisfying the aforementioned cuts in the pion mass.
The only exception are the coefficients $G_E^a$ parametrizing the lattice-spacing dependence of the isovector electric form factor:
our data for $G_E^{u-d}$ is sufficiently precise even at low pion masses to allow a determination of $G_E^a$, so that we can leave it as a free parameter.

Since the number of configurations and hence the number of Jackknife samples differ between ensembles (\cf \cref{tab:ensembles}), we use parametric bootstrap to resample the distributions on each ensemble.
With all mean values for the form factors entering a specific fit, the nucleon and the pion mass, as well as their covariance matrix, we draw \num{10000} random samples from a corresponding multivariate Gaussian distribution.
The covariance matrix one can build from these samples is consistent with the original covariance matrix.
Moreover, the parametric bootstrap procedure enables us to account for the errors of the scale parameters $t_0^\mathrm{sym}/a^2$ and $\sqrt{t_{0, \mathrm{phys}}}$, as well as $M_\rho / M_\pi$, which are external to this analysis.
Hence, we create an independent random Gaussian distribution for $\sqrt{t_{0, \mathrm{phys}}}$, for $t_0^\mathrm{sym}/a^2$ at each value of $\beta$, and for $M_\rho / M_\pi$ on each ensemble.

From the fits in the isovector and isoscalar channels, we reconstruct the form factors and all derived observables for the proton and neutron.
For this purpose, we build the appropriate linear combinations of the B$\chi$PT formulae and plug in the LECs as determined from the fits to the isovector and isoscalar form factors.
This is the more natural approach both from the perspective of lattice QCD and of chiral perturbation theory:
The form factors in the isospin basis are our primary\footnote{The form factors in the isospin basis are also secondary quantities in the sense that they are functions of gauge averages due to the reweighting and the ratio method [\cf \cref{eq:ratio}].} lattice observables, while the proton and neutron form factors can only be obtained indirectly as linear combinations of them [\cf \cref{eq:eff_ff_p_n}].
For the B$\chi$PT fits, the isospin basis is also advantageous because of the clear separation of the contributing resonances in the isovector and isoscalar channels, so that there are no common fit parameters between the two of them.

\begin{figure*}[p]
    \includegraphics[width=\textwidth]{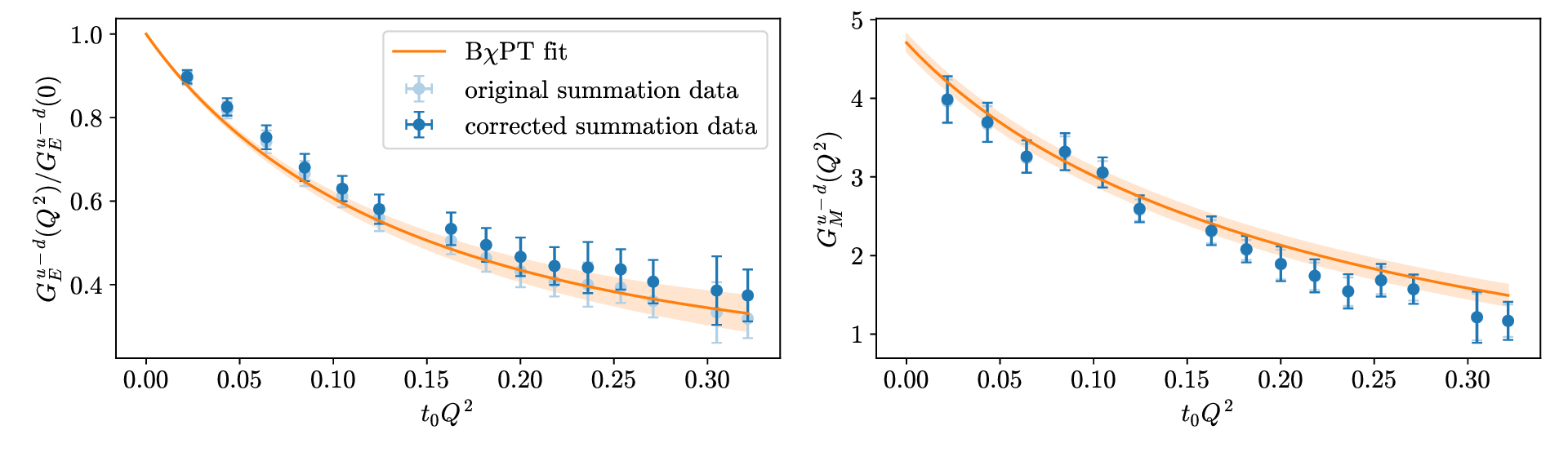}
    \includegraphics[width=\textwidth]{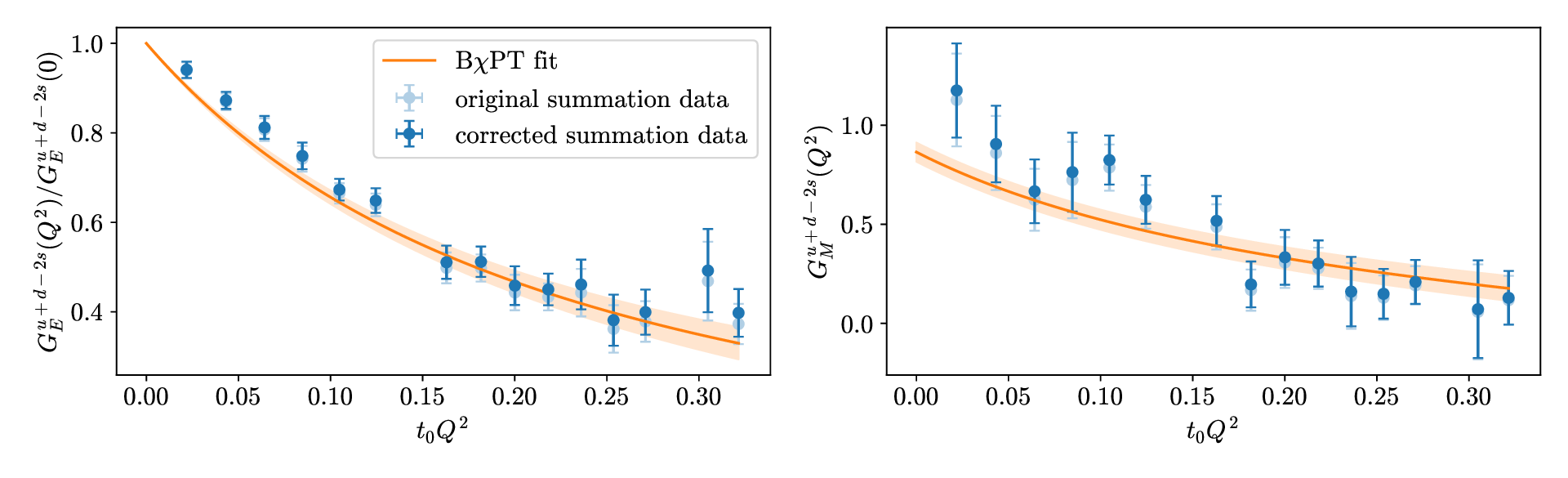}
    \hrule
    \includegraphics[width=\textwidth]{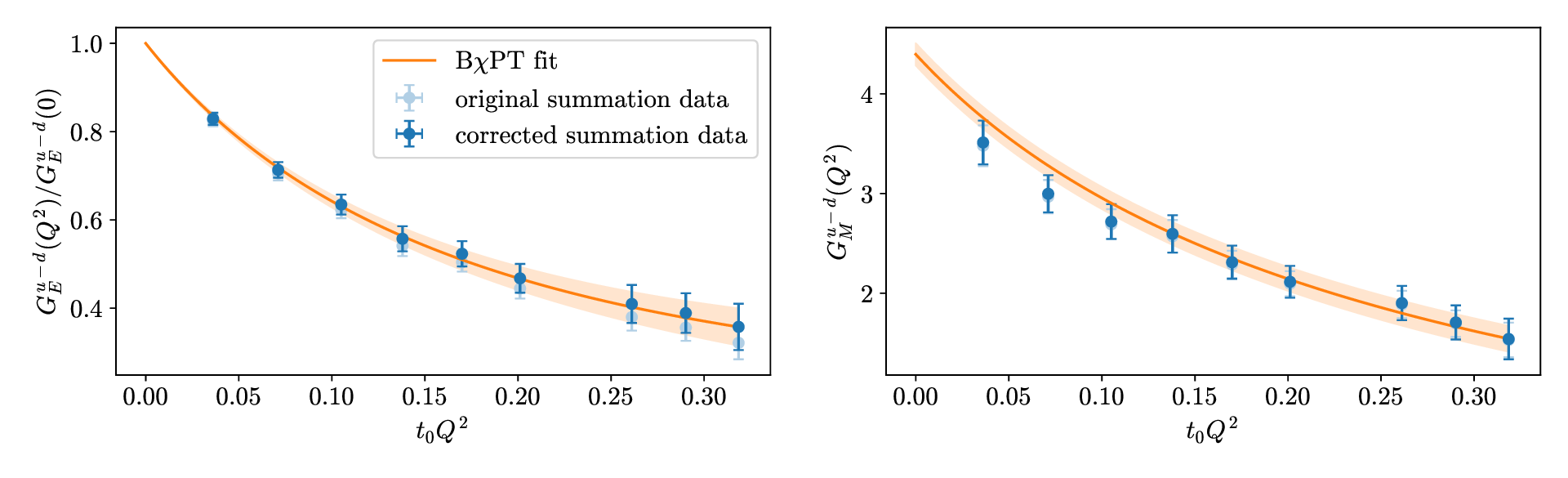}
    \includegraphics[width=\textwidth]{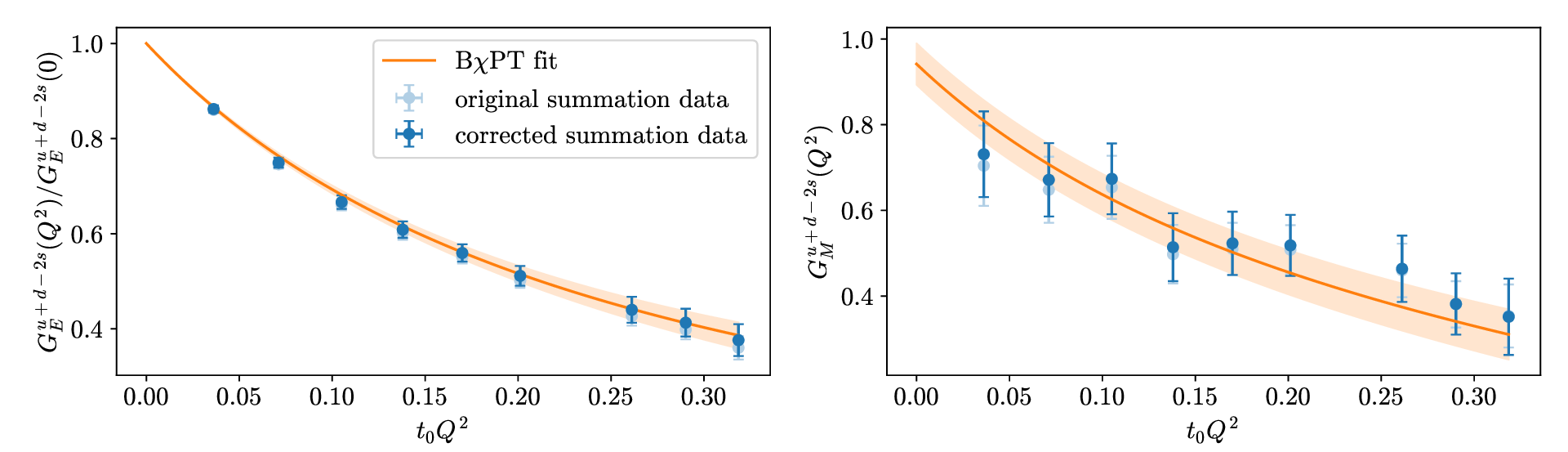}
    \caption{Isovector and isoscalar electromagnetic form factors on the ensembles E250 (upper panel) and E300 (lower panel) as a function of $Q^2$.
      Our original lattice data as obtained from the summation method using the window average are represented by the faint blue points, while the opaque ones have been corrected for the continuum and infinite-volume limit.
      The orange curves and bands depict direct fits with $M_{\pi, \mathrm{cut}} = \qty{0.23}{GeV}$ and $Q^2_\mathrm{cut} = \qty{0.6}{GeV^2}$, evaluated at the pion mass of the respective ensemble, zero lattice spacing, and infinite volume.}
    \label{fig:bchpt_fits}
\end{figure*}

The quality of the direct fits is illustrated in \cref{fig:bchpt_fits} for our two most chiral ensembles E250 and E300.
The fits shown here correspond to the additive model of \cref{eq:GE_add,eq:GM_add} employed to parametrize discretization and finite-volume effects, with $M_{\pi, \mathrm{cut}} = \qty{0.23}{GeV}$ and $Q^2_\mathrm{cut} = \qty{0.6}{GeV^2}$.
In general, the fits describe the data very well.
We observe that the error of the fits is significantly reduced compared to the data points on E250, but only slightly on E300.
The latter is also the case on all other ensembles entering the displayed fits (D200, D450, and C101, which are not shown in \cref{fig:bchpt_fits}).
We conclude that the error reduction on E250 is due to the global fit, \ie the inclusion of several ensembles in one fit, with the data at larger pion masses being more precise than near $M_{\pi, \mathrm{phys}}$.

For the electric form factors on E250, we find a slight deviation between the fit and the data, which is mostly absent in all other cases.
Nevertheless, the p-values of the shown fits are acceptable, with $p > 0.15$ in both channels, because the data points are highly correlated, so that actually fewer degrees of freedom deviate than it naively appears from the plots in \cref{fig:bchpt_fits}.
Also, the fits are more consistent with the data on most of the other ensembles.
We take this as an indication that the form factors on E250 are more heavily affected by correlated statistical fluctuations which are not sufficiently suppressed by our choice of window parameters in the summation method.
In \cref{sec:appendix_E250}, we discuss this point in more detail, concluding that it is in fact not caused by residual excited-state contamination.
Regarding the isovector form factors, the fact that the fit lies somewhat below the data for $G_E^{u-d}$ and somewhat above for $G_M^{u-d}$ on E250, but to a much lesser degree so on ensembles with heavier pion masses, has already been noticed in Ref.\@ \cite{Djukanovic2021}, and is qualitatively confirmed by this updated study.

Note that the curves shown in \cref{fig:bchpt_fits} only correspond to one specific model and are thus not be interpreted as our definitive results for the form factors.
These can be found in \cref{fig:bchpt_fits_model_average} for the proton and neutron.

\subsection{Crosscheck with \texorpdfstring{$z$}{z}-expansion}
\label{sec:zexp}
As an alternative to the direct fits, one can treat the parametrization of the $Q^2$-dependence and the chiral, continuum, and infinite-volume extrapolation as two separate steps.
A model-independent description of the $Q^2$-dependence of the form factors can be achieved by the $z$-expansion \cite{Hill2010},
\begin{align}
    \label{eq:zexp_GE}
    \frac{G_E(Q^2)}{G_E(0)} &= \sum_{k=0}^n a_k z(Q^2)^k , \\
    \label{eq:zexp_GM}
    G_M(Q^2) &= \sum_{k=0}^n b_k z(Q^2)^k ,
\end{align}
with
\begin{equation}
    z(Q^2) = \frac{\sqrt{\tau_\mathrm{cut} + Q^2} - \sqrt{\tau_\mathrm{cut} - \tau_0}}{\sqrt{\tau_\mathrm{cut} + Q^2} + \sqrt{\tau_\mathrm{cut} - \tau_0}} .
    \label{eq:z}
\end{equation}
The parameter $\tau_0$ (not to be confused with the gradient flow scale $t_0$, hence the slightly uncommon nomenclature) is the value of $-Q^2$ which is mapped to $z = 0$.
In this work, we employ $\tau_0 = 0$.
On each ensemble, we set $\tau_\mathrm{cut} = 9 M_\pi^2$ for the isoscalar channel and $\tau_\mathrm{cut} = 4 M_\pi^2$ for the remaining channels, respectively, where $M_\pi$ is the pion mass on the respective ensemble (\cf \cref{sec:appendix_masses}).
For the purpose of the $z$-expansion analysis, we use the form factors of the proton and neutron obtained as described in \cref{sec:appendix_formfactors}.
In analogy to the direct fits, we fit the normalized ratio $G_E(Q^2) / G_E(0)$ and enforce the normalization by fixing $a_0 = 1$.
For the exceptional case of the neutron, where $G_E^n(0) = 0$, we do not normalize $G_E^n(Q^2)$, exclude the point at $Q^2 = 0$, and set $a_0 = 0$.

Using the same strategy to set priors on the coefficients $a_k$ and $b_k$ as in our earlier publication Ref.\@ \cite{Djukanovic2021}, we observe with our updated, more precise data that such priors impose too strict constraints on the $Q^2$-behavior of the form factors.
In particular, the fits with priors tend to follow more closely the data points at large $Q^2$ than those at low $Q^2$, which is undesirable for an extraction of the radii and the magnetic moment.
Hence, we resort to fits without priors, going up to order $n = 2$.
This represents a compromise between the fit function not being unduly stiff and avoiding overfitting especially on ensembles with a bad resolution in $Q^2$.

The errors on the nucleon and pion masses as well as those from the scale setting are included in the analysis using the same parametric bootstrap procedure as for the direct fits.
In order to account for the correlation between $G_E$ and $G_M$, the two form factors are fitted simultaneously even though they do not share any common fit parameters.
However, the different channels (proton, neutron, isovector, and isoscalar) are treated separately, in order to mirror the analysis with the direct fits as closely as possible.

We find that the $z$-expansion describes the data generally very well.
In particular for $G_M$, however, the form factor is in some cases very flat around $Q^2 = 0$.
This seems to be partly due to fluctuations in the data, against which the $z$-expansion is not sufficiently stable.
Consequently, it can be assumed that the magnetic moments and radii determined from these fits are considerably too small.

Using the $z$-expansion, we obtain a set of results for the electromagnetic radii and the magnetic moment on each ensemble.
They are listed in \cref{sec:appendix_zexp} for two different cuts in the momentum transfer ($Q^2 \leq \qty{0.6}{GeV^2}$ and $Q^2 \leq \qty{0.7}{GeV^2}$).
We note that the lowest cut in $Q^2$ which is applicable for all ensembles required for a chiral and continuum extrapolation is in fact roughly \qty{0.6}{GeV^2}.
The ensembles dedicated to the study of finite-volume effects (H105 and S201) do not even have enough data points to permit a $z$-expansion with these momentum cuts.
As the finite-volume dependence is not sufficiently constrained by the $z$-expansion data on the remaining ensembles, we neglect it for the purpose of this crosscheck.

For the chiral and continuum extrapolation of the above datasets, we employ fit formulae inspired by heavy-baryon chiral perturbation theory (HB$\chi$PT) \cite{Goeckeler2005}.
We only take the leading-order dependence on the pion mass into account, since any higher-order coefficients are very poorly constrained by our data, and add terms $\propto a^2$ in order to account for discretization effects.
This leads us to the following \ansaetze for the isovector, proton, and neutron channels,
\begin{align}
    \label{eq:rE2_CCF}
    \frac{\langle r_E^2 \rangle}{t_0} &= A + D \ln(\sqrt{t_0} M_\pi) + E \frac{a^2}{t_0} , \\
    \label{eq:rM2_CCF}
    \frac{\langle r_M^2 \rangle}{t_0} &= A + \frac{D}{\sqrt{t_0} M_\pi} + E \frac{a^2}{t_0} , \\
    \label{eq:muM_CCF}
    \mu_M &= A + B \sqrt{t_0} M_\pi + E \frac{a^2}{t_0} .
\end{align}
The terms in the one-loop HB$\chi$PT expressions for the electromagnetic radii and the magnetic moment with the pion-mass dependence shown in \cref{eq:rE2_CCF,eq:rM2_CCF,eq:muM_CCF} do not contribute in the isoscalar channel \cite{Bernard1998,Goeckeler2005,Bauer2012}.
In the absence of any concrete higher-order HB$\chi$PT results, we employ the following \ansatz for all three isoscalar observables,
\begin{equation}
    A + C t_0 M_\pi^2 + E \frac{a^2}{t_0} .
    \label{eq:CCF_iss}
\end{equation}

The extrapolated results at the physical point are collected in \cref{sec:appendix_CCF} for two different cuts each in the pion mass ($M_\pi \leq \qty{0.27}{GeV}$ and $M_\pi \leq \qty{0.3}{GeV}$) and the momentum transfer ($Q^2 \leq \qty{0.6}{GeV^2}$ and $Q^2 \leq \qty{0.7}{GeV^2}$).
The numbers for $\langle r_E^2 \rangle$ are, with the slight exception of the neutron, stable within their errors and compare well to the results of the direct fits [\cf \cref{eq:rE2_isv_final,eq:rE2_iss_final,eq:rE2_p_final,eq:rE2_n_final} below].
The magnetic radii exhibit somewhat more variation while having a considerably larger error compared with the direct fits.
This is mostly due to the latter fitting $G_E$ and $G_M$ together with common fit parameters, thus leveraging the knowledge that both form factors are governed by the same underlying physics.
For the magnetic moments, the agreement with the direct fits tends to be worse than for the radii:
they are (except in the isoscalar channel) significantly smaller in magnitude than those obtained from the direct fits [\cf \cref{eq:muM_isv_final,eq:muM_iss_final,eq:muM_p_final,eq:muM_n_final}], which are in turn well compatible with the experimentally very precisely known values (\cf \cref{fig:comparison}).

\begin{figure*}[htb]
    \begin{minipage}{0.5\textwidth}
        \includegraphics[width=\textwidth]{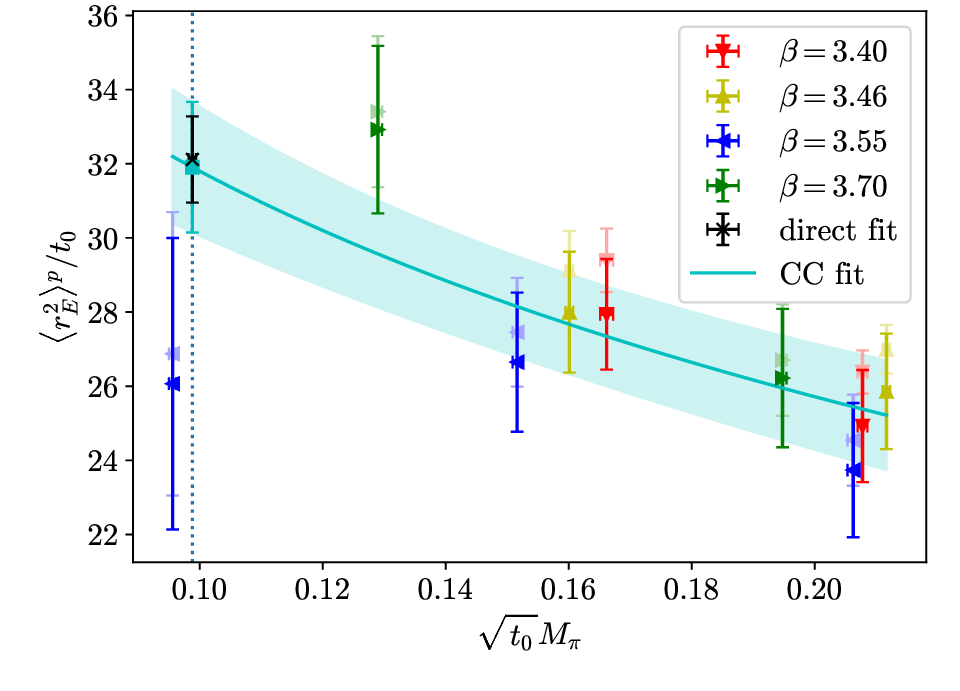}
    \end{minipage}%
    \begin{minipage}{0.5\textwidth}
        \includegraphics[width=\textwidth]{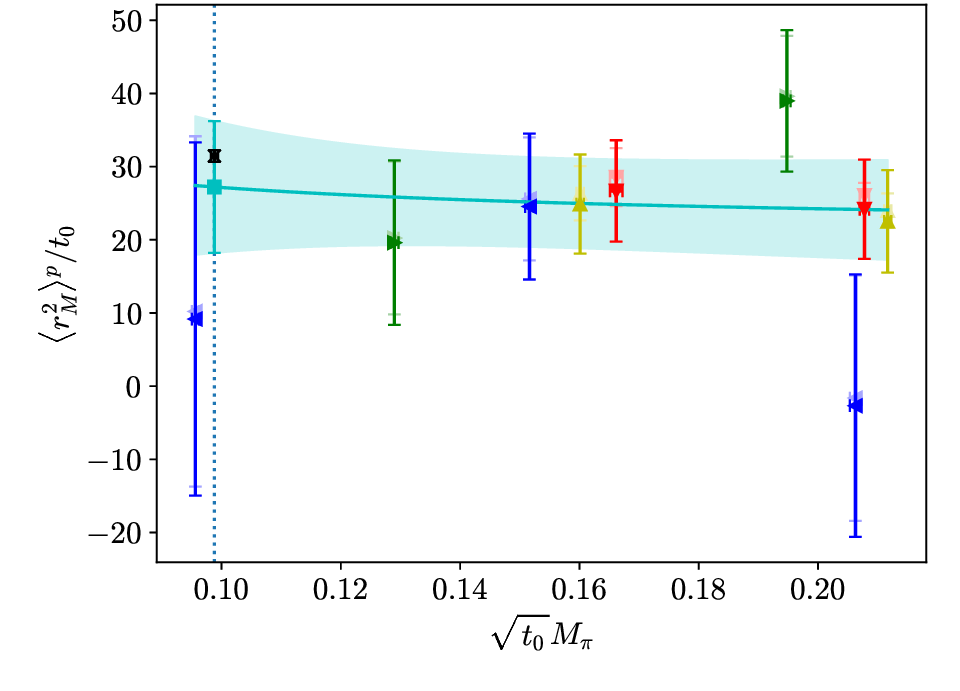}
    \end{minipage}
    \begin{minipage}{0.5\textwidth}
        \includegraphics[width=\textwidth]{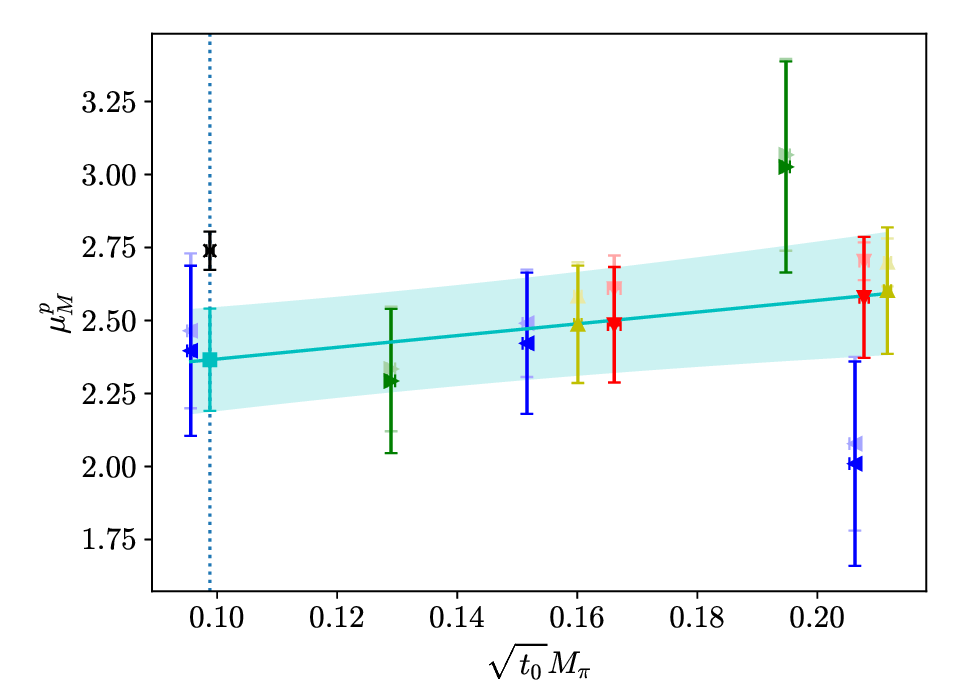}
    \end{minipage}
    \caption{Electromagnetic radii and magnetic moment of the proton as a function of the pion mass.
      The faint symbols represent our original lattice data obtained from a $z$-expansion with $Q^2_\mathrm{cut} = \qty{0.6}{GeV^2}$, while the opaque ones have been corrected for the continuum limit.
      The cyan lines and bands depict an extrapolation fit (CC fit) according to \cref{eq:rE2_CCF,eq:rM2_CCF,eq:muM_CCF}.
      Its results at the physical point are shown as cyan squares and the model-averaged results of the direct fits as black crosses [\cf \cref{eq:rE2_p_final,eq:rM2_p_final,eq:muM_p_final}], with a dotted vertical line at the physical pion mass (in units of $\sqrt{t_0}$) to guide the eye.}
    \label{fig:CCF}
\end{figure*}

For illustration, we display in \cref{fig:CCF} the extrapolation for the proton using $M_{\pi, \mathrm{cut}} = \qty{0.3}{GeV}$ and $Q^2_\mathrm{cut} = \qty{0.6}{GeV^2}$.
We note that individual ensembles can deviate rather strongly from the fit curve, which is most apparent for the magnetic observables on J303 and N200.
This is probably due to the low momentum resolution on these two ensembles, which implies a long extrapolation to $Q^2 = 0$, where the radius and the magnetic moment are defined, but no lattice data is available.
We also note that the relative weights in the extrapolation fit do not reflect the number of $Q^2$-points entering the $z$-expansion, which is different on each ensemble.
In this sense the two-step process, first performing $z$-expansion fits and subsequently extrapolating, masks the relative paucity of data points at small momentum transfer for some ensembles.

We conclude that the direct fits are superior to the analysis using a $z$-expansion followed by a chiral and continuum extrapolation, in particular for the description of the magnetic form factor:
they are more stable against fluctuations on individual momenta or ensembles and take more information about the physical properties of the form factors into account, which helps in reducing the errors.
In cases where the two-step procedure based on the $z$-expansion is stable and trustworthy, both methods give consistent results.
Therefore, we do not find any evidence that the functional forms employed by the direct B$\chi$PT fits introduce a systematic bias.
We also remark that there is no meaningful possibility of averaging the results from the direct B$\chi$PT fits with those from the $z$-expansion because the latter are, in particular for the magnetic quantities, much less precise and simply not competitive.

\section{Model average and final results}
\label{sec:model_average}
For the reasons stated above, we favor the analysis based on the direct fits and hence restrict our presentation of the final results to this method.
Within this approach, we have no strong \apriori preference for one specific setup, and thus determine our final results and total errors from averages over different fit models and kinematic cuts.
For this purpose, we employ weights derived from the Akaike Information Criterion (AIC) \cite{Akaike1973,Akaike1974}.

The applicability of the AIC in its original form is called into question in the presence of Bayesian priors for some fit parameters or when only subsets of data are fitted as a result of applying kinematical cuts. Therefore, we apply the modified criterion BAIC proposed in Ref.\@ \cite{Neil2022}.
It reads
\begin{equation}
    \mathrm{BAIC}_i = \chi^2_{\mathrm{noaug, min}, i} + 2n_{f, i} + 2n_{c, i} ,
    \label{eq:BAIC}
\end{equation}
where $\chi^2_{\mathrm{noaug, min}, i}$ denotes the minimum of the $\chi^2$-function for the $i$-th model excluding the contribution of the priors.
$n_{f, i}$ is the corresponding number of fit parameters and $n_{c, i}$ the number of cut data points.
This criterion takes the goodness of fit into account, while at the same time penalizing a reduction of the degrees of freedom that may result either from the introduction of further fit parameters or from cutting away data points.
For the proton and neutron observables, which are derived from two separate fits to the isovector and isoscalar form factors, the BAIC is obtained as the sum of the BAIC values of both fits (which would be the BAIC resulting from a combined fit with the cross-correlations between the two channels set to zero\footnote{Estimating these correlations is not feasible because the size of the resulting covariance matrices would be much larger than the number of available configurations on some ensembles.}).
For the weighting of the different models, one can use \cite{Burnham2004,Borsanyi2015,Neil2022}
\begin{equation}
    w_i^\mathrm{BAIC} = \frac{e^{-\mathrm{BAIC}_i/2}}{\sum_j e^{-\mathrm{BAIC}_j/2}} .
    \label{eq:BAIC_weights}
\end{equation}

When computing these weights for our set of models, it turns out that the BAIC strongly prefers the fits with the least stringent cut in $Q^2$.
This is due to the relatively large number of data points which is cut away by lowering $Q^2_\mathrm{cut}$ in a fit across several ensembles.
The effect is enhanced by our two most chiral ensembles E250 and E300, which feature a comparatively large density of $Q^2$-points.
Since the radii and the magnetic moment are defined in terms of the low-$Q^2$ behavior of the form factors, a stricter cut in $Q^2$ is theoretically better motivated for an extraction of these quantities.
Hence, we employ \cref{eq:BAIC_weights} for each value of $Q^2_\mathrm{cut}$ at a time to weight the remaining variations, \ie the pion-mass cut and the modelling of lattice artefacts.
Using these separately normalized BAICs, we finally apply a flat weight function to the estimates originating from the different $Q^2_\mathrm{cut}$.
This prescription, which we dub $\overline{\mathrm{BAIC}}$, ensures that the stricter cuts in $Q^2$, and thus our low-momentum data, have a strong influence on our final results.

In order to estimate the statistical and systematic uncertainties of our model averages, we adopt the procedure from Ref.\@ \cite{Borsanyi2021}, which we briefly sketch in the following.
To this end, one treats the model-averaged quantity as a random variable with a cumulative distribution function (CDF) adding up from the weighted CDFs of the individual models,
\begin{equation}
    P^x(y) = \sum_{i = 1}^{N_M} \frac{w_i^{\overline{\mathrm{BAIC}}}}{N_B} \sum_{n = 1}^{N_B} \Theta(y - x_{i, n}) .
    \label{eq:model_average_CDF}
\end{equation}
Here, the outer sum runs over our $N_M = 4 \times 2 \times 7 = 56$ models with the associated weights $w_i^{\overline{\mathrm{BAIC}}}$ computed as explained above.
The inner sum runs over the $N_B = \num{10000}$ bootstrap samples obtained from our resampled analysis (\cf \cref{sec:bchpt_fits}).
$\Theta$ denotes the Heaviside step function and $x_{i, n}$ the estimate for the observable $x$ on the $n$-th sample and using the $i$-th model.
Due to the large number of bootstrap samples $N_B$, the distribution in \cref{eq:model_average_CDF} is effectively smoothed in spite of being a sum of step functions.
The final value and the total error are easily read off from this distribution as the median and the quantiles which would correspond to the central $1\,\sigma$ of an effective Gaussian distribution, respectively.
In order to isolate the statistical and systematic errors, one can scale the width of the bootstrap distributions entering \cref{eq:model_average_CDF} by a factor of $\lambda$.
Under the assumption that such a rescaling of the errors of the individual model results only affects the statistical, but not the systematic error, one can separate these two contributions as demonstrated in Ref.\@ \cite{Borsanyi2021}.
We use $\lambda = 2$ as in Ref.\@ \cite{Borsanyi2021}, but we remark that the results of this method are essentially independent of the choice of $\lambda$ for our data as long as $\lambda \gtrsim 2$.

\begin{figure*}[htb]
    \includegraphics[width=\textwidth]{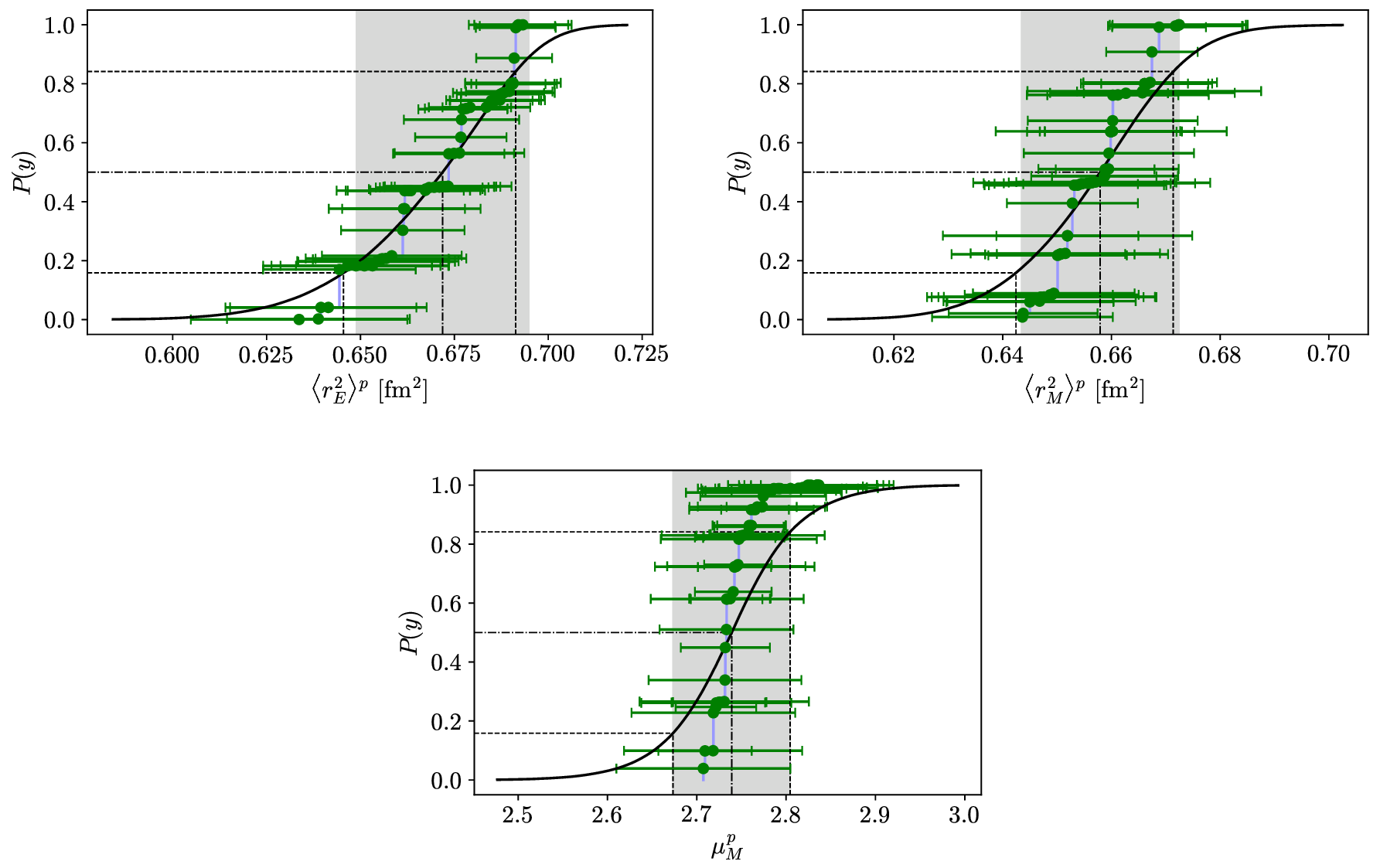}
    \caption{Cumulative distribution function of the electromagnetic radii and the magnetic moment of the proton for all fitted models.
      The green points depict the central values and errors of the individual fit results.
      The thick black line shows the weighted CDF according to \cref{eq:model_average_CDF}.
      For comparison, we also include a CDF based on the central values only, \ie $P^x(y) = \sum_{i = 1}^{N_M} w_i^{\overline{\mathrm{BAIC}}} \Theta(y - x_i)$, which is displayed by the light blue line.
      The dashed-dotted and dashed lines indicate the median and the central \qty{68}{\percent} quantiles, respectively.
      The gray bands, on the other hand, depict the symmetrized errors quoted in \cref{eq:rE2_p_final,eq:rM2_p_final,eq:muM_p_final}.}
    \label{fig:model_average}
\end{figure*}

The collection of results for the electromagnetic radii and the magnetic moment of the proton together with the CDF obtained as explained above is displayed in \cref{fig:model_average}.
One can see that approximately the expected fraction of results lie within the \qty{68}{\percent} quantiles of the averaged distribution.
Moreover, the symmetrized errors as shown by the gray bands agree well with the (generally non-symmetric) quantiles of the distributions, which are indicated by the dashed lines.
Since this statement holds for all four channels, we quote the symmetrized errors together with our final results,
{
    \allowdisplaybreaks
    \begin{align}
        \label{eq:rE2_isv_final}
        \langle r_E^2 \rangle^{u-d}    &= \qty{0.785(22)(26)}{fm^2} , \\
        \label{eq:rM2_isv_final}
        \langle r_M^2 \rangle^{u-d}    &= \qty{0.663(11)(8)}{fm^2} , \\
        \label{eq:muM_isv_final}
        \mu_M^{u-d}                    &= \num{4.62(10)(7)} , \\[\defaultaddspace]
        \label{eq:rE2_iss_final}
        \langle r_E^2 \rangle^{u+d-2s} &= \qty{0.554(18)(13)}{fm^2} , \\
        \label{eq:rM2_iss_final}
        \langle r_M^2 \rangle^{u+d-2s} &= \qty{0.657(30)(31)}{fm^2} , \\
        \label{eq:muM_iss_final}
        \mu_M^{u+d-2s}                 &= \num{2.47(11)(10)} , \\[\defaultaddspace]
        \label{eq:rE2_p_final}
        \langle r_E^2 \rangle^p        &= \qty{0.672(14)(18)}{fm^2} , \\
        \label{eq:rM2_p_final}
        \langle r_M^2 \rangle^p        &= \qty{0.658(12)(8)}{fm^2} , \\
        \label{eq:muM_p_final}
        \mu_M^p                        &= \num{2.739(63)(18)} , \\[\defaultaddspace]
        \label{eq:rE2_n_final}
        \langle r_E^2 \rangle^n        &= \qty{-0.115(13)(7)}{fm^2} , \\
        \label{eq:rM2_n_final}
        \langle r_M^2 \rangle^n        &= \qty{0.667(11)(16)}{fm^2} , \\
        \label{eq:muM_n_final}
        \mu_M^n                        &= \num{-1.893(39)(58)} .
    \end{align}
}%
We find that we can obtain the magnetic radii of the proton and neutron to a very similar precision to their respective electric radii.

We have compared the above numbers to the results of two alternative averaging strategies: the BAIC weights of \cref{eq:BAIC_weights} applied to all variations, \ie also the cut in $Q^2$, or a naive (flat) average imposing a p-value cut at \qty{1}{\percent}.
For the latter, we have used the average statistical uncertainty, and the variance determined from the spread of the fit results as a systematic error estimate \cite{Carrasco2014}.
While this method is robust, it is also very conservative and susceptible to overestimating the true errors.
The \enquote{plain} BAIC, on the other hand, drastically underestimates the systematic error for observables which display a non-negligible dependence on $Q^2_\mathrm{cut}$.
For these reasons, we adopt the model averaging procedure $\overline{\mathrm{BAIC}}$ explained above.
We note, however, that the results of all three methods are compatible within errors.

\begin{figure*}[htb]
    \includegraphics[width=\textwidth]{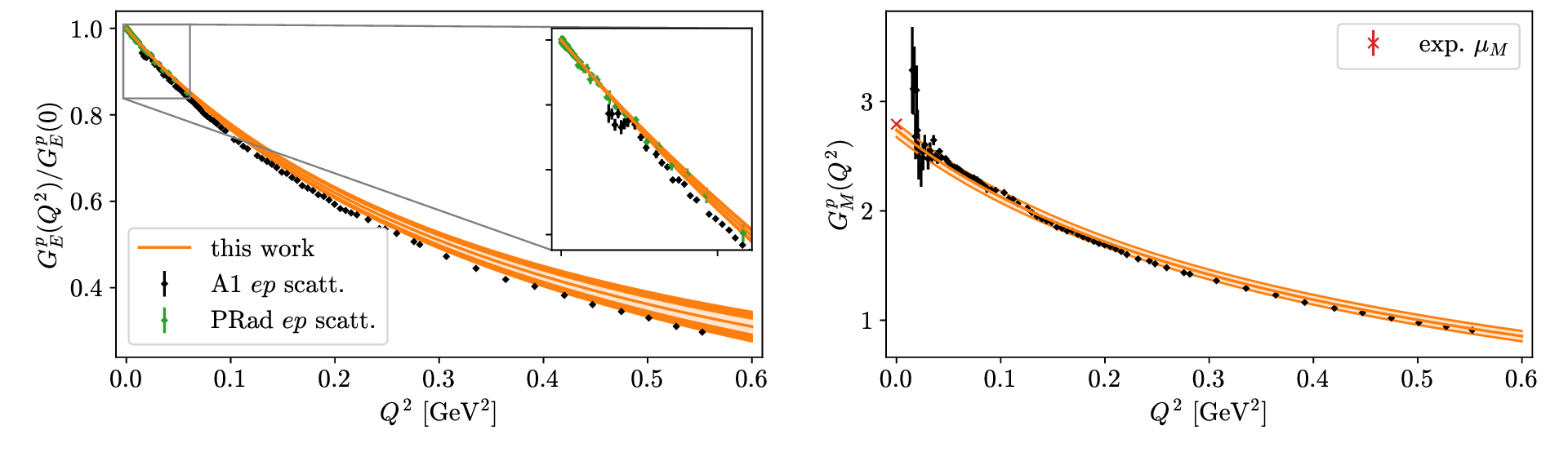}
    \includegraphics[width=\textwidth]{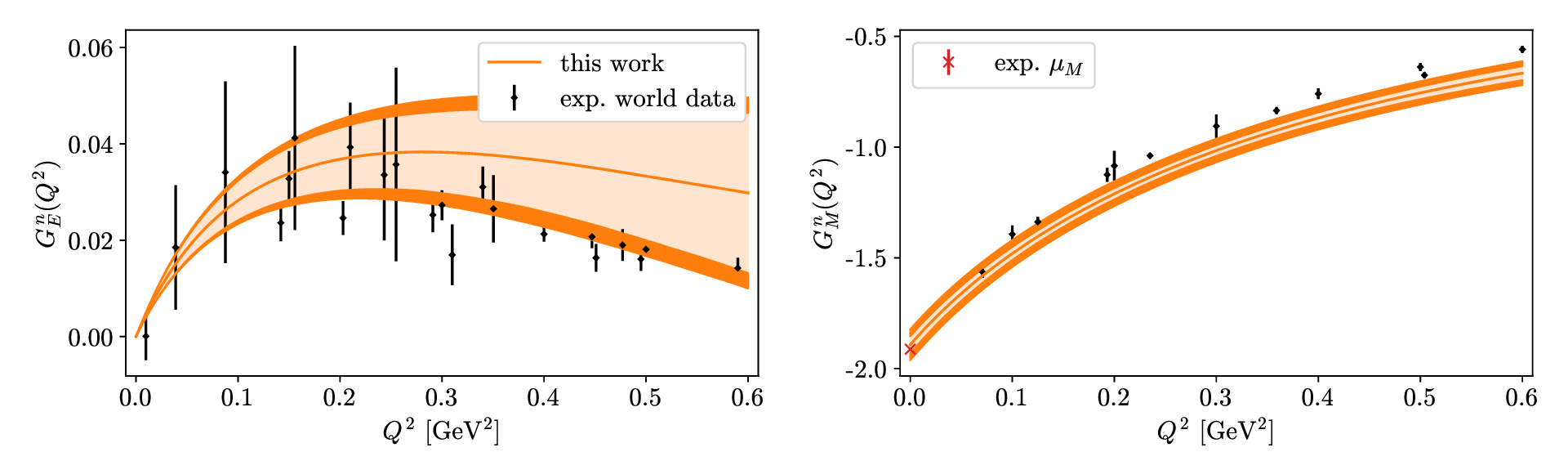}
    \caption{Electromagnetic form factors of the proton and neutron at the physical point as a function of $Q^2$.
      The orange curves and bands correspond to our final results with their full uncertainties obtained as model averages over the different direct fits.
      The light orange bands indicate the statistical uncertainties only.
      For the proton, the black diamonds represent the experimental $ep$-scattering data from A1 \cite{Bernauer2014} obtained using Rosenbluth separation, and the green diamonds the corresponding data from PRad \cite{Xiong2019}.
      For the neutron, the black diamonds show the experimental world data collected in Ref.\@ \cite{Ye2018}.
      The experimental values of the magnetic moments \cite{Workman2022} are depicted by red crosses.}
    \label{fig:bchpt_fits_model_average}
\end{figure*}

In analogy to the radii and the magnetic moments, one can also average the form factors evaluated at the physical point and at particular values of $Q^2$ over the model variations.
The results are plotted in \cref{fig:bchpt_fits_model_average} for the proton and neutron and are directly compared to experimental data.
For the proton, one can observe a moderate deviation in the slope of the electric form factor between our result and that of the A1 Collaboration \cite{Bernauer2014} over the whole range of $Q^2$.
As shown in the inset, the slope of our electric form factor at low $Q^2$ is much closer to that of the PRad experiment \cite{Xiong2019} than to that of Ref.\@ \cite{Bernauer2014}.
The magnetic form factor, on the other hand, agrees well with that of Ref.\@ \cite{Bernauer2014}.
For the neutron, we compare with the collected experimental world data \cite{Ye2018}, which are largely compatible with our curves within our quoted errors.
Only the slope of our magnetic form factor differs somewhat from experiment.
Furthermore, our results reproduce within their errors the experimental values of the magnetic moments both of the proton and of the neutron \cite{Workman2022}.

Our updated radii and magnetic moment in the isovector channel [\cf \cref{eq:rE2_isv_final,eq:rM2_isv_final,eq:muM_isv_final}] agree well with our previously published results \cite{Djukanovic2021}, with similar errors on the electric radius and the magnetic moment, and an improved error on the magnetic radius.

\begin{figure*}[htb]
    \includegraphics[width=\textwidth]{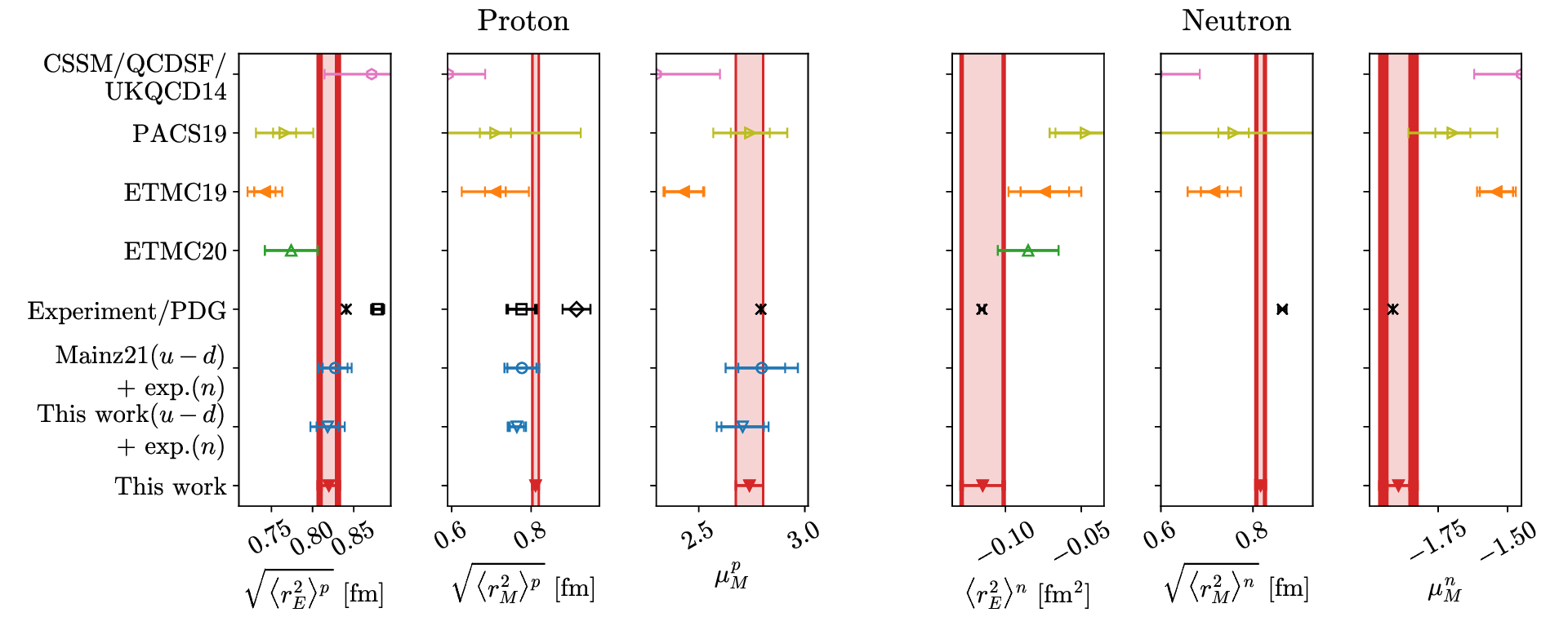}
    \caption{Comparison of our best estimates (red downward-pointing triangle) for the electromagnetic radii and the magnetic moments of the proton and neutron with other lattice calculations, \ie Mainz21 \cite{Djukanovic2021} (blue circle), ETMC20 \cite{Alexandrou2020} (green upward-pointing triangle), ETMC19 \cite{Alexandrou2019} (orange leftward-pointing triangle), PACS19 \cite{Shintani2019} (yellow rightward-pointing triangle), and CSSM/QCDSF/UKQCD14 \cite{Shanahan2014a,Shanahan2014} (pink hexagon).
      Only studies with filled markers, \ie ETMC19 and this work, include disconnected contributions and hence represent a full lattice calculation.
      The Mainz21 values for the proton have been computed by combining their isovector results with the PDG values for the neutron \cite{Workman2022}.
      We also show this estimate using our updated isovector results from this work (blue downward-pointing triangle).
      The experimental values for the neutron and for $\mu_M^p$ are taken from PDG \cite{Workman2022} (black cross).
      The two data points for $\sqrt{\langle r_E^2 \rangle^p}$ represent the values from PDG \cite{Workman2022} (cross) and Mainz/A1 \cite{Bernauer2014} (square), respectively.
      The two data points for $\sqrt{\langle r_M^2 \rangle^p}$, on the other hand, depict the reanalysis of Ref.\@ \cite{Lee2015} either using the world data excluding that of Ref.\@ \cite{Bernauer2014} (diamond) or using only that of Ref.\@ \cite{Bernauer2014} (square).
      For ease of comparison, the red bands show our final results with the full uncertainty, with the light bands indicating the statistical errors.}
    \label{fig:comparison}
\end{figure*}

In \cref{fig:comparison}, we compare our results for the proton and neutron [\cf \cref{eq:rE2_p_final,eq:rM2_p_final,eq:muM_p_final,eq:rE2_n_final,eq:rM2_n_final,eq:muM_n_final}] to recent lattice determinations and to the experimental values.
We remark that the only other complete lattice study including disconnected contributions is Ref.\@ \cite{Alexandrou2019}, which, however, does not perform a continuum and infinite-volume extrapolation.
Our estimates for the electric radii of the proton and neutron are larger in magnitude than the results of Refs.\@ \cite{Alexandrou2019,Alexandrou2020,Shintani2019}, while Ref.\@ \cite{Shanahan2014} quotes an even larger central value for $\sqrt{\langle r_E^2 \rangle^p}$.
We stress that any difference between our estimate and previous lattice calculations is not related to our preference for direct fits to the form factors, as opposed to the more traditional analysis via the $z$-expansion.
In fact, the $z$-expansion approach yields similar values for our data.
Furthermore, we obtain results for the magnetic moments of the proton and neutron, as well as for $\sqrt{\langle r_M^2 \rangle^n}$, which are considerably larger in magnitude than that of Refs.\@ \cite{Alexandrou2019,Shanahan2014a}, while being compatible with that of Ref.\@ \cite{Shintani2019}.
This improves the agreement with the experimental values \cite{Workman2022}.
In the case of the magnetic moments, the latter are very precisely known and are reproduced by our estimates within our quoted uncertainties.
For $\sqrt{\langle r_M^2 \rangle^n}$, we observe nevertheless a $3.2\,\sigma$ tension between our result and the PDG value (after combining all errors in quadrature).
On the level of the form factor $G_M^n$ evaluated at any particular value of $Q^2$, however, the discrepancy is much smaller, as can be seen from \cref{fig:bchpt_fits_model_average} (bottom right).
For $\sqrt{\langle r_M^2 \rangle^p}$, our result is only about $1.2$ combined standard deviations larger than that of Ref.\@ \cite{Alexandrou2019}.
We note that our results for the isoscalar radii [\cf \cref{eq:rE2_iss_final,eq:rM2_iss_final}] are larger than those of Ref.\@ \cite{Alexandrou2019} by a greater amount, while $\mu_M^{u+d-2s}$ compares well between our study and Ref.\@ \cite{Alexandrou2019}.

For the electric and magnetic radii of the proton, the experimental situation is much less clear than for the magnetic moment.
As is the case for most of the other recent lattice calculations \cite{Alexandrou2020,Alexandrou2019,Shintani2019}, our result for $\sqrt{\langle r_E^2 \rangle^p}$ is much closer to the PDG value \cite{Workman2022}, which is completely dominated by muonic hydrogen spectroscopy, than to the A1 $ep$-scattering result \cite{Bernauer2014}:
While we only observe a very mild $1.5\,\sigma$ tension with the former, we disagree at the $3.7\,\sigma$ level with the latter (after combining all errors in quadrature).
We note that we achieve an even better $0.6 \sigma$ agreement with the recent $ep$-scattering experiment by PRad \cite{Xiong2019}, which has also yielded a small electric radius of the proton.
For $\sqrt{\langle r_M^2 \rangle^p}$, on the other hand, our estimate is well compatible with the value inferred from the A1 experiment by the analyses \cite{Bernauer2014,Lee2015} and exhibits a sizable $2.8\,\sigma$ tension with the other collected world data \cite{Lee2015}.
As can be seen from \cref{fig:bchpt_fits_model_average} (top right), the good agreement with A1 is not only observed in the magnetic radius, but also for the $Q^2$-dependence of the magnetic form factor over the whole range of $Q^2$ under study.
We note that the dispersive analysis of the Mainz/A1 and PRad data in Ref.\@ \cite{Lin2021a} has yielded a significantly larger magnetic radius [$\sqrt{\langle r_M^2 \rangle^p} = \qty{0.847(4)(4)}{fm}$] than the $z$-expansion-based analysis of the Mainz/A1 data in Ref.\@ \cite{Lee2015}.
The former value also exhibits a $3.4\,\sigma$ tension with our result, which is partly due to its substantially smaller error compared to Ref.\@ \cite{Lee2015}.
Possible reasons for this discrepancy include unaccounted-for isospin-breaking effects.

Our statistical and systematic error estimates for the electric radii and magnetic moments are commensurate with the other lattice studies, while being substantially smaller for the magnetic radii.
We remark that the missing data point at $Q^2 = 0$ complicates the extraction of the magnetic low-$Q^2$ observables in most recent lattice determinations, especially for $z$-expansion fits on individual ensembles.
The direct approach has, additionally to combining information from several ensembles and from $G_E$ and $G_M$, less freedom and by itself allows for considerably less variation in the form factors at low $Q^2$.
We believe this to be responsible, in large part, for the small errors we find in the magnetic radii.

\section{Conclusions}
\label{sec:conclusions}
In this paper, we have investigated the electromagnetic form factors of the proton and neutron in lattice QCD with $2 + 1$ flavors of dynamical quarks at the physical point including both quark-connected and -disconnected contributions.
For the precise and effective computation of the latter, we have made use of a split-even estimator, \ie the one-end trick \cite{McNeile2006,Boucaud2008,Giusti2019}.
Systematic effects due to excited states have been accounted for by using the summation method with a conservative choice for the window in the minimal source-sink separation over which the summation fits have been averaged.
By matching our lattice results with the predictions from covariant baryon chiral perturbation theory, we have performed simultaneous fits to the $Q^2$-, pion-mass, lattice-spacing, and finite-volume dependence of the form factors in the isospin basis.
From these fits, the electromagnetic radii and magnetic moments of the proton and neutron have been extracted.
We thus obtain the first complete lattice results for these quantities which have a full error budget, \ie all relevant systematic effects are taken into account.
Our final estimates can be found in \cref{eq:rE2_isv_final,eq:rM2_isv_final,eq:muM_isv_final,eq:rE2_iss_final,eq:rM2_iss_final,eq:muM_iss_final,eq:rE2_p_final,eq:rM2_p_final,eq:muM_p_final,eq:rE2_n_final,eq:rM2_n_final,eq:muM_n_final}.

As an important benchmark, we reproduce the experimentally very precisely known magnetic moments within our quoted uncertainties.
Our result for the electric (charge) radius of the proton is much closer to the value extracted from muonic hydrogen spectroscopy \cite{Antognini2013} and recent $ep$-scattering experiments \cite{Xiong2019} than to the A1 $ep$-scattering result \cite{Bernauer2014}.
For the magnetic radius, on the other hand, our estimate is compatible with the analyses \cite{Bernauer2014,Lee2015} of the A1 data, while being in tension with the other collected world data \cite{Lee2015}.
In summary, we contribute additional evidence to suggest that lattice calculations agree with the emerging consensus about the experimental value of the electric proton radius \cite{Hammer2020,Tiesinga2021,Antognini2022}.
Meanwhile, the results for the magnetic proton radius require further investigation.

For lattice studies of the electromagnetic form factors, the excited-state contamination remains an important source of systematic uncertainty.
Using the summation method, the signal gets lost in the exponentially growing noise very quickly after the plateau region is reached.
This renders firm statements about the exact location of the plateau impossible.
A promising strategy to tackle this issue, besides drastically increasing statistics at large source-sink separations, is to perform a dedicated study of the excitation spectrum.
To improve on the systematic error due to the continuum and infinite-volume extrapolation, a larger range of lattice spacings and volumes at or near the physical pion mass is necessary.
Because of concerns regarding the algorithmic stability of our simulations, the production of coarse and light ensembles is not feasible in our current setup.
We are, however, working on the production of a fine ensemble at the physical pion mass, which would help to further constrain both the chiral interpolation and the continuum extrapolation.
Moreover, its large volume implies a high density of $Q^2$-points, which is crucial for an accurate extraction of the radii.
We plan to update our analysis including this new ensemble in a future publication.

\begin{acknowledgments}
    This research is partly supported by the Deutsche Forschungsgemeinschaft (DFG, German Research Foundation) through project HI 2048/1-2 (project No.\@ 399400745) and through the Cluster of Excellence \enquote{Precision Physics, Fundamental Interactions and Structure of Matter} (PRISMA${}^+$ EXC 2118/1) funded within the German Excellence Strategy (project ID 39083149).
    Calculations for this project were partly performed on the HPC clusters \enquote{Clover} and \enquote{HIMster2} at the Helmholtz Institute Mainz.
    Other parts were conducted using the supercomputer \enquote{Mogon 2} offered by Johannes Gutenberg University Mainz (\url{https://hpc.uni-mainz.de}), which is a member of the AHRP (Alliance for High Performance Computing in Rhineland Palatinate, \url{https://www.ahrp.info}) and the Gauss Alliance e.V.
    The authors also gratefully acknowledge the John von Neumann Institute for Computing (NIC) and the Gauss Centre for Supercomputing e.V. (\url{https://www.gauss-centre.eu}) for funding this project by providing computing time on the GCS Supercomputer JUWELS at Jülich Supercomputing Centre (JSC) through projects CHMZ21, CHMZ36, NUCSTRUCLFL, and GCSNUCL2PT.

    Our programs use the QDP++ library \cite{Edwards2005} and deflated SAP+GCR solver from the openQCD package \cite{Luescher2013}, while the contractions have been explicitly checked using the Quark Contraction Tool \cite{Djukanovic2020}.
    We thank Simon Kuberski for providing the improved reweighting factors \cite{Kuberski2023} for the gauge ensembles used in our calculation.
    Moreover, we are grateful to our colleagues in the CLS initiative for sharing the gauge field configurations on which this work is based.
\end{acknowledgments}

\bibliography{literature.bib}

\appendix
\section{Pion and nucleon masses}
\label{sec:appendix_masses}
Here, we tabulate the nucleon masses on the gauge ensembles listed in \cref{tab:ensembles} which we employ in the kinematical prefactors entering the effective form factors and in the definition of $Q^2$, as well as the corresponding pion masses used for the extrapolation to the physical point and for setting $\tau_\mathrm{cut}$ in the $z$-expansion fits.

Note that the values differ from the the ones used in our earlier studies \cite{Djukanovic2021,Harris2019}.
With significantly increased statistics on many ensembles, we now aim for an analysis which is as self-contained as possible.
Hence, the nucleon and pion masses are extracted from the two-point functions of the nucleon and pion at zero momentum, respectively, employing a setup that matches the one of our main analysis as closely as possible.

For the nucleon, we use $\Gamma^p = \frac{1}{2}(1 + \gamma_0)$, and the highest available statistics in terms of sources.
Since we want to utilize much larger source-sink separations than for the construction of the disconnected part of the three-point function, however, we need to ensure that boundary effects are sufficiently suppressed on oBC ensembles.
Hence, we impose a minimal temporal distance of \qty{4}{fm} that the source of the nucleon propagating towards the boundary has to keep from the latter.
Afterwards, we average the two-point functions of the forward- and backward-propagating nucleon.
On pBC ensembles, such a constraint is obviously not required, and all sources can be used for the forward- as well as the backward-propagating nucleon.
On the oBC ensemble S201, all sources are placed on a single timeslice which is less than \qty{4}{fm} from the lower temporal boundary of the lattice, so that only the forward-propagating nucleon can be used.

The two-point functions of the pion have only been measured on a subset of the sources employed for the nucleon because they are already much more precise.

\begin{table}[htbp]
    \caption{Pion and nucleon masses used in this study (in units of $\sqrt{t_0}$).
      The quoted errors include the error of $\sqrt{t_0^\mathrm{sym}}/a$ \cite{Bruno2017}.}
    \begin{ruledtabular}
        \begin{tabular}{lll}
            ID   & \multicolumn{1}{c}{$\sqrt{t_0} M_\pi$} & \multicolumn{1}{c}{$\sqrt{t_0} M_N$} \\ \hline
            C101 & 0.1662(10)                             & 0.7160(35)                           \\
            N101 & 0.20777(78)                            & 0.7555(22)                           \\
            H105 & 0.2073(19)                             & 0.7704(35)                           \\
            D450 & 0.16010(62)                            & 0.7160(40)                           \\
            N451 & 0.21167(60)                            & 0.7728(22)                           \\
            E250 & 0.09560(59)                            & 0.6882(21)                           \\
            D200 & 0.15162(74)                            & 0.7261(23)                           \\
            N200 & 0.20626(93)                            & 0.7796(27)                           \\
            S201 & 0.2163(13)                             & 0.8323(49)                           \\
            E300 & 0.12898(70)                            & 0.7151(23)                           \\
            J303 & 0.19476(64)                            & 0.7667(22)
        \end{tabular}
    \end{ruledtabular}
\end{table}

\clearpage

\section{Additional crosschecks on the excited-state analysis}
\label{sec:appendix_twostate}
In this appendix, we provide additional crosschecks on our excited-state analysis in order to ensure that ground-state dominance is reached by our preferred procedure which is described in \cref{sec:excited_states}.

\subsection{Summation method}
\begin{figure*}[htb]
    \includegraphics[width=\textwidth]{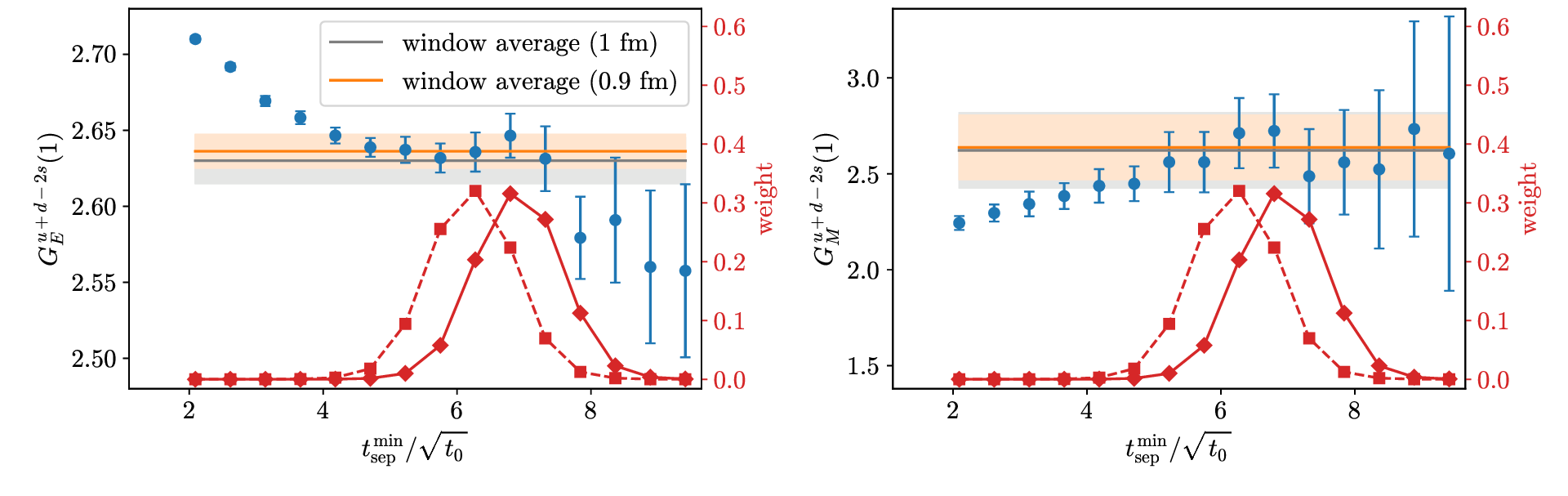}
    \includegraphics[width=\textwidth]{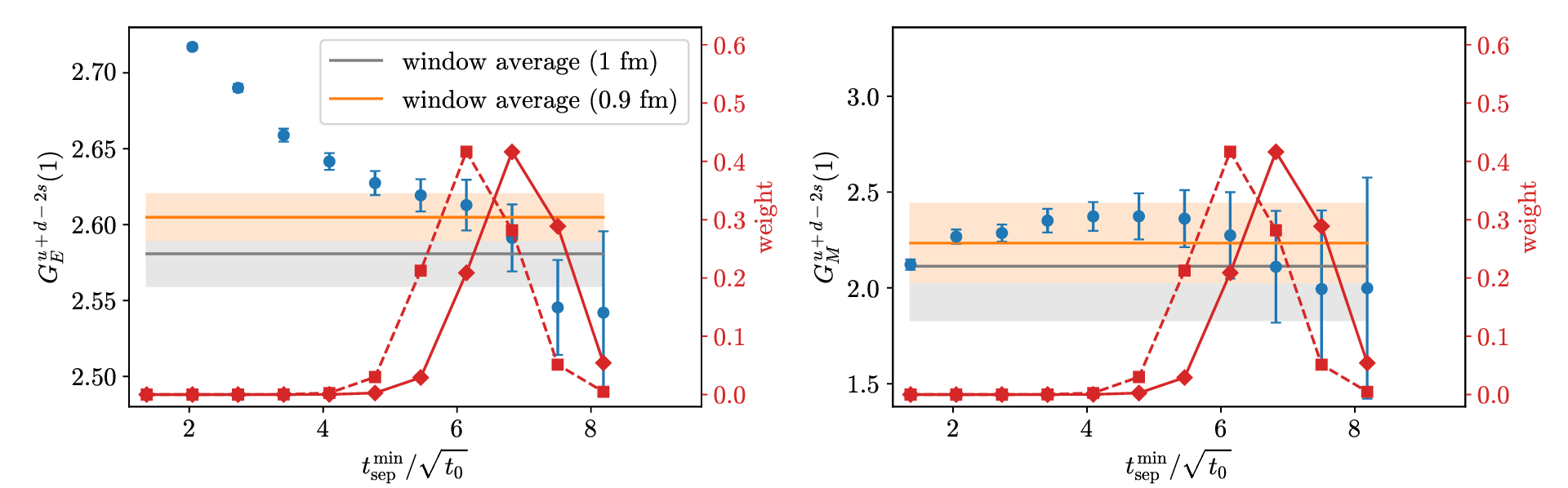}
    \caption{Extension of \cref{fig:window_average} with a different choice of window parameters.
      Upper panel: D450, lower panel: E300.
      The gray band corresponds to the result obtained with the weights shown by the diamonds connected by the solid line (centered at \qty{1}{fm}, our preferred choice).
      The orange band corresponds to the result obtained with the weights shown by the squares connected by the dashed line (centered at \qty{0.9}{fm}, as in Refs.\@ \cite{Djukanovic2022,Agadjanov2023}).}
    \label{fig:summation_stability_window_loc}
\end{figure*}

As mentioned in \cref{sec:excited_states}, the choice of the parameters defining the window average in \cref{eq:window_average} is not unique.
In \cref{fig:summation_stability_window_loc}, we compare two different options corresponding to windows centered at $t_\mathrm{sep}^\mathrm{min} \approx \qty{0.9}{fm}$ and \qty{1}{fm}, respectively.
We observe that both windows lead to compatible (in most cases actually very well compatible) results.
Furthermore, the window located at larger $t_\mathrm{sep}^\mathrm{min}$, which is the one we have chosen for our main analysis, is not only more conservative regarding the suppression of excited states, but also yields a larger error.

Another tunable parameter of the summation method is the number of timeslices $t_\mathrm{skip}$ omitted at both ends of the operator insertion time [\cf \cref{eq:summation}].
Contrary to the naive expectation, a larger value of $t_\mathrm{skip}$ actually leads to a larger excited-state contamination in the summed ratio at fixed $t_\mathrm{sep}$.
This is due to the factor $\propto \exp[-\Delta (t_\mathrm{sep} - t_\mathrm{skip})]$ in the two-state truncated version of the summation method\footnote{This expression can be obtained by summing \cref{eq:twostate_fit_eff_ff} below over $t$ according to \cref{eq:summation}.} \cite{Djukanovic2021},
\begin{align}
    S_{E, M}(Q^2; t_\mathrm{sep}) &= r_{00}(Q^2) \left[ 1 - \frac{\rho(Q^2)}{2} e^{-\Delta(Q^2) t_\mathrm{sep}} - \frac{\rho(0)}{2} e^{-\Delta(0) t_\mathrm{sep}} \right] \frac{1}{a} (t_\mathrm{sep} + a - 2t_\mathrm{skip}) \nonumber \\
    &\quad {} + \left[ r_{01}(Q^2) + r_{00}(Q^2) \frac{\rho(Q^2)}{2} \right] \frac{e^{-\Delta(Q^2) (t_\mathrm{skip} - a)} - e^{-\Delta(Q^2) (t_\mathrm{sep} - t_\mathrm{skip})}}{e^{a\Delta(Q^2)} - 1} \nonumber \\
    &\quad {} + \left[ r_{10}(Q^2) + r_{00}(Q^2) \frac{\rho(0)}{2} \right] \frac{e^{-\Delta(0) (t_\mathrm{skip} - a)} - e^{-\Delta(0) (t_\mathrm{sep} - t_\mathrm{skip})}}{e^{a\Delta(0)} - 1} \nonumber \\
    &\quad {} + r_{11}(Q^2) \frac{e^{-\Delta(Q^2) (t_\mathrm{skip} - a) - \Delta(0) (t_\mathrm{sep} - t_\mathrm{skip})} - e^{-\Delta(Q^2) (t_\mathrm{sep} - t_\mathrm{skip}) - \Delta(0) (t_\mathrm{skip} - a)}}{e^{a\Delta(Q^2)} - e^{a\Delta(0)}} + \ldots
    \label{eq:summation_NLO}
\end{align}
Here, $\Delta(Q^2)$ and $\Delta(0)$ are energy gaps, while the factors $\rho(Q^2)$ and $\rho(0)$ are defined in terms of the overlaps in the two-point functions entering the ratio \cref{eq:ratio} [for the exact definition of the latter, see below \cref{eq:twostate_fit_C2}].
$r_{00}(Q^2) = G_{E, M}(Q^2)$ is the ground-state form factor.

\begin{figure*}[htb]
    \includegraphics[width=\textwidth]{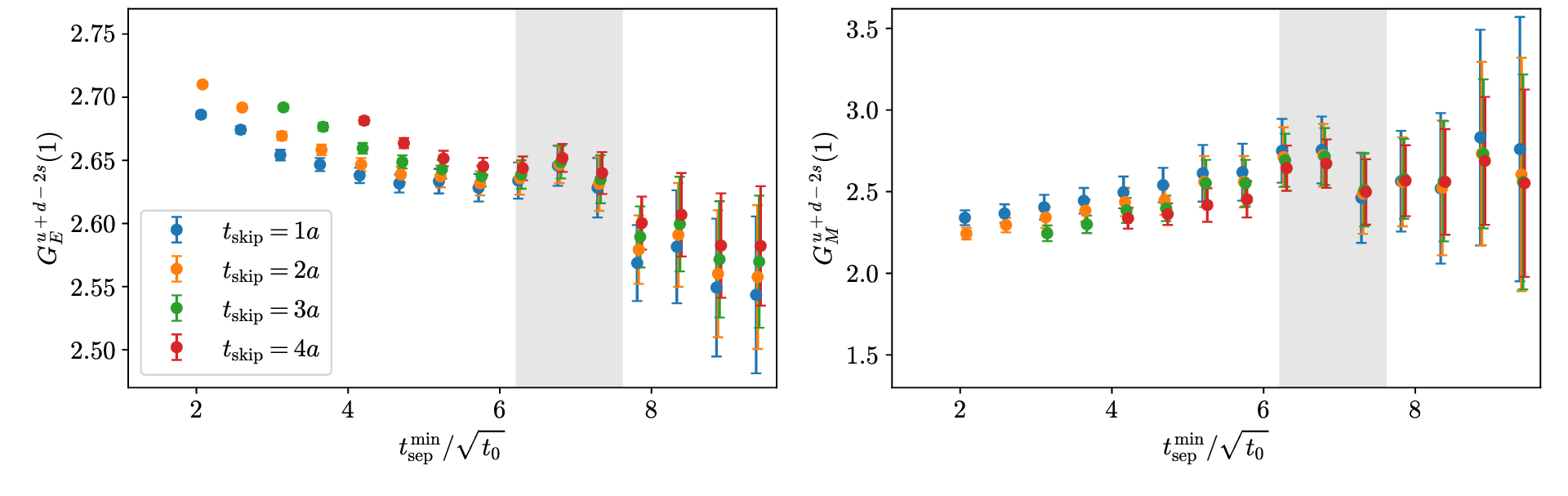}
    \includegraphics[width=\textwidth]{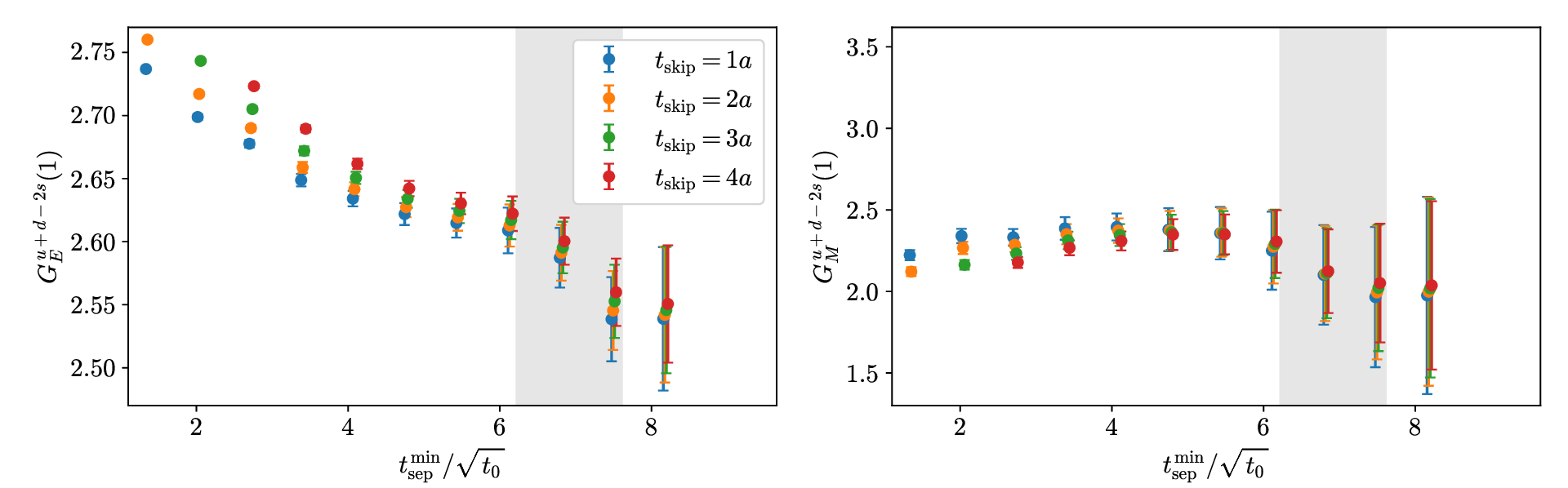}
    \caption{Extension of \cref{fig:window_average} (which only shows $t_\mathrm{skip} = 2a$) to other values of $t_\mathrm{skip}$.
      Upper panel: D450, lower panel: E300.
      The vertical gray bands indicate the interval $t_\mathrm{sep}^\mathrm{min} \in [t_w^\mathrm{low}, t_w^\mathrm{up}]$ in which the window of \cref{eq:window_average} has its largest support.}
    \label{fig:summation_stability_t_sep_min_t_skip}
\end{figure*}

If the limit $t_\mathrm{sep} \to \infty$ is taken at fixed $t_\mathrm{skip}$, all extractions should, independently of $t_\mathrm{skip}$, converge to the same ground-state form factor.
Both aforementioned trends are actually observed in stability plots for the results of the summation method as a function of $t_\mathrm{sep}^\mathrm{min}$ and $t_\mathrm{skip}$, which can be found in \cref{fig:summation_stability_t_sep_min_t_skip}.
In the region where the window of \cref{eq:window_average} has its largest support (indicated in gray in \cref{fig:summation_stability_t_sep_min_t_skip}), all (not unreasonably large) values of $t_\mathrm{skip}$ already agree well.
Therefore, using only one value of $t_\mathrm{skip}$ is perfectly adequate to obtain a reliable estimate of the ground-state form factors and their uncertainties.
It can also be seen from \cref{fig:summation_stability_t_sep_min_t_skip} that the errors tend to get smaller with rising $t_\mathrm{skip}$ because less of the somewhat more noisy data at the borders of the insertion time are employed to build the summed ratios.
Consequently, $t_\mathrm{skip} = 2a$ appears to be a good compromise between not increasing the excited-state contamination in the summed ratio due to the effect mentioned before and excluding some of the potentially slightly less reliable data close to the source or sink.
This is the value we have employed for our main analysis.

\subsection{Two-state fits}
Even if the results of the summation method with the window average tuned as explained above appear to yield very conservative error estimates, a comparison with a completely different approach is desirable in order to exclude any systematic bias introduced by relying exclusively on the summation method.
Therefore, we perform two-state fits to the effective form factors according to
\begin{align}
    G_{E, M}^\mathrm{eff}(Q^2; t_\mathrm{sep}, t) &= r_{00}(Q^2) \begin{aligned}[t]
        &\left\{ 1 + \frac{\rho(Q^2)}{2} \left[ e^{-\Delta(Q^2) (t_\mathrm{sep} - t)} - e^{-\Delta(Q^2) t_\mathrm{sep}} \right] \right. \\
        &\:\: \left. {} + \frac{\rho(0)}{2} \left[ e^{-\Delta(0) t} - e^{-\Delta(0) t_\mathrm{sep}} \right] \right\}
    \end{aligned} \nonumber \\
    &\quad {} + r_{01}(Q^2) e^{-\Delta(Q^2) t} + r_{10}(Q^2) e^{-\Delta(0) (t_\mathrm{sep} - t)}  + r_{11}(Q^2) e^{-\Delta(Q^2) t} e^{-\Delta(0) (t_\mathrm{sep} - t)} ,
    \label{eq:twostate_fit_eff_ff}
\end{align}
in order to extract the ground-state form factors $r_{00}(Q^2) = G_{E, M}(Q^2)$.
We fit the electric and magnetic effective form factors together, with the energy gaps $\Delta(Q^2)$, $\Delta(0)$ and the overlap factors $\rho(Q^2)$, $\rho(0)$ as common fit parameters.

To achieve stable fits, priors on the energy gaps and also on the overlap factors are required.
To determine these, we perform two-state fits to the two-point functions,
\begin{equation}
    \Braket{C_2(\mathbf{p}'; t_\mathrm{sep})} = c_0(\mathbf{p}'^2) e^{-E_0(\mathbf{p}') t_\mathrm{sep}} + c_1(\mathbf{p}'^2) e^{-E_1(\mathbf{p}') t_\mathrm{sep}} .
    \label{eq:twostate_fit_C2}
\end{equation}
From these fits, we extract the energy gaps $\Delta(Q^2) = \sum_{\tilde{\mathbf{p}} \in \mathfrak{p}} [E_1(\tilde{\mathbf{p}}^2) - E_0(\tilde{\mathbf{p}}^2)] / \sum_{\tilde{\mathbf{p}} \in \mathfrak{p}} 1$ and the overlap factors $\rho(Q^2) = \sum_{\tilde{\mathbf{p}} \in \mathfrak{p}} [c_1(\tilde{\mathbf{p}}^2)/c_0(\tilde{\mathbf{p}}^2)] / \sum_{\tilde{\mathbf{p}} \in \mathfrak{p}} 1$, \ie we average over equivalent three-momenta.
The systematic uncertainty originating from the choice of fit ranges is accounted for by averaging over all reasonable options using AIC weights [\cf \cref{eq:BAIC,eq:BAIC_weights}] and adding Gaussian noise to the Jackknife distribution according to the systematic covariance matrix [the second term in eq.~(7) in Ref.\@ \cite{Segner2023}, which is a straightforward generalization of eq.~(7) in Ref.\@ \cite{Neil2022}].
Both the two-state fits to the two-point functions and those to the effective form factors are performed using the \textsc{VarPro} method \cite{Golub1973} to eliminate the need for initial guesses for the prefactors $c_j$ and $r_{jk}$, respectively.

As the $\rho$-factors, which are defined by the overlaps in the two-point function, only enter the spectral representation of the ratio used to derive \cref{eq:twostate_fit_eff_ff} via the expansion of the two-point function, we directly take the values obtained from the fits to \cref{eq:twostate_fit_C2} as priors for them.
Here, we increase the width of the priors by multiplying the error of $\rho$ by a conservative factor of 3.

The energy gaps, on the other hand, also enter the terms in \cref{eq:twostate_fit_eff_ff} which originate from the expansion of the three-point function.
Hence, the situation is less clear than for the $\rho$-factors because the three-point function might have a stronger overlap with different excited states than the two-point function.
This is intimately connected with the issue that our calculation, as almost certainly all other current lattice calculations of nucleon matrix elements, is not in the regime where a single excited state dominates the excited-state contamination.
Since it is impossible to fit more than one excited state without putting unduly strict priors on the energy gaps (which would correspond to making an \apriori assumption about what the states are which couple most strongly, and not really letting the data decide on this), the latter must be regarded as effective gaps summarizing the contribution of several excited states.
Therefore, we employ relatively loose priors set to the range between $2M_\pi$ and the energy gap obtained from fitting the two-point function.
This comes of course at the expense of less stable fits to the effective form factors.
But we stress again that it is necessary in order to not introduce a systematic bias by assuming a particular value of the gap.

\begin{figure*}[htb]
    \includegraphics[width=\textwidth]{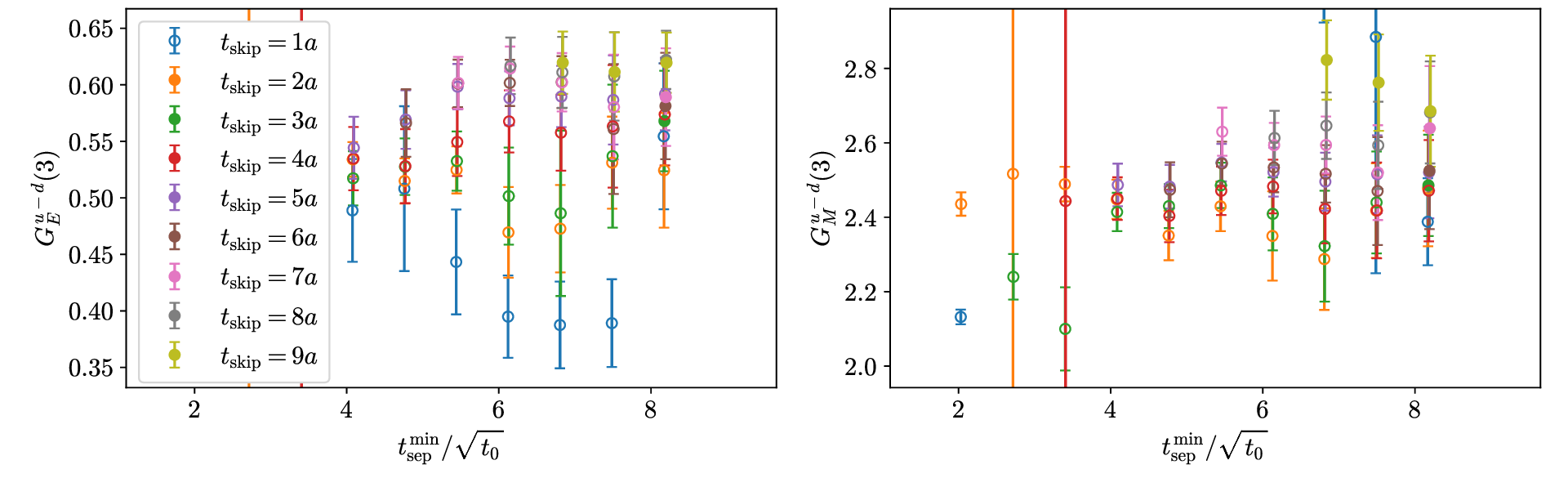}
    \hrule
    \includegraphics[width=\textwidth]{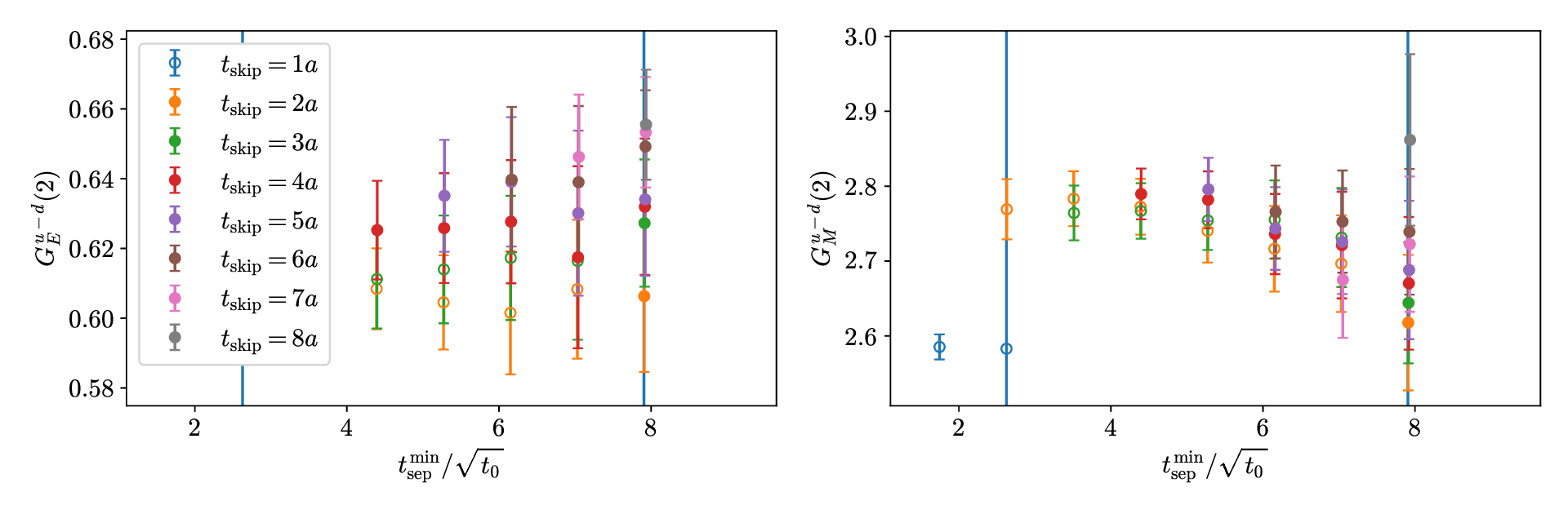}
    \hrule
    \includegraphics[width=\textwidth]{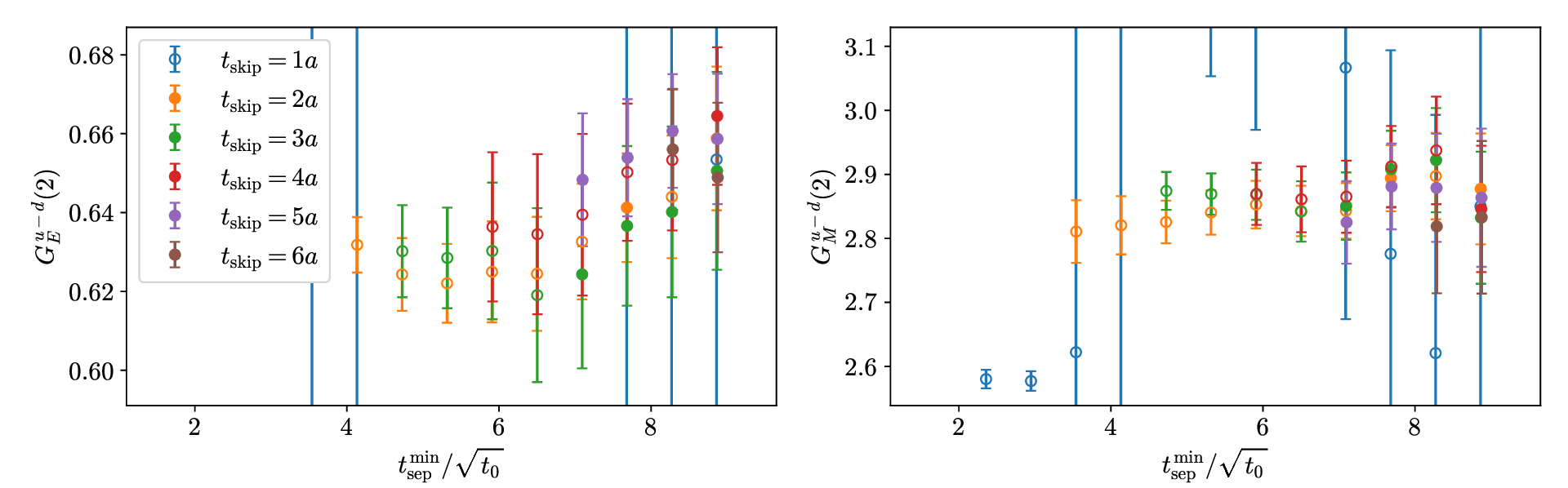}
    \caption{Isovector electromagnetic form factors at $Q^2 \approx \qty{0.2}{GeV^2}$ on the ensembles E300 (upper panel), D200 (middle panel), and C101 (lower panel) as a function of the minimal source-sink separation entering the fits to \cref{eq:twostate_fit_eff_ff} and for different numbers of timeslices skipped from the borders. Open circles refer to fits with a p-value less than \qty{5}{\percent}. Seemingly missing points lie, due to convergence issues, outside of the plotted range.}
    \label{fig:twostate_stability_t_sep_min_t_skip}
\end{figure*}

To determine the range of points which should enter the two-state fits to the effective form factors, we compare different choices of $t_\mathrm{sep}^\mathrm{min}$ and $t_\mathrm{skip}$ in \cref{fig:twostate_stability_t_sep_min_t_skip}.
We find that both parameters need to be set to relatively large values in order to obtain stable fit results and p-values which are acceptable at least in the majority of cases.
Our final choices are $t_\mathrm{sep}^\mathrm{min} \gtrsim 6.9\sqrt{t_0}$, which corresponds to the peak of the window used in the summation method, and $t_\mathrm{skip} \gtrsim 2.6\sqrt{t_0} \approx \qty{0.4}{fm}$.
The latter is realized by $t_\mathrm{skip} = 8a$ on E300, $6a$ on D200, and $5a$ on C101.
We remark that these values correspond to omitting about half of the data points even at our largest source-sink separation.

\begin{figure*}[htb]
    \begin{minipage}{0.5\textwidth}
        \includegraphics[width=\textwidth]{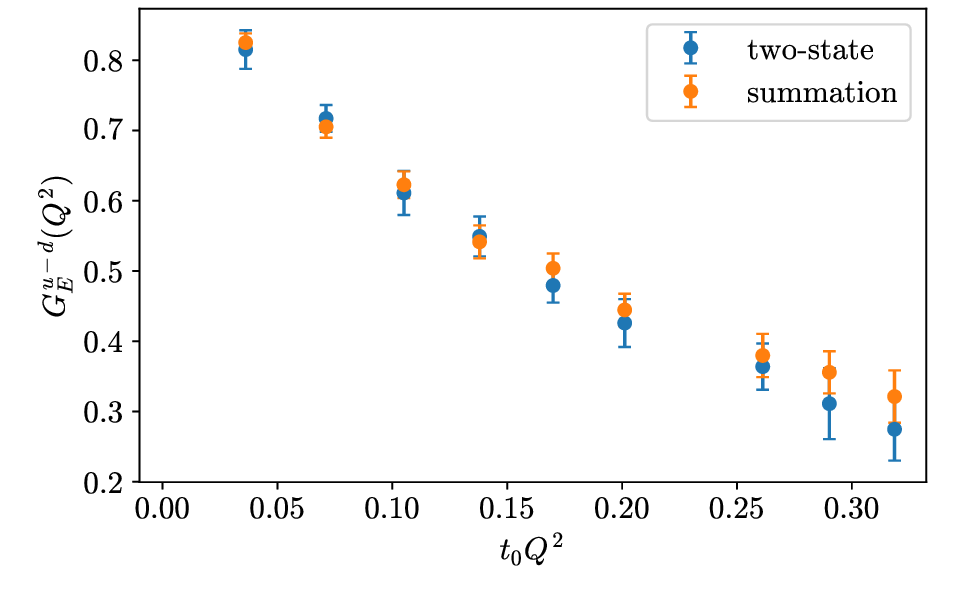}
    \end{minipage}%
    \begin{minipage}{0.5\textwidth}
        \includegraphics[width=\textwidth]{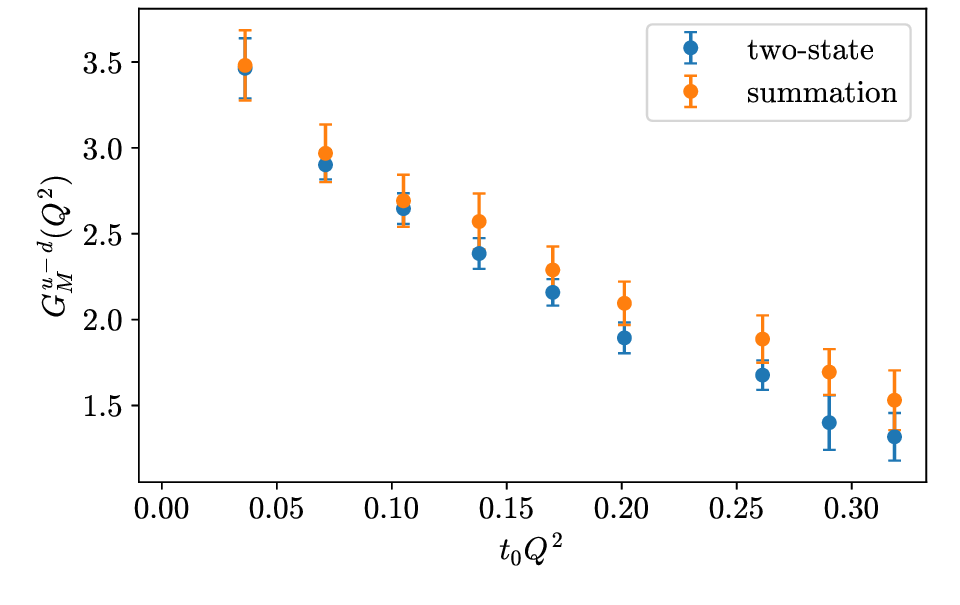}
    \end{minipage}
    \begin{minipage}{0.5\textwidth}
        \includegraphics[width=\textwidth]{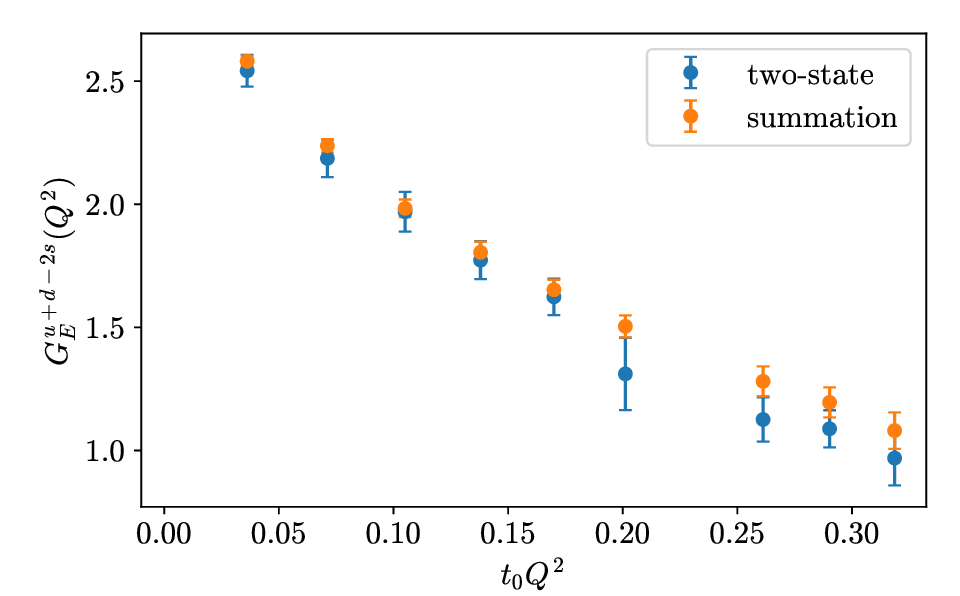}
    \end{minipage}%
    \begin{minipage}{0.5\textwidth}
        \includegraphics[width=\textwidth]{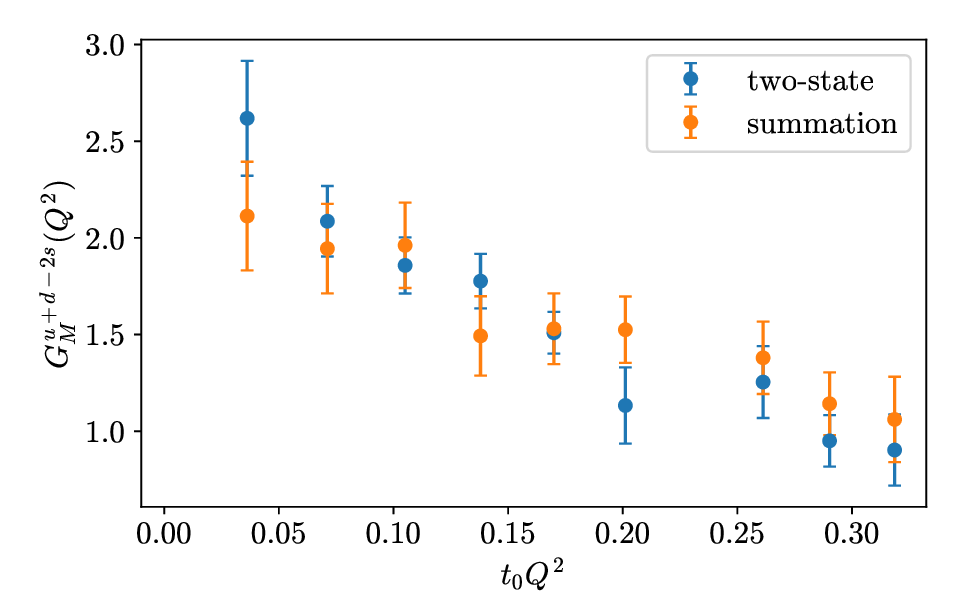}
    \end{minipage}
    \caption{Isovector and isoscalar electromagnetic form factors on the ensemble E300 as a function of $Q^2$.
      The blue points originate from two-state fits to the effective form factors according to \cref{eq:twostate_fit_eff_ff}, while the orange ones have been obtained from the summation method using the window average.}
    \label{fig:E300_twostate_vs_summation}
\end{figure*}

\begin{figure*}[htb]
    \begin{minipage}{0.5\textwidth}
        \includegraphics[width=\textwidth]{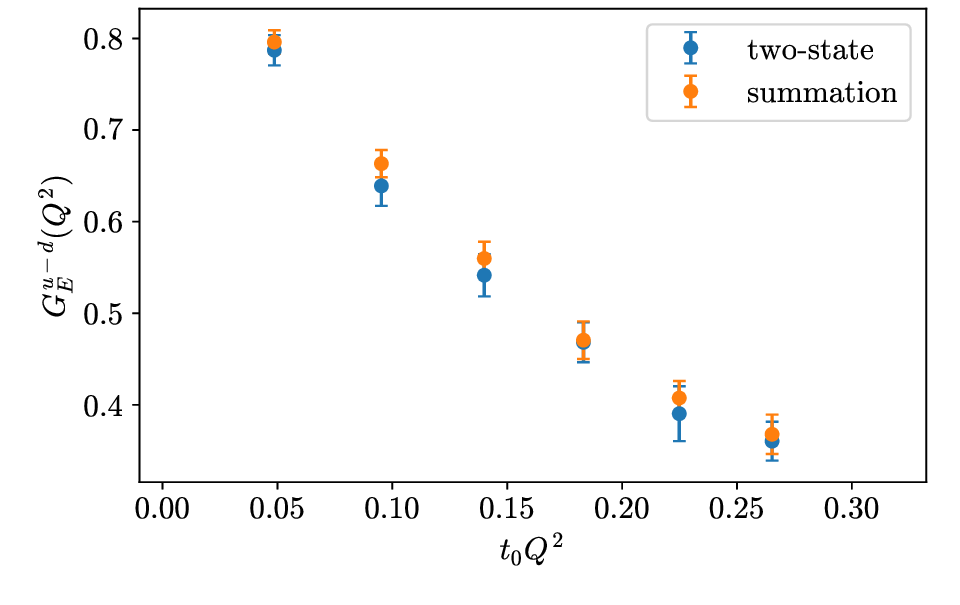}
    \end{minipage}%
    \begin{minipage}{0.5\textwidth}
        \includegraphics[width=\textwidth]{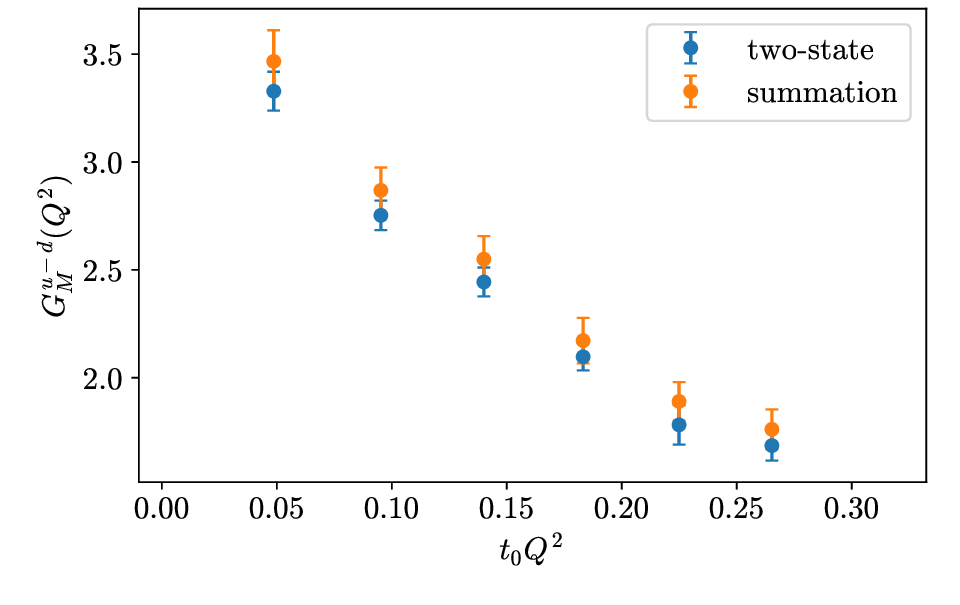}
    \end{minipage}
    \begin{minipage}{0.5\textwidth}
        \includegraphics[width=\textwidth]{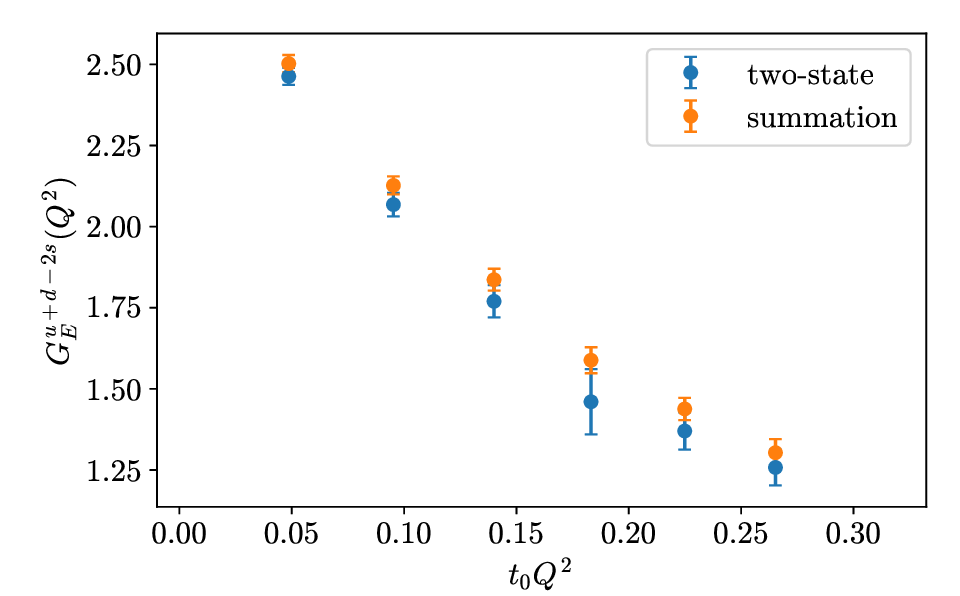}
    \end{minipage}%
    \begin{minipage}{0.5\textwidth}
        \includegraphics[width=\textwidth]{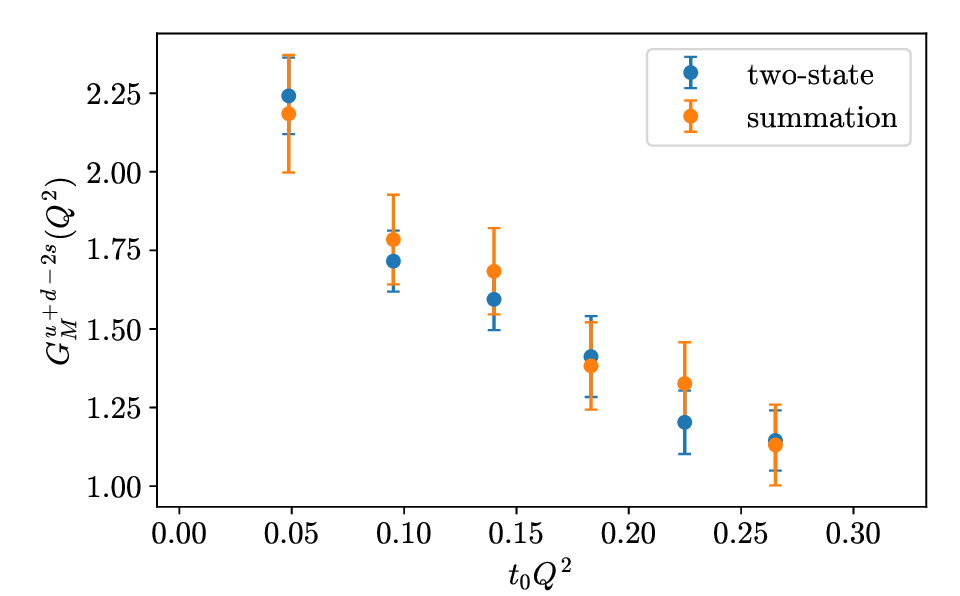}
    \end{minipage}
    \caption{Same as \cref{fig:E300_twostate_vs_summation} for ensemble D200.}
    \label{fig:D200_twostate_vs_summation}
\end{figure*}

\begin{figure*}[htb]
    \begin{minipage}{0.5\textwidth}
        \includegraphics[width=\textwidth]{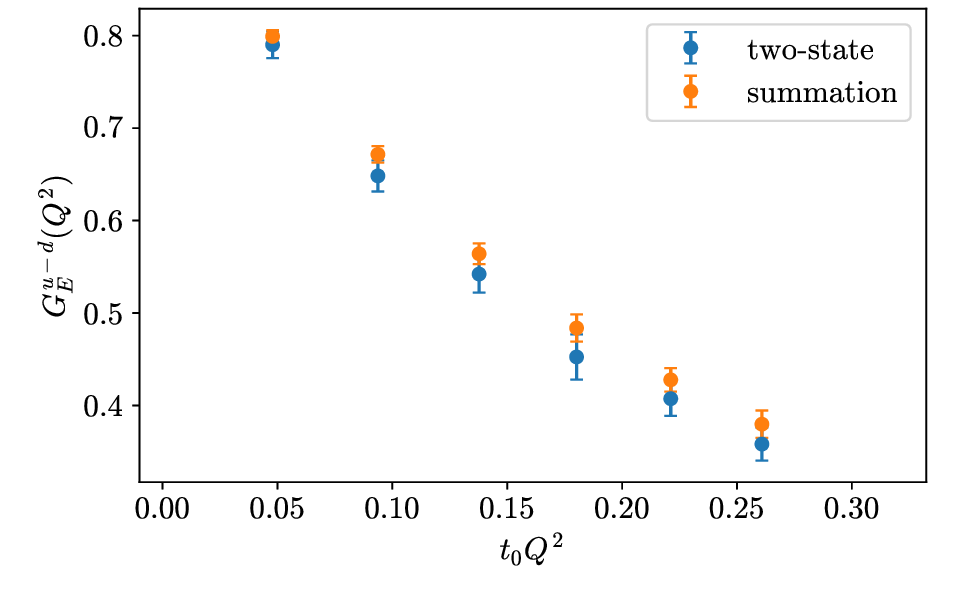}
    \end{minipage}%
    \begin{minipage}{0.5\textwidth}
        \includegraphics[width=\textwidth]{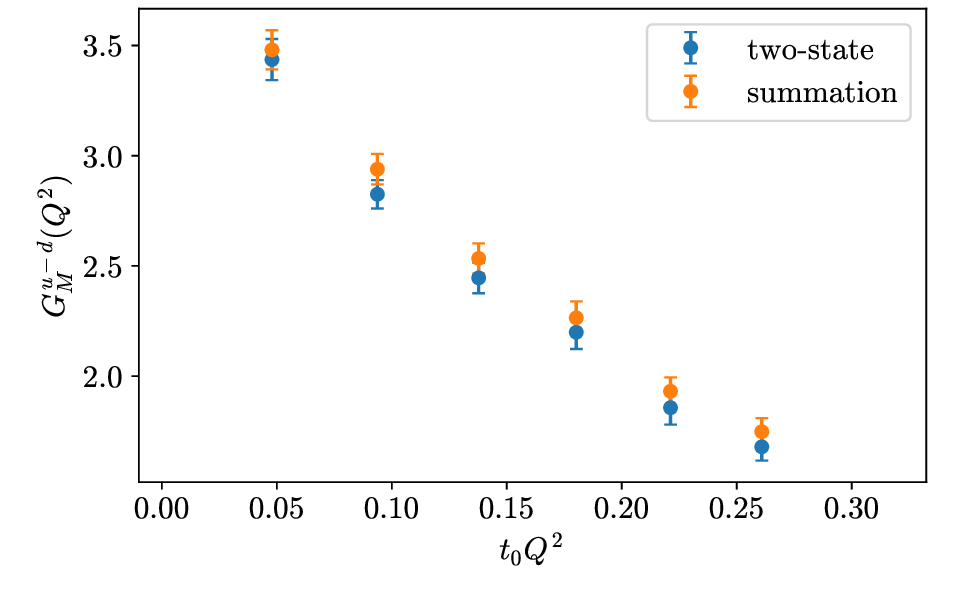}
    \end{minipage}
    \begin{minipage}{0.5\textwidth}
        \includegraphics[width=\textwidth]{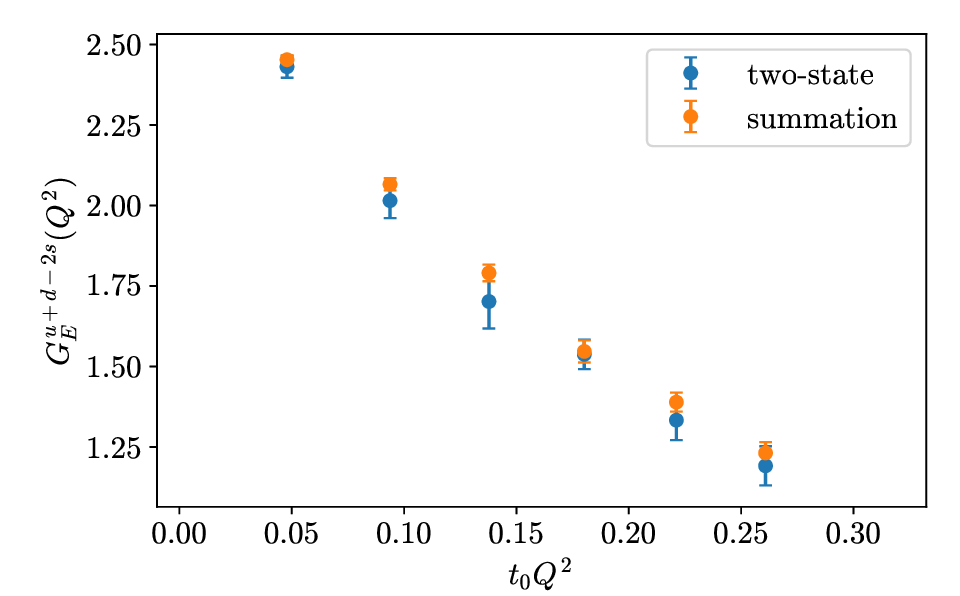}
    \end{minipage}%
    \begin{minipage}{0.5\textwidth}
        \includegraphics[width=\textwidth]{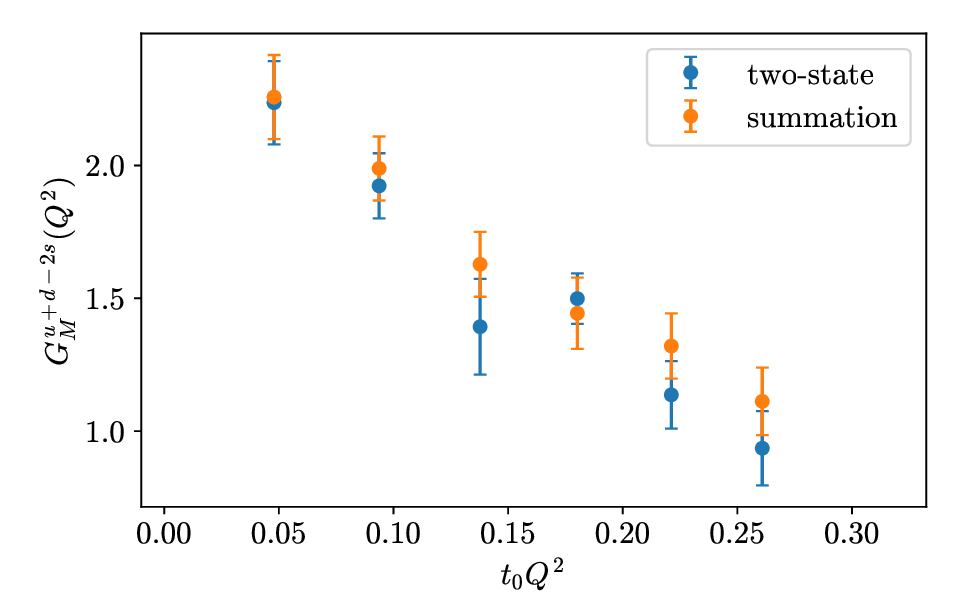}
    \end{minipage}
    \caption{Same as \cref{fig:E300_twostate_vs_summation} for ensemble C101.}
    \label{fig:C101_twostate_vs_summation}
\end{figure*}

In \cref{fig:E300_twostate_vs_summation,fig:D200_twostate_vs_summation,fig:C101_twostate_vs_summation}, we show plots of the $Q^2$-dependence of $G_E$ and $G_M$ on the ensembles E300, D200, and C101, comparing the two-state and the summation method.
These plots reveal that the two-state fits in general yield smaller errors than the summation method, in particular for the magnetic form factor.
We remark that many other lattice studies of nucleon form factors have observed a similar trend in the errors \cite{Ottnad2020}.
Besides, the comparison in \cref{fig:E300_twostate_vs_summation,fig:D200_twostate_vs_summation,fig:C101_twostate_vs_summation} does not permit the conclusion that either method introduces a directed, systematic bias.

To further quantify the effect on the resulting radii of choosing either the summation method or the two-state fits for the extraction of the ground-state form factors, we have performed B$\chi$PT fits on individual ensembles (the ones mentioned above) for both datasets.
This may not be the exact same method used to obtain our final results, but it employs the same functional forms and permits a relatively straightforward comparison of the two datasets on the level of individual ensembles.
We stress that for the purpose of the subsequent comparison, we have subjected both datasets to exactly the same procedure.
While we observe some variations in the results and the correlated difference of the radii extracted either from the summation or the two-state data can be larger than $1\,\sigma$, there is absolutely no clear pattern to see.
On the contrary, the variations appear to be completely random in nature.

\begin{figure*}[htb]
    \includegraphics[width=\textwidth]{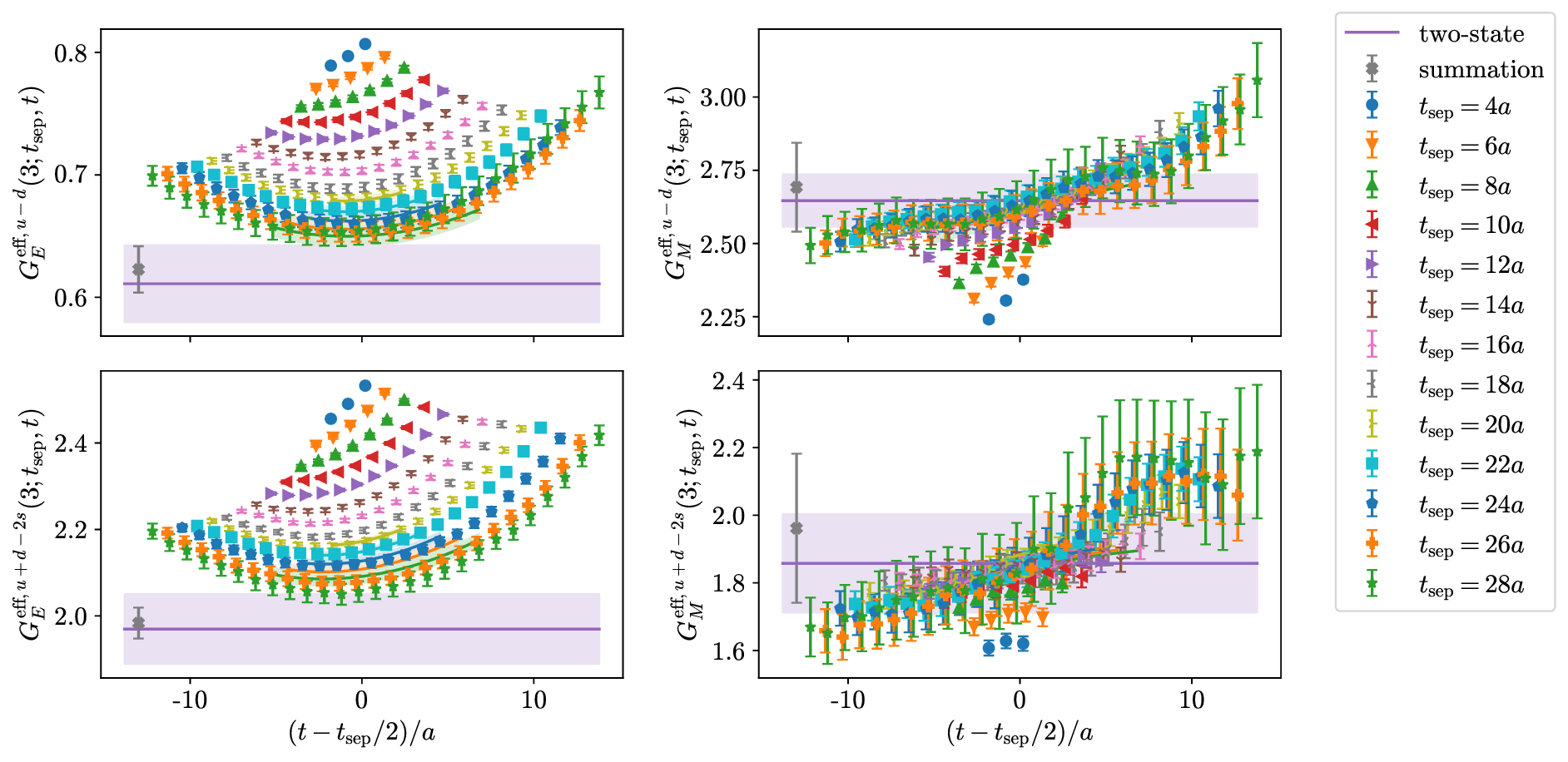}
    \caption{Isovector and isoscalar effective form factors for ensemble E300 and $Q^2 \approx \qty{0.196}{GeV^2}$.
      The data points are horizontally displaced for better visibility.
      The curves represent the two-state fits in their respective fit intervals, and the horizontal bands their extrapolation to $t_\mathrm{sep}, t \to \infty$.}
    \label{fig:eff_ff_E300_q_3}
\end{figure*}

\begin{figure*}[htb]
    \includegraphics[width=\textwidth]{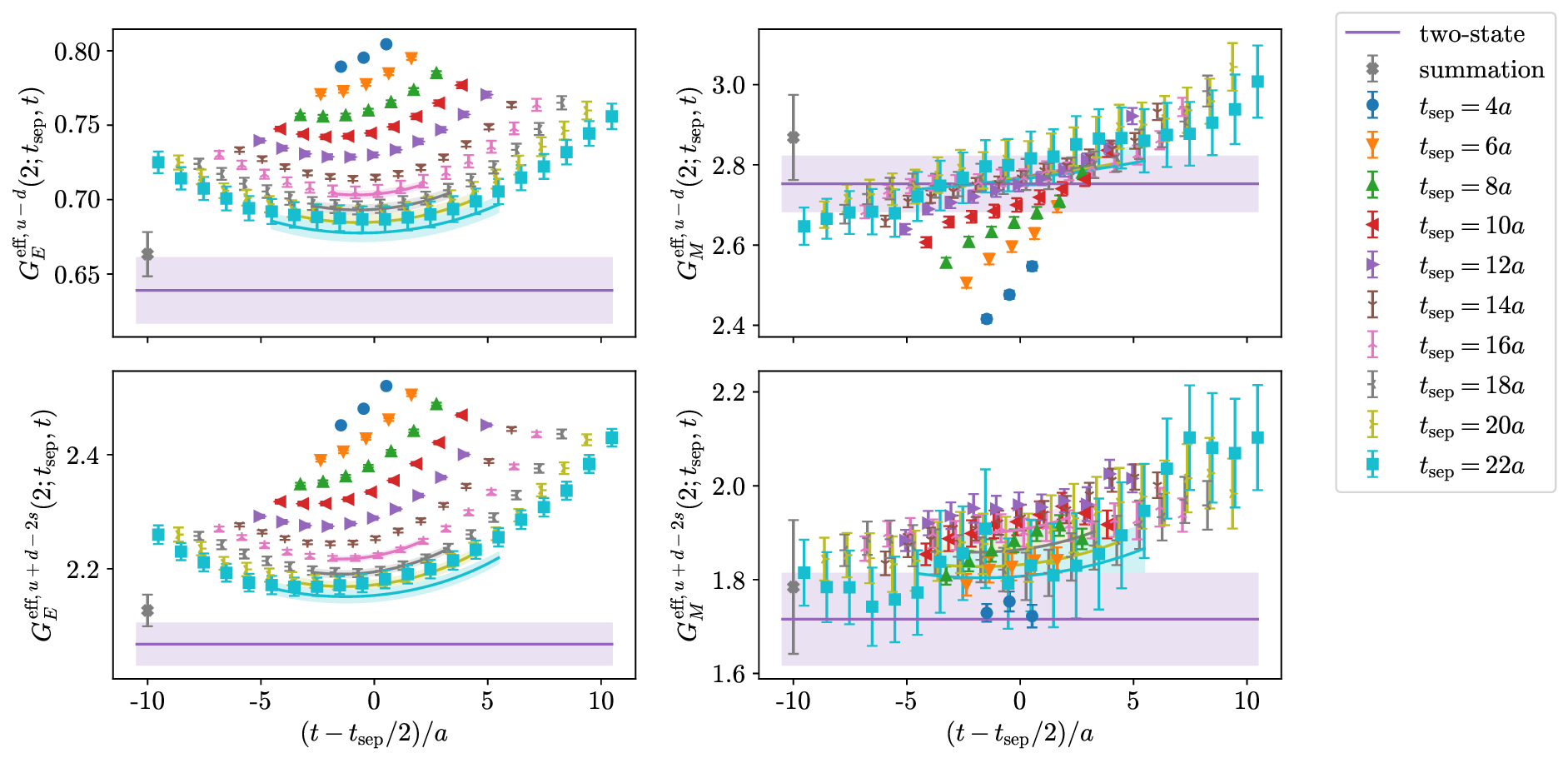}
    \caption{Same as \cref{fig:eff_ff_E300_q_3} for ensemble D200 and $Q^2 \approx \qty{0.177}{GeV^2}$.}
    \label{fig:eff_ff_D200_q_2}
\end{figure*}

\begin{figure*}[htb]
    \includegraphics[width=\textwidth]{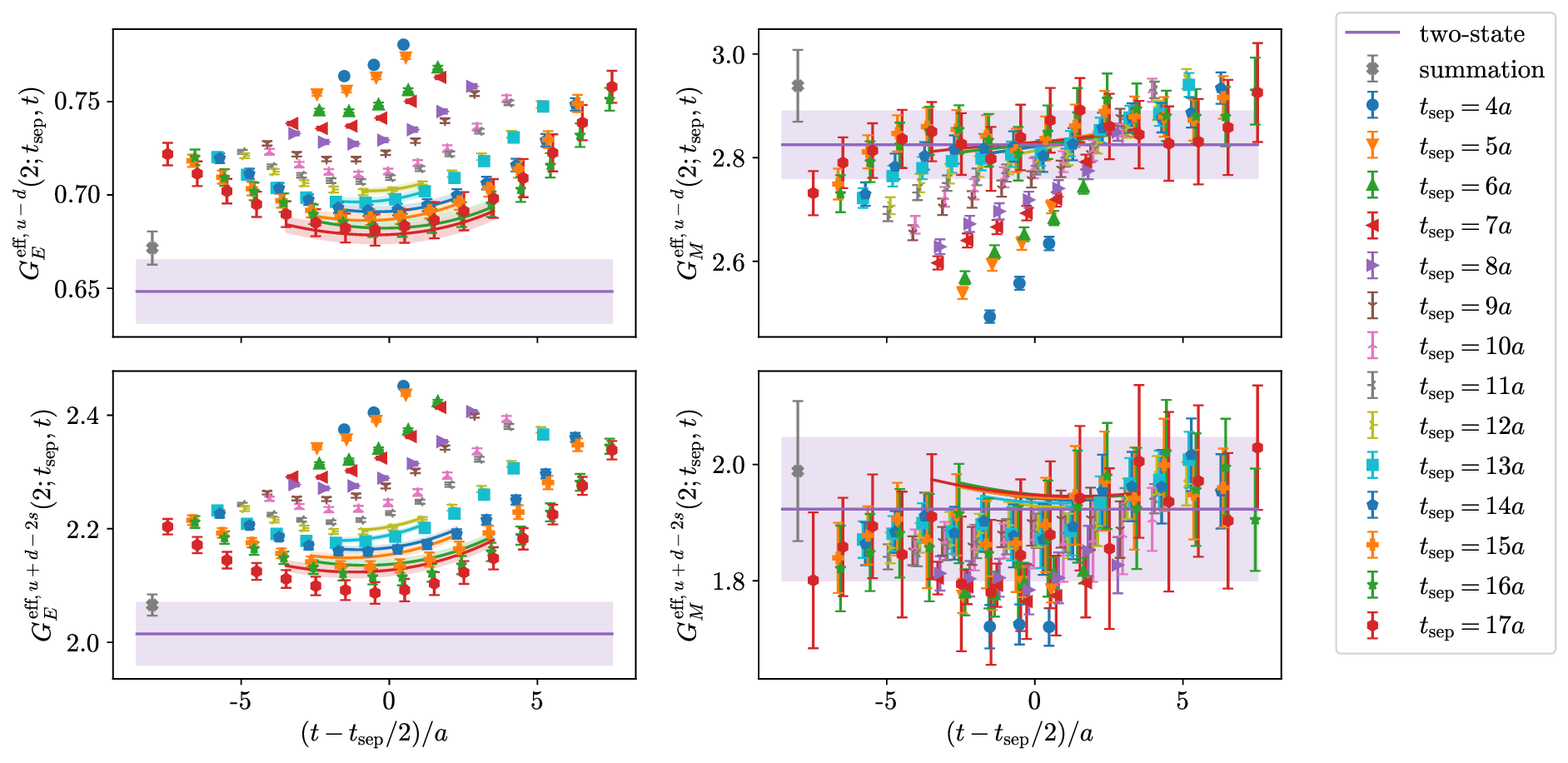}
    \caption{Same as \cref{fig:eff_ff_E300_q_3} for ensemble C101 and $Q^2 \approx \qty{0.174}{GeV^2}$.}
    \label{fig:eff_ff_C101_q_2}
\end{figure*}

Any judgment about the reliability of a method to extract the ground-state form factors should also be based on plots of the effective form factors themselves.
These can be found in \cref{fig:eff_ff_E300_q_3,fig:eff_ff_D200_q_2,fig:eff_ff_C101_q_2} for the three aforementioned ensembles and $Q^2 \approx \qty{0.2}{GeV^2}$.
They show that the two-state fits in many cases miss the data (even if the p-value is decent) and/or lead to an unrealistically large correction compared to the largest source-sink separation we have computed.
They also demonstrate that the summation method yields entirely plausible values for the ground-state form factors.

To summarize the discussion, we did not find any indication that the summation method introduces a systematic bias compared to two-state fits.
In contrast to the former, the latter require the use of priors.
We have observed that the choice of the location as well as of the width of the priors on the energy gaps strongly influences both the central values of the resulting ground-state form factors and their errors.
Furthermore, with the relatively broad priors which we have finally adopted, we observe some instabilities in the two-state fits, mostly on ensembles with less statistics than the ones shown here and at higher momenta than those included in the analysis.
The opportunity to avoid the use of priors in this very sensitive and crucial step of the analysis is our main reason for preferring the summation method.

\clearpage

\section{Form factors on E250}
\label{sec:appendix_E250}
As mentioned in the main text describing \cref{fig:bchpt_fits}, the form factors on our near-physical pion mass ensemble E250 seem to exhibit more pronounced statistical fluctuations than on other ensembles.
Here, we investigate this point in more detail.

\begin{figure*}[htb]
    \includegraphics[width=\textwidth]{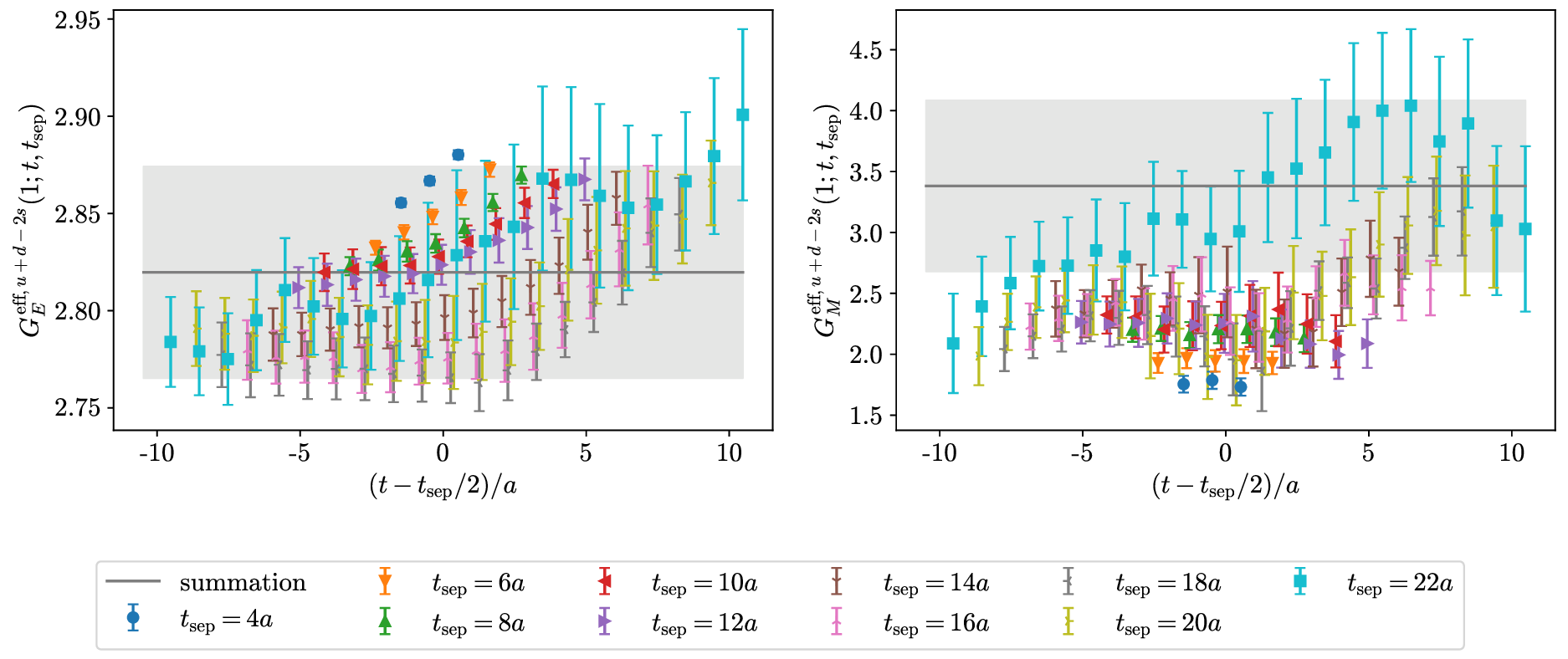}
    \caption{Isoscalar effective form factors at the first non-vanishing momentum on the ensemble E250 ($Q^2 \approx \qty{0.041}{GeV^2}$).
      The meaning of the points and bands is analogous to \cref{fig:effective_formfactors}.}
    \label{fig:effective_formfactors_E250}
\end{figure*}

In \cref{fig:effective_formfactors_E250}, the isoscalar effective form factors are shown for the first non-vanishing momentum on E250.
It is obvious that the largest two source-sink separations ($t_\mathrm{sep} = 20a$ and $22a$) represent an upwards fluctuation.
In the electric form factor, this is particularly clear because excited-state effects always have a positive sign here, so that the effective form factor is expected to monotonically decrease with rising $t_\mathrm{sep}$.
Moreover, doubling the statistics for the disconnected part from the original 398 to now 796 configurations has had (almost) no effect on the largest two source-sink separations as far as the errors are concerned, while reducing the errors of the disconnected contribution substantially for the lower values of $t_\mathrm{sep}$.
This indicates likewise that fluctuations are still dominant for the largest two source-sink separations.

\begin{figure*}[htb]
    \includegraphics[width=\textwidth]{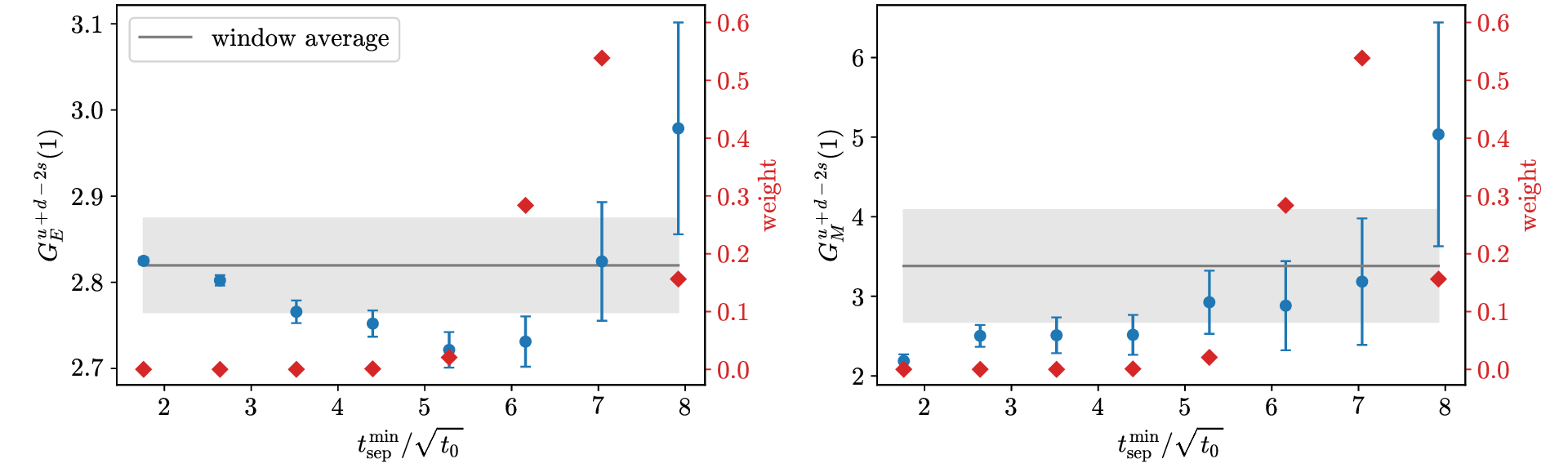}
    \caption{Isoscalar electromagnetic form factors at the first non-vanishing momentum on the ensemble E250 as a function of the minimal source-sink separation entering the summation fit.
      The meaning of the points and bands is analogous to \cref{fig:window_average}.}
    \label{fig:window_average_E250}
\end{figure*}

The upwards fluctuation is also clearly visible in the $t_\mathrm{sep}^\mathrm{min}$-plots for the form factors extracted from the summation method (\cf \cref{fig:window_average_E250}).
Here, mostly the extractions with the largest two values of $t_\mathrm{sep}^\mathrm{min}$ are affected, as in these cases the influence of the effective form factors at large $t_\mathrm{sep}$ on the summation fit becomes sizable.
Due to our choice for the window, these points have a significant effect on the averaged result as well.
This means that on E250, the window average with our values for the parameters $t_w^\mathrm{up, low}$ is not able to suppress this statistical fluctuation sufficiently.
On the other hand, excited-state effects are also expected to be stronger at lower pion masses.
Hence, it does not appear reasonable to adjust the window to lower values of $t_\mathrm{sep}^\mathrm{min}$ on E250.
Besides, the lowest point for $G_E^{u+d-2s}$ in \cref{fig:window_average_E250} is still within $2\,\sigma$ of our averaged result, so that these two values are not incompatible with each other, and our error is not grossly underestimated.

As can be seen from \cref{fig:bchpt_fits}, our direct fits are stable against such fluctuations on single ensembles:
the fit curves lie below the data at the first $\sim 6$ $Q^2$-points for the isoscalar form factors on E250, \ie the fit follows much more closely the trend determined by the other ensembles than this obvious fluctuation on E250.

\section{Form factor data}
\label{sec:appendix_formfactors}
In this appendix, we present the results of extracting the electromagnetic form factors with the summation method using the window average, as described in \cref{sec:excited_states}, for every gauge ensemble listed in \cref{tab:ensembles}.
The effective form factors of the proton and neutron are put together from the isovector and isoscalar combinations according to
\begin{equation}
    G_{E, M}^{\mathrm{eff}, p} = \frac{1}{6} \left( G_{E, M}^{\mathrm{eff}, u+d-2s} + 3G_{E, M}^{\mathrm{eff}, u-d} \right) , \quad G_{E, M}^{\mathrm{eff}, n} = \frac{1}{6} \left( G_{E, M}^{\mathrm{eff}, u+d-2s} - 3G_{E, M}^{\mathrm{eff}, u-d} \right) .
    \label{eq:eff_ff_p_n}
\end{equation}
To obtain the corresponding numbers in the tables below, we subsequently apply the summation method and the window average.
The electric form factors are normalized by $G_E(0)$, except for the neutron, where $G_E^n(0) = 0$.
To clarify the notation, we have marked the table columns to which the normalization has been applied with a hat.

\begin{table*}[htbp]
    \caption{Electric and magnetic form factors on C101.}
    \begin{ruledtabular}
\begin{tabular}{lllllllll}
            $Q^2$ [GeV${}^2$] & \multicolumn{1}{c}{$\hat{G}_E^{u-d}$} & \multicolumn{1}{c}{$G_M^{u-d}$} & \multicolumn{1}{c}{$\hat{G}_E^{u+d-2s}$} & \multicolumn{1}{c}{$G_M^{u+d-2s}$} & \multicolumn{1}{c}{$\hat{G}_E^p$} & \multicolumn{1}{c}{$G_M^p$} & \multicolumn{1}{c}{$G_E^n$} & \multicolumn{1}{c}{$G_M^n$} \\ \hline
            0.089             & 0.7991(66) & 3.480(89) & 0.8175(44) & 2.26(16) & 0.8075(45) & 2.116(61) & 0.0092(33) & -1.368(40) \\
            0.174             & 0.6715(89) & 2.939(69) & 0.6885(63) & 1.99(12) & 0.6777(60) & 1.804(48) & 0.0096(47) & -1.137(31) \\
            0.256             & 0.564(11) & 2.534(68) & 0.5968(86) & 1.63(12) & 0.5792(78) & 1.538(47) & 0.0189(58) & -0.996(32) \\
            0.335             & 0.484(15) & 2.264(75) & 0.516(11) & 1.44(13) & 0.495(11) & 1.367(52) & 0.0140(79) & -0.886(35) \\
            0.412             & 0.428(13) & 1.931(63) & 0.4630(98) & 1.32(12) & 0.4376(93) & 1.180(43) & 0.0164(65) & -0.748(31) \\
            0.485             & 0.380(15) & 1.748(62) & 0.410(11) & 1.11(13) & 0.392(11) & 1.061(43) & 0.0174(73) & -0.692(32) \\
            0.627             & 0.318(20) & 1.513(81) & 0.329(17) & 1.06(16) & 0.324(15) & 0.945(56) & 0.0050(99) & -0.566(39) \\
            0.695             & 0.276(21) & 1.470(82) & 0.301(17) & 0.89(15) & 0.292(16) & 0.887(54) & 0.011(10) & -0.584(40) \\
            0.761             & 0.245(24) & 1.35(10) & 0.262(21) & 0.93(21) & 0.254(19) & 0.826(70) & 0.003(11) & -0.517(54) \\
            0.825             & 0.234(25) & 1.275(98) & 0.231(24) & 0.70(22) & 0.232(21) & 0.754(70) & -0.001(12) & -0.523(53) \\
            0.888             & 0.258(49) & 1.07(16) & 0.272(41) & 0.43(33) & 0.272(40) & 0.60(11) & 0.011(21) & -0.483(82)
        \end{tabular}
    \end{ruledtabular}
\end{table*}

\begin{table*}[htbp]
    \caption{Electric and magnetic form factors on N101.}
    \begin{ruledtabular}
        \begin{tabular}{lllllllll}
            $Q^2$ [GeV${}^2$] & \multicolumn{1}{c}{$\hat{G}_E^{u-d}$} & \multicolumn{1}{c}{$G_M^{u-d}$} & \multicolumn{1}{c}{$\hat{G}_E^{u+d-2s}$} & \multicolumn{1}{c}{$G_M^{u+d-2s}$} & \multicolumn{1}{c}{$\hat{G}_E^p$} & \multicolumn{1}{c}{$G_M^p$} & \multicolumn{1}{c}{$G_E^n$} & \multicolumn{1}{c}{$G_M^n$} \\ \hline
            0.089             & 0.8090(52) & 3.583(70) & 0.8372(29) & 2.42(10) & 0.8227(36) & 2.189(46) & 0.0128(24) & -1.405(32) \\
            0.175             & 0.6753(63) & 3.065(55) & 0.7129(34) & 1.997(82) & 0.6940(43) & 1.866(36) & 0.0174(29) & -1.199(25) \\
            0.258             & 0.5763(78) & 2.657(51) & 0.6199(43) & 1.711(84) & 0.5980(53) & 1.612(34) & 0.0206(35) & -1.045(24) \\
            0.338             & 0.4790(92) & 2.311(58) & 0.5360(60) & 1.56(11) & 0.5058(66) & 1.400(42) & 0.0263(41) & -0.908(28) \\
            0.415             & 0.4277(91) & 2.074(52) & 0.4802(59) & 1.349(77) & 0.4538(65) & 1.265(32) & 0.0259(40) & -0.806(25) \\
            0.491             & 0.381(12) & 1.882(52) & 0.4385(65) & 1.153(75) & 0.4102(81) & 1.136(32) & 0.0264(53) & -0.739(25) \\
            0.635             & 0.296(14) & 1.534(62) & 0.358(10) & 1.039(95) & 0.327(11) & 0.936(42) & 0.0321(58) & -0.593(27) \\
            0.704             & 0.272(16) & 1.456(59) & 0.3345(99) & 0.982(91) & 0.303(11) & 0.885(41) & 0.0290(67) & -0.574(25) \\
            0.772             & 0.249(17) & 1.383(77) & 0.295(13) & 1.14(12) & 0.269(13) & 0.879(50) & 0.0252(73) & -0.504(37) \\
            0.838             & 0.238(19) & 1.281(76) & 0.279(16) & 0.90(10) & 0.257(14) & 0.789(49) & 0.0262(87) & -0.495(33) \\
            0.903             & 0.239(32) & 1.30(11) & 0.265(21) & 1.05(14) & 0.257(24) & 0.828(71) & 0.021(13) & -0.470(45)
        \end{tabular}
    \end{ruledtabular}
\end{table*}

\begin{table*}[htbp]
    \caption{Electric and magnetic form factors on H105.}
    \begin{ruledtabular}
        \begin{tabular}{lllllllll}
            $Q^2$ [GeV${}^2$] & \multicolumn{1}{c}{$\hat{G}_E^{u-d}$} & \multicolumn{1}{c}{$G_M^{u-d}$} & \multicolumn{1}{c}{$\hat{G}_E^{u+d-2s}$} & \multicolumn{1}{c}{$G_M^{u+d-2s}$} & \multicolumn{1}{c}{$\hat{G}_E^p$} & \multicolumn{1}{c}{$G_M^p$} & \multicolumn{1}{c}{$G_E^n$} & \multicolumn{1}{c}{$G_M^n$} \\ \hline
            0.196             & 0.632(29) & 2.71(20) & 0.660(17) & 1.58(25) & 0.646(20) & 1.59(12) & 0.013(12) & -1.112(88) \\
            0.378             & 0.428(29) & 2.31(17) & 0.482(18) & 1.32(16) & 0.452(21) & 1.298(95) & 0.022(12) & -0.923(78) \\
            0.548             & 0.304(39) & 1.75(16) & 0.369(24) & 1.39(22) & 0.330(27) & 1.12(10) & 0.033(16) & -0.649(71) \\
            0.708             & 0.147(51) & 1.30(25) & 0.274(38) & 0.95(32) & 0.214(38) & 0.75(15) & 0.058(23) & -0.53(11) \\
            0.859             & 0.165(41) & 1.08(16) & 0.246(30) & 0.72(19) & 0.208(33) & 0.650(97) & 0.035(16) & -0.436(74) \\
            1.003             & 0.126(41) & 0.64(14) & 0.199(36) & 0.65(20) & 0.160(35) & 0.425(91) & 0.035(17) & -0.223(65) \\
            1.402             & 0.028(68) & 0.03(29) & -0.012(66) & 0.20(33) & 0.002(56) & 0.04(18) & -0.033(35) & -0.00(13) \\
            1.526             & 0.06(13) & -0.91(42) & -0.04(10) & -0.40(68) & -0.004(96) & -0.60(28) & -0.054(53) & 0.28(20) \\
            1.645             & -0.061(97) & -0.42(36) & -0.06(11) & 0.14(60) & -0.067(88) & -0.22(25) & -0.017(55) & 0.28(17)
        \end{tabular}
    \end{ruledtabular}
\end{table*}

\begin{table*}[htbp]
    \caption{Electric and magnetic form factors on D450.}
    \begin{ruledtabular}
        \begin{tabular}{lllllllll}
            $Q^2$ [GeV${}^2$] & \multicolumn{1}{c}{$\hat{G}_E^{u-d}$} & \multicolumn{1}{c}{$G_M^{u-d}$} & \multicolumn{1}{c}{$\hat{G}_E^{u+d-2s}$} & \multicolumn{1}{c}{$G_M^{u+d-2s}$} & \multicolumn{1}{c}{$\hat{G}_E^p$} & \multicolumn{1}{c}{$G_M^p$} & \multicolumn{1}{c}{$G_E^n$} & \multicolumn{1}{c}{$G_M^n$} \\ \hline
            0.065             & 0.8520(75) & 3.73(13) & 0.8767(50) & 2.62(20) & 0.8657(52) & 2.305(85) & 0.0083(37) & -1.411(58) \\
            0.127             & 0.727(10) & 3.26(11) & 0.7668(64) & 2.15(17) & 0.7489(71) & 1.973(72) & 0.0164(47) & -1.269(50) \\
            0.188             & 0.631(13) & 2.92(11) & 0.6768(85) & 1.85(16) & 0.6561(90) & 1.738(71) & 0.0192(60) & -1.153(51) \\
            0.247             & 0.566(14) & 2.69(10) & 0.628(11) & 1.92(17) & 0.597(10) & 1.667(67) & 0.0268(72) & -1.016(49) \\
            0.304             & 0.506(15) & 2.403(90) & 0.557(10) & 1.52(15) & 0.534(11) & 1.444(59) & 0.0247(66) & -0.972(44) \\
            0.360             & 0.458(17) & 2.211(93) & 0.506(11) & 1.39(15) & 0.487(12) & 1.316(63) & 0.0260(74) & -0.880(43) \\
            0.468             & 0.392(21) & 1.746(92) & 0.445(15) & 1.13(18) & 0.420(16) & 1.020(64) & 0.025(10) & -0.715(46) \\
\end{tabular}
    \end{ruledtabular}
\end{table*}

\begin{table*}[htbp]
    \caption{Electric and magnetic form factors on N451.}
    \begin{ruledtabular}
        \begin{tabular}{lllllllll}
            $Q^2$ [GeV${}^2$] & \multicolumn{1}{c}{$\hat{G}_E^{u-d}$} & \multicolumn{1}{c}{$G_M^{u-d}$} & \multicolumn{1}{c}{$\hat{G}_E^{u+d-2s}$} & \multicolumn{1}{c}{$G_M^{u+d-2s}$} & \multicolumn{1}{c}{$\hat{G}_E^p$} & \multicolumn{1}{c}{$G_M^p$} & \multicolumn{1}{c}{$G_E^n$} & \multicolumn{1}{c}{$G_M^n$} \\ \hline
            0.114             & 0.7753(62) & 3.509(66) & 0.7930(40) & 2.461(90) & 0.7843(46) & 2.169(41) & 0.0089(26) & -1.336(30) \\
            0.222             & 0.6211(78) & 2.939(55) & 0.6520(45) & 2.034(76) & 0.6362(53) & 1.813(35) & 0.0157(35) & -1.126(25) \\
            0.326             & 0.5147(94) & 2.508(54) & 0.5555(60) & 1.649(80) & 0.5344(64) & 1.530(34) & 0.0189(46) & -0.973(25) \\
            0.426             & 0.447(13) & 2.220(60) & 0.4911(82) & 1.540(92) & 0.4683(92) & 1.365(39) & 0.0219(54) & -0.861(29) \\
            0.522             & 0.389(12) & 1.971(55) & 0.4346(84) & 1.419(79) & 0.4102(85) & 1.223(36) & 0.0210(59) & -0.752(24) \\
            0.615             & 0.340(14) & 1.779(60) & 0.393(11) & 1.210(89) & 0.366(10) & 1.094(39) & 0.0246(69) & -0.685(27) \\
            0.792             & 0.284(23) & 1.537(91) & 0.330(17) & 1.02(11) & 0.305(18) & 0.942(56) & 0.0209(97) & -0.593(42) \\
            0.877             & 0.229(22) & 1.307(85) & 0.302(19) & 0.89(11) & 0.265(18) & 0.800(53) & 0.036(10) & -0.504(38) \\
            0.960             & 0.214(32) & 1.27(14) & 0.281(24) & 1.16(21) & 0.246(26) & 0.822(92) & 0.030(12) & -0.453(67) \\
            1.040             & 0.195(33) & 1.16(14) & 0.276(27) & 1.00(21) & 0.235(26) & 0.735(86) & 0.039(14) & -0.427(64) \\
            1.119             & 0.147(57) & 0.95(21) & 0.246(52) & 0.35(28) & 0.198(50) & 0.54(14) & 0.047(22) & -0.404(92)
        \end{tabular}
    \end{ruledtabular}
\end{table*}

\begin{table*}[htbp]
    \caption{Electric and magnetic form factors on E250.}
    \begin{ruledtabular}
        \begin{tabular}{lllllllll}
            $Q^2$ [GeV${}^2$] & \multicolumn{1}{c}{$\hat{G}_E^{u-d}$} & \multicolumn{1}{c}{$G_M^{u-d}$} & \multicolumn{1}{c}{$\hat{G}_E^{u+d-2s}$} & \multicolumn{1}{c}{$G_M^{u+d-2s}$} & \multicolumn{1}{c}{$\hat{G}_E^p$} & \multicolumn{1}{c}{$G_M^p$} & \multicolumn{1}{c}{$G_E^n$} & \multicolumn{1}{c}{$G_M^n$} \\ \hline
            0.041             & 0.894(16) & 3.96(28) & 0.939(18) & 3.38(70) & 0.919(14) & 2.51(21) & 0.022(11) & -1.41(16) \\
            0.081             & 0.818(20) & 3.67(23) & 0.869(19) & 2.58(56) & 0.844(16) & 2.36(15) & 0.030(11) & -1.35(14) \\
            0.119             & 0.742(27) & 3.24(18) & 0.807(25) & 1.87(47) & 0.771(23) & 1.94(13) & 0.029(13) & -1.31(11) \\
            0.158             & 0.666(30) & 3.30(21) & 0.742(29) & 2.17(58) & 0.694(23) & 2.02(15) & 0.027(19) & -1.27(15) \\
            0.195             & 0.612(27) & 3.04(16) & 0.665(23) & 2.36(35) & 0.633(20) & 1.91(11) & 0.016(16) & -1.125(91) \\
            0.232             & 0.559(31) & 2.58(14) & 0.639(25) & 1.76(33) & 0.592(23) & 1.582(99) & 0.029(18) & -1.015(76) \\
            0.303             & 0.505(33) & 2.30(15) & 0.498(34) & 1.46(34) & 0.497(27) & 1.36(11) & -0.005(21) & -0.925(75) \\
\end{tabular}
    \end{ruledtabular}
\end{table*}

\begin{table*}[htbp]
    \caption{Electric and magnetic form factors on D200.}
    \begin{ruledtabular}
        \begin{tabular}{lllllllll}
            $Q^2$ [GeV${}^2$] & \multicolumn{1}{c}{$\hat{G}_E^{u-d}$} & \multicolumn{1}{c}{$G_M^{u-d}$} & \multicolumn{1}{c}{$\hat{G}_E^{u+d-2s}$} & \multicolumn{1}{c}{$G_M^{u+d-2s}$} & \multicolumn{1}{c}{$\hat{G}_E^p$} & \multicolumn{1}{c}{$G_M^p$} & \multicolumn{1}{c}{$G_E^n$} & \multicolumn{1}{c}{$G_M^n$} \\ \hline
            0.091             & 0.801(13) & 3.47(14) & 0.8390(58) & 2.18(19) & 0.8201(78) & 2.094(91) & 0.0192(63) & -1.379(65) \\
            0.177             & 0.668(14) & 2.87(11) & 0.7131(80) & 1.78(14) & 0.6904(96) & 1.725(66) & 0.0221(68) & -1.147(49) \\
            0.261             & 0.563(18) & 2.55(11) & 0.616(10) & 1.68(14) & 0.590(12) & 1.551(64) & 0.0257(81) & -1.000(51) \\
            0.341             & 0.474(20) & 2.17(11) & 0.533(12) & 1.38(14) & 0.504(14) & 1.316(65) & 0.0284(96) & -0.857(49) \\
            0.419             & 0.410(18) & 1.890(90) & 0.482(11) & 1.33(13) & 0.448(12) & 1.162(56) & 0.0360(87) & -0.725(43) \\
            0.494             & 0.370(21) & 1.761(93) & 0.437(14) & 1.13(13) & 0.406(15) & 1.070(59) & 0.034(10) & -0.691(41) \\
            0.638             & 0.293(32) & 1.66(12) & 0.381(22) & 0.93(17) & 0.339(25) & 0.990(75) & 0.043(12) & -0.663(56) \\
            0.707             & 0.306(33) & 1.37(11) & 0.353(22) & 0.94(14) & 0.331(25) & 0.846(70) & 0.024(13) & -0.532(47) \\
            0.774             & 0.261(38) & 1.50(17) & 0.347(30) & 1.20(23) & 0.305(30) & 0.95(11) & 0.046(16) & -0.553(71) \\
            0.839             & 0.286(38) & 1.33(17) & 0.335(31) & 1.12(24) & 0.311(30) & 0.85(11) & 0.025(17) & -0.480(70) \\
            0.904             & 0.202(68) & 0.84(23) & 0.243(58) & 0.49(35) & 0.222(56) & 0.51(15) & 0.013(30) & -0.33(10)
        \end{tabular}
    \end{ruledtabular}
\end{table*}

\begin{table*}[htbp]
    \caption{Electric and magnetic form factors on N200.}
    \begin{ruledtabular}
        \begin{tabular}{lllllllll}
            $Q^2$ [GeV${}^2$] & \multicolumn{1}{c}{$\hat{G}_E^{u-d}$} & \multicolumn{1}{c}{$G_M^{u-d}$} & \multicolumn{1}{c}{$\hat{G}_E^{u+d-2s}$} & \multicolumn{1}{c}{$G_M^{u+d-2s}$} & \multicolumn{1}{c}{$\hat{G}_E^p$} & \multicolumn{1}{c}{$G_M^p$} & \multicolumn{1}{c}{$G_E^n$} & \multicolumn{1}{c}{$G_M^n$} \\ \hline
            0.159             & 0.731(14) & 3.12(13) & 0.7317(77) & 2.19(17) & 0.7313(97) & 1.925(83) & 0.0011(69) & -1.189(57) \\
            0.308             & 0.576(18) & 2.69(11) & 0.580(10) & 1.85(14) & 0.578(12) & 1.651(69) & 0.0025(81) & -1.035(48) \\
            0.449             & 0.493(25) & 2.23(12) & 0.493(16) & 1.40(15) & 0.493(18) & 1.345(78) & 0.001(11) & -0.880(55) \\
            0.584             & 0.350(36) & 1.67(15) & 0.386(22) & 1.20(18) & 0.369(26) & 1.034(96) & 0.021(14) & -0.632(65) \\
            0.712             & 0.303(32) & 1.52(14) & 0.326(22) & 1.18(16) & 0.315(24) & 0.959(86) & 0.014(13) & -0.564(60) \\
            0.834             & 0.352(42) & 1.63(18) & 0.331(30) & 1.09(19) & 0.342(33) & 1.00(11) & -0.009(16) & -0.635(75) \\
            1.066             & 0.279(81) & 0.83(30) & 0.206(61) & 0.62(34) & 0.248(66) & 0.51(19) & -0.026(28) & -0.33(13) \\
            1.176             & 0.297(96) & 0.98(36) & 0.291(81) & 0.54(35) & 0.293(83) & 0.57(22) & -0.010(30) & -0.42(15) \\
            1.283             & -0.07(17) & 1.24(94) & 0.17(17) & 1.9(1.1) & -0.00(14) & 0.92(62) & 0.165(77) & -0.36(36) \\
            1.387             & 0.10(18) & 0.92(84) & 0.30(19) & 1.68(99) & 0.19(17) & 0.74(56) & 0.071(60) & -0.20(31)
        \end{tabular}
    \end{ruledtabular}
\end{table*}

\begin{table*}[htbp]
    \caption{Electric and magnetic form factors on S201.}
    \begin{ruledtabular}
        \begin{tabular}{lllllllll}
            $Q^2$ [GeV${}^2$] & \multicolumn{1}{c}{$\hat{G}_E^{u-d}$} & \multicolumn{1}{c}{$G_M^{u-d}$} & \multicolumn{1}{c}{$\hat{G}_E^{u+d-2s}$} & \multicolumn{1}{c}{$G_M^{u+d-2s}$} & \multicolumn{1}{c}{$\hat{G}_E^p$} & \multicolumn{1}{c}{$G_M^p$} & \multicolumn{1}{c}{$G_E^n$} & \multicolumn{1}{c}{$G_M^n$} \\ \hline
            0.347             & 0.482(45) & 2.31(27) & 0.554(26) & 1.31(30) & 0.518(31) & 1.39(16) & 0.035(20) & -0.91(13) \\
            0.657             & 0.390(49) & 1.40(20) & 0.399(33) & 1.35(33) & 0.400(37) & 0.91(14) & 0.012(21) & -0.551(86) \\
            0.940             & 0.47(17) & 2.23(57) & 0.44(10) & 1.81(57) & 0.47(13) & 1.42(36) & -0.017(53) & -0.80(23) \\
            1.202             & 0.31(23) & 1.70(92) & 0.11(16) & 2.4(1.1) & 0.19(18) & 1.23(62) & 0.028(76) & -0.47(35)
        \end{tabular}
    \end{ruledtabular}
\end{table*}

\begin{table*}[htbp]
    \caption{Electric and magnetic form factors on E300.}
    \begin{ruledtabular}
        \begin{tabular}{lllllllll}
            $Q^2$ [GeV${}^2$] & \multicolumn{1}{c}{$\hat{G}_E^{u-d}$} & \multicolumn{1}{c}{$G_M^{u-d}$} & \multicolumn{1}{c}{$\hat{G}_E^{u+d-2s}$} & \multicolumn{1}{c}{$G_M^{u+d-2s}$} & \multicolumn{1}{c}{$\hat{G}_E^p$} & \multicolumn{1}{c}{$G_M^p$} & \multicolumn{1}{c}{$G_E^n$} & \multicolumn{1}{c}{$G_M^n$} \\ \hline
            0.067             & 0.825(13) & 3.48(20) & 0.8602(71) & 2.11(28) & 0.8422(86) & 2.09(13) & 0.0183(61) & -1.390(85) \\
            0.132             & 0.705(16) & 2.97(17) & 0.7459(89) & 1.94(23) & 0.725(10) & 1.80(11) & 0.0211(72) & -1.167(70) \\
            0.196             & 0.623(19) & 2.69(15) & 0.661(12) & 1.96(22) & 0.639(13) & 1.67(10) & 0.0197(89) & -1.023(65) \\
            0.257             & 0.541(23) & 2.57(16) & 0.602(14) & 1.49(21) & 0.571(16) & 1.54(10) & 0.029(11) & -1.036(70) \\
            0.316             & 0.504(21) & 2.29(14) & 0.551(14) & 1.53(18) & 0.527(15) & 1.399(89) & 0.0245(97) & -0.891(58) \\
            0.374             & 0.445(23) & 2.10(13) & 0.501(15) & 1.52(17) & 0.472(16) & 1.297(82) & 0.030(11) & -0.802(54) \\
            0.486             & 0.380(31) & 1.89(14) & 0.427(20) & 1.38(19) & 0.402(23) & 1.172(89) & 0.021(13) & -0.710(59) \\
\end{tabular}
    \end{ruledtabular}
\end{table*}

\begin{table*}[htbp]
    \caption{Electric and magnetic form factors on J303.}
    \begin{ruledtabular}
        \begin{tabular}{lllllllll}
            $Q^2$ [GeV${}^2$] & \multicolumn{1}{c}{$\hat{G}_E^{u-d}$} & \multicolumn{1}{c}{$G_M^{u-d}$} & \multicolumn{1}{c}{$\hat{G}_E^{u+d-2s}$} & \multicolumn{1}{c}{$G_M^{u+d-2s}$} & \multicolumn{1}{c}{$\hat{G}_E^p$} & \multicolumn{1}{c}{$G_M^p$} & \multicolumn{1}{c}{$G_E^n$} & \multicolumn{1}{c}{$G_M^n$} \\ \hline
            0.149             & 0.726(15) & 3.27(14) & 0.7569(80) & 2.17(18) & 0.741(10) & 1.999(87) & 0.0149(70) & -1.279(58) \\
            0.289             & 0.554(18) & 2.53(11) & 0.596(10) & 1.68(13) & 0.575(12) & 1.547(67) & 0.0204(80) & -0.987(46) \\
            0.422             & 0.469(25) & 2.29(12) & 0.494(14) & 1.44(15) & 0.481(17) & 1.389(73) & 0.011(11) & -0.904(51) \\
            0.548             & 0.418(36) & 2.12(16) & 0.452(24) & 1.36(19) & 0.435(27) & 1.289(98) & 0.016(14) & -0.829(67) \\
            0.669             & 0.377(30) & 1.73(13) & 0.372(20) & 1.15(16) & 0.376(23) & 1.059(81) & -0.002(12) & -0.674(54) \\
            0.784             & 0.334(38) & 1.52(15) & 0.315(26) & 1.12(18) & 0.329(29) & 0.952(92) & -0.003(16) & -0.574(63) \\
            1.003             & 0.320(70) & 1.15(26) & 0.145(46) & 1.06(31) & 0.242(53) & 0.76(17) & -0.050(28) & -0.39(11) \\
            1.107             & 0.295(78) & 1.27(31) & 0.185(58) & 1.31(34) & 0.245(63) & 0.87(20) & -0.016(28) & -0.41(12) \\
            1.208             & 0.34(13) & 1.02(50) & 0.32(11) & 1.10(56) & 0.33(11) & 0.69(32) & -0.010(39) & -0.32(21) \\
            1.306             & 0.09(11) & 0.57(43) & 0.028(76) & 1.11(50) & 0.057(83) & 0.47(27) & 0.010(41) & -0.11(18) \\
            1.401             & 0.66(54) & 1.5(1.8) & 0.27(38) & 2.6(2.1) & 0.57(48) & 1.3(1.3) & -0.08(13) & 0.09(52)
        \end{tabular}
    \end{ruledtabular}
\end{table*}

\clearpage

\section{Direct \texorpdfstring{B$\chi$PT}{BChPT} fits}
\label{sec:appendix_bchpt_fits}
Here, we summarize the results for the physical values of the electromagnetic radii and the magnetic moment obtained from the direct B$\chi$PT fits as explained in \cref{sec:bchpt_fits}, applying various cuts in the pion mass ($M_\pi \leq \qty{0.23}{GeV}$ and $M_\pi \leq \qty{0.27}{GeV}$) and the momentum transfer ($Q^2 \leq \qtyrange[range-phrase = {, \ldots, }, range-units = single]{0.3}{0.6}{GeV^2}$).
The entries with and without an asterisk in the third column refer to the multiplicative model of \cref{eq:GE_mult,eq:GM_mult} and the additive model of \cref{eq:GE_add,eq:GM_add}, respectively.
All variations which are presented here are included in our model average (\cf \cref{sec:model_average}).
Note that the results for the proton and neutron have been obtained from the fits in the isovector and isoscalar channels; for more details, we refer to \cref{sec:bchpt_fits}.

\begin{longtable*}[c]{ccccllll}
    \caption{Results for the direct B$\chi$PT fits to the isovector electromagnetic form factors.} \\*
    \hline\hline \\* [-1.2em]
    $M_{\pi, \mathrm{cut}}$ [GeV] & $Q^2_\mathrm{cut}$ [GeV${}^2$] & correction             & $\langle r_E^2 \rangle$ [fm${}^2$] & \multicolumn{1}{c}{$\langle r_M^2 \rangle$ [fm${}^2$]} & \multicolumn{1}{c}{$\mu_M$} & \multicolumn{1}{c}{p-value} & \multicolumn{1}{c}{$\overline{\mathrm{BAIC}}$ weight} \\* \hline
    \endfirsthead
    \caption[]{\emph{(Continued)}} \\*
    \hline\hline \\* [-1.2em]
    $M_{\pi, \mathrm{cut}}$ [GeV] & $Q^2_\mathrm{cut}$ [GeV${}^2$] & correction             & $\langle r_E^2 \rangle$ [fm${}^2$] & \multicolumn{1}{c}{$\langle r_M^2 \rangle$ [fm${}^2$]} & \multicolumn{1}{c}{$\mu_M$} & \multicolumn{1}{c}{p-value} & \multicolumn{1}{c}{$\overline{\mathrm{BAIC}}$ weight} \\* \hline
    \endhead
    \hline
    \endfoot
    \hline\hline
    \endlastfoot
    0.23                          & 0.3                            & --                     & 0.793(19) & 0.651(13) & 4.610(76) & \num{0.188} & \num{0.0463} \\
    0.23                          & 0.3                            & $a^2$                  & 0.758(28) & 0.653(18) & 4.58(12) & \num{0.214} & \num{0.0319} \\
    0.23                          & 0.3                            & ${}^*a^2$              & 0.742(36) & 0.656(15) & 4.51(15) & \num{0.274} & \num{0.0663} \\
    0.23                          & 0.3                            & $M_\pi L$              & 0.791(20) & 0.654(15) & 4.644(76) & \num{0.165} & \num{0.00610} \\
    0.23                          & 0.3                            & ${}^*M_\pi L$          & 0.795(21) & 0.654(16) & 4.640(76) & \num{0.167} & \num{0.00798} \\
    0.23                          & 0.3                            & $a^2, M_\pi L$         & 0.752(30) & 0.657(22) & 4.61(12) & \num{0.199} & \num{0.00400} \\
    0.23                          & 0.3                            & ${}^*a^2, {}^*M_\pi L$ & 0.737(40) & 0.661(19) & 4.53(15) & \num{0.252} & \num{0.00855} \\
    [\defaultaddspace]
    0.23                          & 0.4                            & --                     & 0.803(16) & 0.6558(93) & 4.625(70) & \num{0.172} & \num{0.0247} \\
\end{longtable*}

\begin{longtable*}[c]{ccccclll}
    \caption{Results for the direct B$\chi$PT fits to the isoscalar electromagnetic form factors.} \\*
    \hline\hline \\* [-1.2em]
    $M_{\pi, \mathrm{cut}}$ [GeV] & $Q^2_\mathrm{cut}$ [GeV${}^2$] & correction             & \multicolumn{1}{c}{$\langle r_E^2 \rangle$ [fm${}^2$]} & $\langle r_M^2 \rangle$ [fm${}^2$] & \multicolumn{1}{c}{$\mu_M$} & \multicolumn{1}{c}{p-value} & \multicolumn{1}{c}{$\overline{\mathrm{BAIC}}$ weight} \\* \hline
    \endfirsthead
    \caption[]{\emph{(Continued)}} \\*
    \hline\hline \\* [-1.2em]
    $M_{\pi, \mathrm{cut}}$ [GeV] & $Q^2_\mathrm{cut}$ [GeV${}^2$] & correction             & \multicolumn{1}{c}{$\langle r_E^2 \rangle$ [fm${}^2$]} & $\langle r_M^2 \rangle$ [fm${}^2$] & \multicolumn{1}{c}{$\mu_M$} & \multicolumn{1}{c}{p-value} & \multicolumn{1}{c}{$\overline{\mathrm{BAIC}}$ weight} \\* \hline
    \endhead
    \hline
    \endfoot
    \hline\hline
    \endlastfoot
    0.23                          & 0.3                            & --                     & 0.561(19) & 0.656(32) & 2.42(11) & \num{0.140} & \num{0.0397} \\
    0.23                          & 0.3                            & $a^2$                  & 0.550(20) & 0.615(42) & 2.58(16) & \num{0.183} & \num{0.0133} \\
    0.23                          & 0.3                            & ${}^*a^2$              & 0.537(22) & 0.592(35) & 2.72(18) & \num{0.198} & \num{0.0181} \\
    0.23                          & 0.3                            & $M_\pi L$              & 0.558(19) & 0.666(34) & 2.42(11) & \num{0.143} & \num{0.00680} \\
    0.23                          & 0.3                            & ${}^*M_\pi L$          & 0.558(21) & 0.656(36) & 2.42(11) & \num{0.120} & \num{0.00451} \\
    0.23                          & 0.3                            & $a^2, M_\pi L$         & 0.546(21) & 0.625(47) & 2.57(16) & \num{0.187} & \num{0.00215} \\
    0.23                          & 0.3                            & ${}^*a^2, {}^*M_\pi L$ & 0.530(25) & 0.588(40) & 2.73(18) & \num{0.180} & \num{0.00211} \\
    [\defaultaddspace]
    0.23                          & 0.4                            & --                     & 0.571(18) & 0.675(26) & 2.452(94) & \num{0.168} & \num{0.0234} \\
\end{longtable*}

\begin{longtable*}[c]{cccllll}
    \caption{Results for the proton from the direct B$\chi$PT fits.} \\*
    \hline\hline \\* [-1.2em]
    $M_{\pi, \mathrm{cut}}$ [GeV] & $Q^2_\mathrm{cut}$ [GeV${}^2$] & correction             & \multicolumn{1}{c}{$\langle r_E^2 \rangle$ [fm${}^2$]} & \multicolumn{1}{c}{$\langle r_M^2 \rangle$ [fm${}^2$]} & \multicolumn{1}{c}{$\mu_M$} & \multicolumn{1}{c}{$\overline{\mathrm{BAIC}}$ weight} \\* \hline
    \endfirsthead
    \caption[]{\emph{(Continued)}} \\*
    \hline\hline \\* [-1.2em]
    $M_{\pi, \mathrm{cut}}$ [GeV] & $Q^2_\mathrm{cut}$ [GeV${}^2$] & correction             & \multicolumn{1}{c}{$\langle r_E^2 \rangle$ [fm${}^2$]} & \multicolumn{1}{c}{$\langle r_M^2 \rangle$ [fm${}^2$]} & \multicolumn{1}{c}{$\mu_M$} & \multicolumn{1}{c}{$\overline{\mathrm{BAIC}}$ weight} \\* \hline
    \endhead
    \hline
    \endfoot
    \hline\hline
    \endlastfoot
    0.23                          & 0.3                            & --                     & 0.677(15) & 0.652(13) & 2.709(52) & \num{0.0597} \\
    0.23                          & 0.3                            & $a^2$                  & 0.654(21) & 0.647(19) & 2.722(84) & \num{0.0138} \\
    0.23                          & 0.3                            & ${}^*a^2$              & 0.639(25) & 0.645(15) & 2.707(97) & \num{0.0390} \\
    0.23                          & 0.3                            & $M_\pi L$              & 0.675(16) & 0.656(15) & 2.726(52) & \num{0.00135} \\
    0.23                          & 0.3                            & ${}^*M_\pi L$          & 0.676(17) & 0.655(16) & 2.724(52) & \num{0.00117} \\
    0.23                          & 0.3                            & $a^2, M_\pi L$         & 0.649(22) & 0.652(23) & 2.734(86) & \num{0.000279} \\
    0.23                          & 0.3                            & ${}^*a^2, {}^*M_\pi L$ & 0.634(29) & 0.649(19) & 2.72(10) & \num{0.000586} \\
    [\defaultaddspace]
    0.23                          & 0.4                            & --                     & 0.687(12) & 0.6587(96) & 2.721(45) & \num{0.0198} \\
\end{longtable*}

\begin{longtable*}[c]{cccllcl}
    \caption{Results for the neutron from the direct B$\chi$PT fits.} \\*
    \hline\hline \\* [-1.2em]
    $M_{\pi, \mathrm{cut}}$ [GeV] & $Q^2_\mathrm{cut}$ [GeV${}^2$] & correction             & $\langle r_E^2 \rangle$ [fm${}^2$] & \multicolumn{1}{c}{$\langle r_M^2 \rangle$ [fm${}^2$]} & $\mu_M$ & \multicolumn{1}{c}{$\overline{\mathrm{BAIC}}$ weight} \\* \hline
    \endfirsthead
    \caption[]{\emph{(Continued)}} \\*
    \hline\hline \\* [-1.2em]
    $M_{\pi, \mathrm{cut}}$ [GeV] & $Q^2_\mathrm{cut}$ [GeV${}^2$] & correction             & $\langle r_E^2 \rangle$ [fm${}^2$] & \multicolumn{1}{c}{$\langle r_M^2 \rangle$ [fm${}^2$]} & $\mu_M$ & \multicolumn{1}{c}{$\overline{\mathrm{BAIC}}$ weight} \\* \hline
    \endhead
    \hline
    \endfoot
    \hline\hline
    \endlastfoot
    0.23                          & 0.3                            & --                     & -0.116(11) & 0.650(16) & -1.901(28) & \num{0.0597} \\
    0.23                          & 0.3                            & $a^2$                  & -0.104(13) & 0.662(20) & -1.863(41) & \num{0.0138} \\
    0.23                          & 0.3                            & ${}^*a^2$              & -0.103(15) & 0.672(18) & -1.800(55) & \num{0.0390} \\
    0.23                          & 0.3                            & $M_\pi L$              & -0.117(11) & 0.651(18) & -1.918(28) & \num{0.00135} \\
    0.23                          & 0.3                            & ${}^*M_\pi L$          & -0.118(12) & 0.654(18) & -1.916(28) & \num{0.00117} \\
    0.23                          & 0.3                            & $a^2, M_\pi L$         & -0.103(13) & 0.664(25) & -1.878(42) & \num{0.000279} \\
    0.23                          & 0.3                            & ${}^*a^2, {}^*M_\pi L$ & -0.104(17) & 0.679(23) & -1.810(58) & \num{0.000586} \\
    [\defaultaddspace]
    0.23                          & 0.4                            & --                     & -0.116(12) & 0.652(11) & -1.904(30) & \num{0.0198} \\
\end{longtable*}

\section{\texorpdfstring{$z$}{z}-expansion results}
\label{sec:appendix_zexp}
This appendix presents the results for the electromagnetic radii and the magnetic moment obtained from $z$-expansion fits on individual ensembles as detailed in \cref{sec:zexp}, applying two different cuts in the momentum transfer ($Q^2 \leq \qty{0.6}{GeV^2}$ and $Q^2 \leq \qty{0.7}{GeV^2}$).

\begin{table*}[htbp]
    \caption{Results for the $z$-expansion fits to the isovector electromagnetic form factors.}
    \begin{ruledtabular}
        \begin{tabular}{lcllll}
            Ensemble & $Q^2_\mathrm{cut}$ [GeV${}^2$] & \multicolumn{1}{c}{$\langle r_E^2 \rangle$ [fm${}^2$]} & \multicolumn{1}{c}{$\langle r_M^2 \rangle$ [fm${}^2$]} & \multicolumn{1}{c}{$\mu_M$} & \multicolumn{1}{c}{p-value} \\ \hline
            C101     & 0.6                            & 0.644(26) & 0.580(75) & 4.21(16) & \num{0.225} \\
            C101     & 0.7                            & 0.643(26) & 0.686(57) & 4.36(15) & \num{0.0358} \\
            N101     & 0.6                            & 0.597(19) & 0.546(31) & 4.385(97) & \num{0.496} \\
            N101     & 0.7                            & 0.595(17) & 0.540(28) & 4.385(96) & \num{0.672} \\
            D450     & 0.6                            & 0.687(33) & 0.564(63) & 4.20(17) & \num{0.00299} \\
            D450     & 0.7                            & 0.691(33) & 0.577(60) & 4.23(16) & \num{0.0114} \\
            N451     & 0.6                            & 0.592(19) & 0.479(50) & 4.34(13) & \num{0.363} \\
\end{tabular}
    \end{ruledtabular}
\end{table*}

\begin{table*}[htbp]
    \caption{Results for the $z$-expansion fits to the isoscalar electromagnetic form factors.}
    \begin{ruledtabular}
        \begin{tabular}{lcllcl}
            Ensemble & $Q^2_\mathrm{cut}$ [GeV${}^2$] & \multicolumn{1}{c}{$\langle r_E^2 \rangle$ [fm${}^2$]} & \multicolumn{1}{c}{$\langle r_M^2 \rangle$ [fm${}^2$]} & $\mu_M$ & \multicolumn{1}{c}{p-value} \\ \hline
            C101     & 0.6                            & 0.549(15) & 0.46(13) & 2.74(24) & \num{0.507} \\
            C101     & 0.7                            & 0.546(15) & 0.542(91) & 2.82(22) & \num{0.399} \\
            N101     & 0.6                            & 0.4861(92) & 0.517(67) & 2.94(15) & \num{0.394} \\
            N101     & 0.7                            & 0.4812(89) & 0.530(57) & 2.96(15) & \num{0.281} \\
            D450     & 0.6                            & 0.517(16) & 0.59(13) & 2.94(25) & \num{0.587} \\
            D450     & 0.7                            & 0.518(15) & 0.57(12) & 2.92(25) & \num{0.706} \\
            N451     & 0.6                            & 0.494(11) & 0.541(65) & 3.16(16) & \num{0.298} \\
\end{tabular}
    \end{ruledtabular}
\end{table*}

\begin{table*}[htbp]
    \caption{Results for the $z$-expansion fits to the proton electromagnetic form factors.}
    \begin{ruledtabular}
        \begin{tabular}{lcclll}
            Ensemble & $Q^2_\mathrm{cut}$ [GeV${}^2$] & $\langle r_E^2 \rangle$ [fm${}^2$] & \multicolumn{1}{c}{$\langle r_M^2 \rangle$ [fm${}^2$]} & \multicolumn{1}{c}{$\mu_M$} & \multicolumn{1}{c}{p-value} \\ \hline
            C101     & 0.6                            & 0.615(19) & 0.598(83) & 2.61(11) & \num{0.683} \\
            C101     & 0.7                            & 0.618(19) & 0.678(70) & 2.66(11) & \num{0.0739} \\
            N101     & 0.6                            & 0.552(14) & 0.545(36) & 2.703(65) & \num{0.181} \\
            N101     & 0.7                            & 0.552(13) & 0.543(33) & 2.701(64) & \num{0.326} \\
            D450     & 0.6                            & 0.609(23) & 0.551(78) & 2.58(12) & \num{0.00248} \\
            D450     & 0.7                            & 0.607(23) & 0.551(77) & 2.58(11) & \num{0.00774} \\
            N451     & 0.6                            & 0.565(15) & 0.502(50) & 2.699(82) & \num{0.104} \\
\end{tabular}
    \end{ruledtabular}
\end{table*}

\begin{table*}[htbp]
    \caption{Results for the $z$-expansion fits to the neutron electromagnetic form factors.}
    \begin{ruledtabular}
        \begin{tabular}{lcllll}
            Ensemble & $Q^2_\mathrm{cut}$ [GeV${}^2$] & \multicolumn{1}{c}{$\langle r_E^2 \rangle$ [fm${}^2$]} & \multicolumn{1}{c}{$\langle r_M^2 \rangle$ [fm${}^2$]} & \multicolumn{1}{c}{$\mu_M$} & \multicolumn{1}{c}{p-value} \\ \hline
            C101     & 0.6                            & -0.031(13) & 0.605(86) & -1.660(73) & \num{0.227} \\
            C101     & 0.7                            & -0.031(12) & 0.691(68) & -1.726(67) & \num{0.0664} \\
            N101     & 0.6                            & -0.0392(69) & 0.546(41) & -1.709(49) & \num{0.454} \\
            N101     & 0.7                            & -0.0371(67) & 0.551(38) & -1.717(49) & \num{0.424} \\
            D450     & 0.6                            & -0.061(14) & 0.45(10) & -1.576(76) & \num{0.581} \\
            D450     & 0.7                            & -0.063(14) & 0.473(96) & -1.577(75) & \num{0.538} \\
            N451     & 0.6                            & -0.0274(76) & 0.442(67) & -1.624(60) & \num{0.870} \\
\end{tabular}
    \end{ruledtabular}
\end{table*}

\clearpage

\section{Chiral and continuum extrapolation}
\label{sec:appendix_CCF}
Here, we summarize the results for the physical values of the electromagnetic radii and the magnetic moment obtained from the HB$\chi$PT-inspired chiral and continuum extrapolation of the $z$-expansion data as explained in \cref{sec:zexp}, applying two different cuts each in the pion mass ($M_\pi \leq \qty{0.27}{GeV}$ and $M_\pi \leq \qty{0.3}{GeV}$) and the momentum transfer ($Q^2 \leq \qty{0.6}{GeV^2}$ and $Q^2 \leq \qty{0.7}{GeV^2}$).
Ensembles which are solely used to study finite-volume effects (H105 and S201) are excluded from all fits.

\begin{table*}[htbp]
    \caption{Results of the chiral and continuum extrapolation for the $z$-expansion extractions.}
    \begin{ruledtabular}
        \begin{tabular}{lcccccl}
            Channel  & $M_{\pi, \mathrm{cut}}$ [GeV] & $Q^2_\mathrm{cut}$ [GeV${}^2$] & $\langle r_E^2 \rangle$ [fm${}^2$] & $\langle r_M^2 \rangle$ [fm${}^2$] & $\mu_M$ & \multicolumn{1}{c}{p-value} \\ \hline
            $u-d$    & 0.27                          & 0.6                            & 0.740(80) & 0.38(27) & 4.09(36) & \num{0.562} \\
            $u-d$    & 0.27                          & 0.7                            & 0.727(79) & 0.54(24) & 4.02(34) & \num{0.364} \\
            $u-d$    & 0.30                          & 0.6                            & 0.739(54) & 0.65(16) & 4.04(26) & \num{0.0573} \\
            $u-d$    & 0.30                          & 0.7                            & 0.750(53) & 0.81(14) & 4.04(25) & \num{0.0676} \\
            [\defaultaddspace]
            $u+d-2s$ & 0.27                          & 0.6                            & 0.557(31) & 0.52(34) & 2.43(46) & \num{0.0557} \\
            $u+d-2s$ & 0.27                          & 0.7                            & 0.551(29) & 0.52(29) & 2.41(45) & \num{0.247} \\
            $u+d-2s$ & 0.30                          & 0.6                            & 0.553(20) & 0.44(23) & 2.40(35) & \num{0.167} \\
            $u+d-2s$ & 0.30                          & 0.7                            & 0.550(19) & 0.47(17) & 2.39(32) & \num{0.350} \\
            [\defaultaddspace]
            $p$      & 0.27                          & 0.6                            & 0.643(56) & 0.23(32) & 2.33(24) & \num{0.414} \\
            $p$      & 0.27                          & 0.7                            & 0.625(53) & 0.37(29) & 2.34(23) & \num{0.453} \\
            $p$      & 0.30                          & 0.6                            & 0.668(37) & 0.57(19) & 2.37(17) & \num{0.131} \\
            $p$      & 0.30                          & 0.7                            & 0.667(35) & 0.70(17) & 2.38(17) & \num{0.150} \\
            [\defaultaddspace]
            $n$      & 0.27                          & 0.6                            & -0.132(37) & 0.54(33) & -1.65(16) & \num{0.687} \\
            $n$      & 0.27                          & 0.7                            & -0.125(36) & 0.80(29) & -1.65(15) & \num{0.272} \\
            $n$      & 0.30                          & 0.6                            & -0.084(24) & 0.75(20) & -1.57(12) & \num{0.0648} \\
            $n$      & 0.30                          & 0.7                            & -0.094(23) & 0.93(17) & -1.59(11) & \num{0.0634}
        \end{tabular}
    \end{ruledtabular}
\end{table*}

\end{document}